\documentclass[aps,pra,reprint,nofootinbib,groupedaddress,citeautoscript,amsmath]{revtex4-2}
\usepackage{graphicx}
\usepackage{latexsym,amssymb,amsmath,bm,bbm}
\usepackage[caption=false]{subfig}
\usepackage{xcolor}
\usepackage{mathrsfs}
\usepackage[colorlinks=true,
urlcolor=blue,
linkcolor=blue,
citecolor=blue
]{hyperref}
\usepackage{esvect}
\usepackage[normalem]{ulem}

\begin{document}
\title{Topological quantum optical states in quasiperiodic cold atomic chains}
\author{B. X. Wang}
\affiliation{Institute of Engineering Thermophysics, School of Mechanical Engineering, Shanghai Jiao Tong University, Shanghai 200240, China}
\affiliation{MOE Key Laboratory for Power Machinery and Engineering, Shanghai Jiao Tong University, Shanghai 200240, China}
\author{C. Y. Zhao}
\email{changying.zhao@sjtu.edu.cn}
\affiliation{Institute of Engineering Thermophysics, School of Mechanical Engineering, Shanghai Jiao Tong University, Shanghai 200240, China}
\affiliation{MOE Key Laboratory for Power Machinery and Engineering, Shanghai Jiao Tong University, Shanghai 200240, China}
\date{\today}

\begin{abstract}
Topological quantum optical states in one-dimensional (1D) quasiperiodic cold atomic chains are studied in this work. We propose that by introducing incommensurate modulations on the interatomic distances of 1D periodic atomic chains, the off-diagonal Aubry-Andr\'e-Harper (AAH) model can be mimicked, although the crucial difference is the existence of long-range dipole-dipole interactions. The discrete band structures with respect to the modulation phase, which plays the role of a dimension extension parameter, are calculated for finite chains beyond the nearest-neighbor approximation. It is found that the present system indeed supports nontrivial topological states localized over the boundaries. Despite the presence of long-range dipole-dipole interactions that leads to an asymmetric band structure, it is demonstrated that this system inherits the topological properties of two-dimensional integer quantum Hall systems. The spectral position, for both real and imaginary frequencies, and number of these topologically protected edge states are still governed by the gap-labeling theorem and characterized by the topological invariant, namely, the (first) Chern number, indicating the validity of bulk-boundary correspondence.  Due to the fractal spectrum arising from the quasiperiodicity in a substantially wide range of system parameters, our system provides a large number of topological gaps and optical states readily for practical use. It is also revealed that a substantial proportion of the topological edge states are highly subradiant with extremely low decay rates, which therefore offer an appealing route for controlling the emission of external quantum emitters and achieving high-fidelity quantum state storage.
\end{abstract}

\maketitle
\section{Introduction} 
Topological phases of matter have received a great deal of attention in recent years, since they can support unidirectional edge states that are immune against backscattering from disorder and imperfections. They have been demonstrated for electronic \cite{hasanRMP2010}, optical \cite{ozawa2018topological}, acoustic \cite{heNaturephys2016}, cold atomic \cite{atalaNaturephys2013} and mechanical \cite{susstrunkScience2015} systems. Among them, topological photonics is one of the most fast-growing subfields in achieving the analog of topological phases of matter \cite{khanikaevNaturemat2013,luNPhoton2014,khanikaevNPhoton2017}. Specially designed topological photonic systems are able to create topologically protected optical states \cite{luNPhoton2014,khanikaevNPhoton2017,wangJAP2020}, which show promising applications in novel photonic devices, such as unidirectional waveguides \cite{poliNComms2015}, optical isolators \cite{el-GanainyOL2015,KarkiPRApplied2019}, topological lasers \cite{partoPRL2018,zengNature2020,shaoNaturenano2020} and topological sensors \cite{wangPRM2020}, etc. 

Conventionally, topological optical states are created by using dielectric and metallic materials with artificially carved micro/nanostructures \cite{ozawa2018topological}. In the meanwhile, topological optical states can be engineered in ultracold atom settings by utilizing the state-of-art versatile control of light-atom interactions \cite{cohentannoudji2011advances}, especially using the cooperative optical states in cold atom arrays \cite{rui2020asubradiant,bettlesPRL2016,yelinPRL2017,wangOE2017}. Current laser cooling and trapping technologies allow the creation of almost arbitrary geometries of atom arrays in a relatively large scale \cite{roatiNature2008,schreiberScience2015,bordiaNaturephys2017}. On this basis, topological optical states in cold atomic arrays loaded in optical lattices have been theoretically proposed recently \cite{perczelPRA2017,bettlesPRA2017,yelinPRL20172,wang2018topological}.  The advantage is that the quantum nature of these topological optical states is promising for high-fidelity quantum state transfer and quantum information storage \cite{wangOptica2019,guimondPRL2019}. In addition, the intrinsic optical nonlinearity of ultracold atoms can be exploited to induce strong photon-photon interactions and thus provides a route for achieving many-body topological photonic states such as the fractional quantum Hall effect for photons \cite{umucalilarPRL2012,carusottoRMP2013,maghrebiPRA2015,roushanNaturephys2017}. For instance, Perczel \textit{et al.} \cite{perczelPRL2020} proposed a two-dimensional lattice of nonlinear quantum emitters/cold atoms with optical transitions embedded in a photonic crystal slab to realize fractional quantum Hall states and fractional topological insulators in a topological quantum optical setting.

By now topological optical states are mainly studied for one- (1D) and two-dimensional (2D) cold atomic arrays by mimicking the 1D Su-Schrieffer-Heeger (SSH) model \cite{wang2018topological} and 2D quantum Hall system \cite{bettlesPRA2017,yelinPRL20172}, respectively, both of which are periodic structures. On the other hand, it is known that quasicrystals, an intermediate phase with long-range order between periodic and fully disordered lattices, can also exhibit nontrivial topological properties \cite{tanesePRL2014,levy2015topological,bandresPRX2016,dareauPRL2017,babouxPRB2017,krausPRL2012,krausPRL2012b,ganeshanPRL2013,verbinPRL2013}. A paradigmatic example of quasiperiodic lattices is the Aubry-Andr\'e-Harper (AAH) model \cite{krausPRL2012,krausPRL2012b,ganeshanPRL2013,verbinPRL2013,poshakinskiyPRL2014}, which is a 1D tight-binding lattice model with on-site (namely, diagonal) or/and hopping terms (off-diagonal) being cosine modulated. When the cosine modulation is incommensurate (commensurate) with the lattice, this system becomes quasiperiodic (periodic). Due to this modulation, the AAH model interestingly possesses nontrivial topological properties that can be mapped to the 2D integer quantum Hall system (more precisely, the Harper-Hofstadter model in square lattice in the presence of a perpendicular magnetic field), without the need to apply a magnetic field \cite{ozawa2018topological,krausPRL2012}. In particular, although the system is 1D, the modulation phase shift $\phi$ plays the role of momentum (wavenumber) in a perpendicular synthetic dimension, leading to a dimensional extension \cite{krausPRL2012,krausPRL2012b,chalopinNaturephys2020}.  A notable example is the realization of the Hofstadter butterfly in the 1D AAH lattice by varying the modulation periodicity \cite{amitPRB2018,rajagopalPRL2019,niCommPhys2019}. Therefore, this model provides a playground for studying profound quantum topological phase transitions and topological states in 1D.

In this work, we show that topological optical states can be realized in 1D quasiperiodic cold atomic chains by introducing incommensurate modulations on the interatomic distances, as an extension of the off-diagonal AAH model. We calculate the discrete band structures with respect to the modulation phase $\phi$ beyond the nearest-neighbor approximation, since the Hamiltonian of the present system demonstrates substantial long-range dipole-dipole interactions, vastly different from the conventional AAH model. In spite of this significant difference, we find the present system still inherits the topological properties of 2D integer quantum Hall systems, and the spectral position (for both real and imaginary frequencies) and number of these topologically protected edge states are actually governed by the gap-labeling theorem and characterized by the topological invariant, namely, the (first) Chern number, indicating the validity of bulk-boundary correspondence. Due to the fractal nature of the spectrum, the present system provides a large number of topological gaps and optical states for practical use. Moreover, by investigating the imaginary parts (decay rates of eigenstates) of the band structure, we reveal a substantial proportion of the topological edge states are highly subradiant, which therefore provide an appealing route for controlling the emission of external quantum emitters and achieving high-fidelity quantum information storage. We expect the present theoretical proposal can offer possibilities for engineering quantum states of light and matter.

\section{Model}
Consider a 1D quasiperiodic chain composed of two-level ultracold atoms aligned along the $x$-axis. The cold atoms are assumed to be well-trapped in their positions upon interaction with photons, and the tunneling of atoms between sites is negligible \cite{perczelPRA2017,blochNaturephys2005,cohentannoudji2011advances}. The quasiperiodicity is introduced via incommensurate modulations of the spacings between cold atoms with the distance between adjacent atoms given by 
\begin{equation}\label{modulation}
x_{n+1}-x_{n}=d[1+\eta\cos(2\pi \beta n+\phi)],
\end{equation}
where $x_n$ denotes the position of the $n$-th atom, $d$ introduces the on-average interatomic distance (or the periodic lattice constant before modulation), $\eta$ determines the amplitude of the quasiperiodic distance modulation, $\beta$ is an irrational number that controls the quasiperiodicity and $\phi$ stands for the modulation phase that corresponds to the momentum (wavenumber) in a synthetic orthogonal dimension that will be discussed in detail below. 

The two-level atom, for simplicity, is assumed to have three degenerate excited states denoted by $|e_{\alpha}\rangle$ polarized along different directions, where $\alpha=x,y,z$ stands for the Cartesian coordinates, with a ground state denoted by $|g\rangle$. By applying the single excitation approximation (which is valid for sufficiently weakly excited system) \cite{kaiserJMO2011,kaiserFP2012,guerinPRL2016}, we can work in the subspace spanned by the ground states $|G\rangle\equiv|g...g\rangle$ and the single excited states $|i\rangle\equiv|g...e_i...g\rangle$ of the atoms \cite{kaiserJMO2011,kaiserFP2012,guerinPRL2016}. Moreover, by adiabatically eliminating the photonic degrees of freedom in the reservoir (i.e., the quantized electromagnetic field), we obtain the effective Hamiltonian describing light-atom interactions in the absence of any external driving field as \cite{antezzaPRL2009,antezzaPRA2009,kaiserJMO2011,kaiserFP2012,guerinPRL2016,perczelPRA2017,yelinPRL2017,yelinPRL20172,wangOE2017} 
\begin{equation}\label{Hamiltonian}
\begin{split}
\mathcal{H}&=\hbar\sum_{i=1}^N\sum_{\alpha=x,y,z}(\omega_0-i\gamma/2)|e_{i,\alpha}\rangle\langle e_{i,\beta}|\\&+\frac{3\pi\hbar\gamma c}{\omega_0}\sum_{i=1,i\neq j}\sum_{\alpha,\beta=x,y,z}G_{\alpha\beta}(\mathbf{r}_j,\mathbf{r}_i)|e_{i,\alpha}\rangle\langle e_{j,\beta}|,
\end{split}
\end{equation}
which acts on the single excited states of the atoms. Here $\hbar$ is the Planck's constant, $\omega_0$ is angular frequency of the dipole transition from $|g\rangle$ to $|e\rangle$ in a single atom in free space with a radiative linewidth of $\gamma$, and $c$ is the speed of light in vacuum. $G_{\alpha\beta}(\mathbf{r}_j,\mathbf{r}_i)$ is the free-space dyadic Green's function describing the propagation of photons emitting from the $i$-th atom to $j$-th atom, where $\mathbf{r}_j$ and $\mathbf{r}_i$ indicate their positions \cite{svidzinskyPRA2010,guerinPRL2016}:
\begin{equation}
\begin{split}
G_{\alpha\beta}(\mathbf{r}_j,\mathbf{r}_i)&=-\frac{\exp{(ikr)}}{4\pi r}\Big[\Big(1+\frac{i}{kr}-\frac{1}{(kr)^2}\Big)\delta_{\alpha\beta}\\&+\Big(-1-\frac{3i}{kr}+\frac{3}{(kr)^2}\Big)\hat{r}_{\alpha}\hat{r}_{\beta}\Big]
\end{split}
\end{equation}
where $k=\omega_0/c$ is the wavenumber in vacuum, $r=|\mathbf{r}|$, $\mathbf{r}=\mathbf{r}_j-\mathbf{r}_i$, and $\hat{r}_\alpha=r_\alpha/r$.

In this low-excitation picture of light-atom interactions, since the quasiperiodic modulation of interatomic distances in Eq. (\ref{modulation}) leads to quasiperiodic ``hopping" amplitudes of excited states along the atom chain, the present Hamiltonian is quite similar to the conventional AAH model with off-diagonal modulations (i.e., modulations over the inter-site hopping amplitudes for electrons). However, due to the long-range  photon-mediated dipole-dipole interactions between the excited states which can induce long-range hoppings (see the power-law decaying interacting terms in the Green's function, in which the $1/r$ term is the far-field one and $1/r^2$ and $1/r^3$ are near-field ones) as well as the retardation effect of electromagnetic fields (the $\exp{(ikr)}$ phase factor), the present system exhibits more complexities than the conventional off-diagonal AAH model with only nearest-neighbor hoppings, as will be seen below. 

\begin{figure*}[htbp]
	\centering
	\subfloat{
		\includegraphics[width=0.4\linewidth]{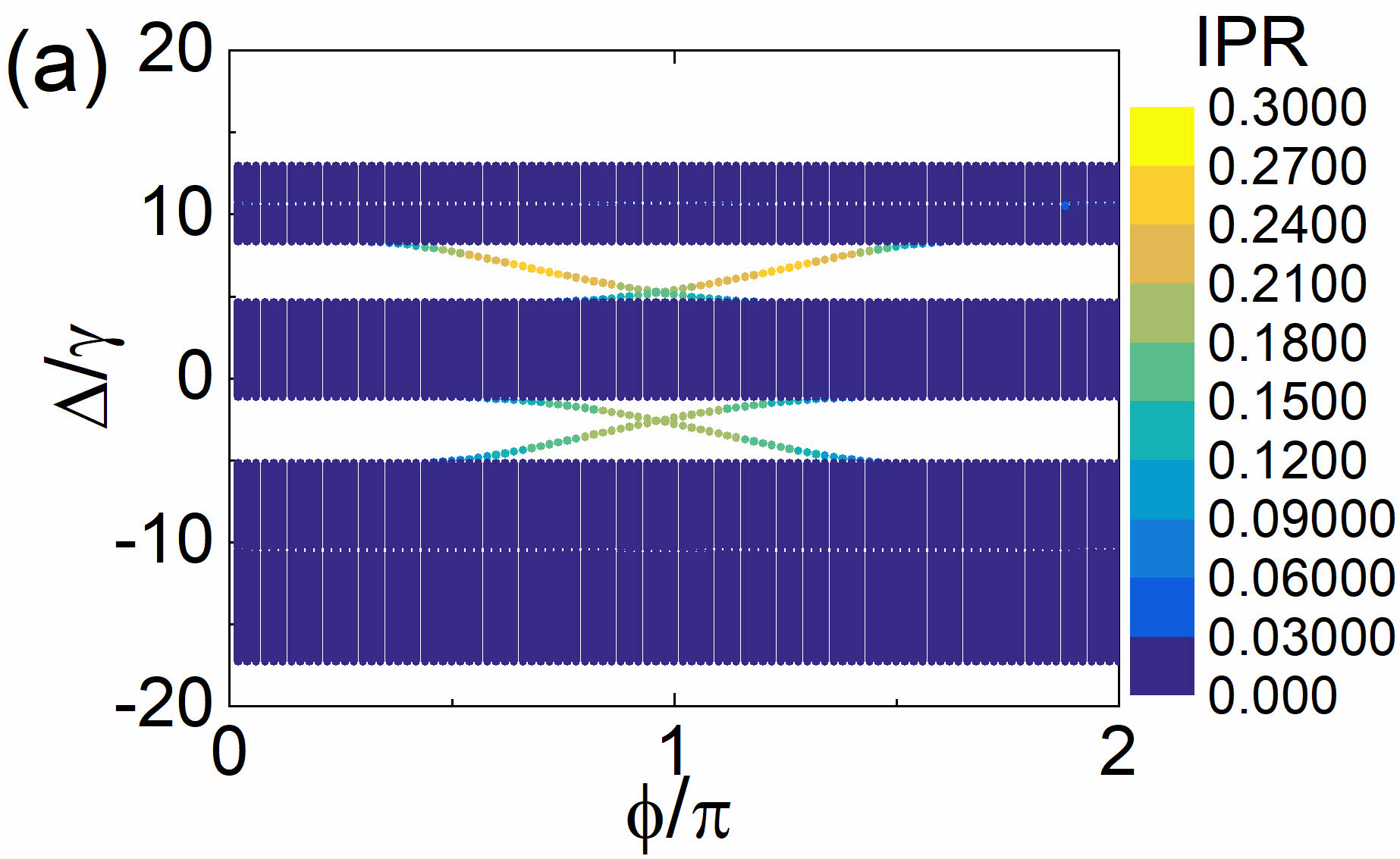}\label{incommd01eta01}
	}
	\subfloat{
		\includegraphics[width=0.4\linewidth]{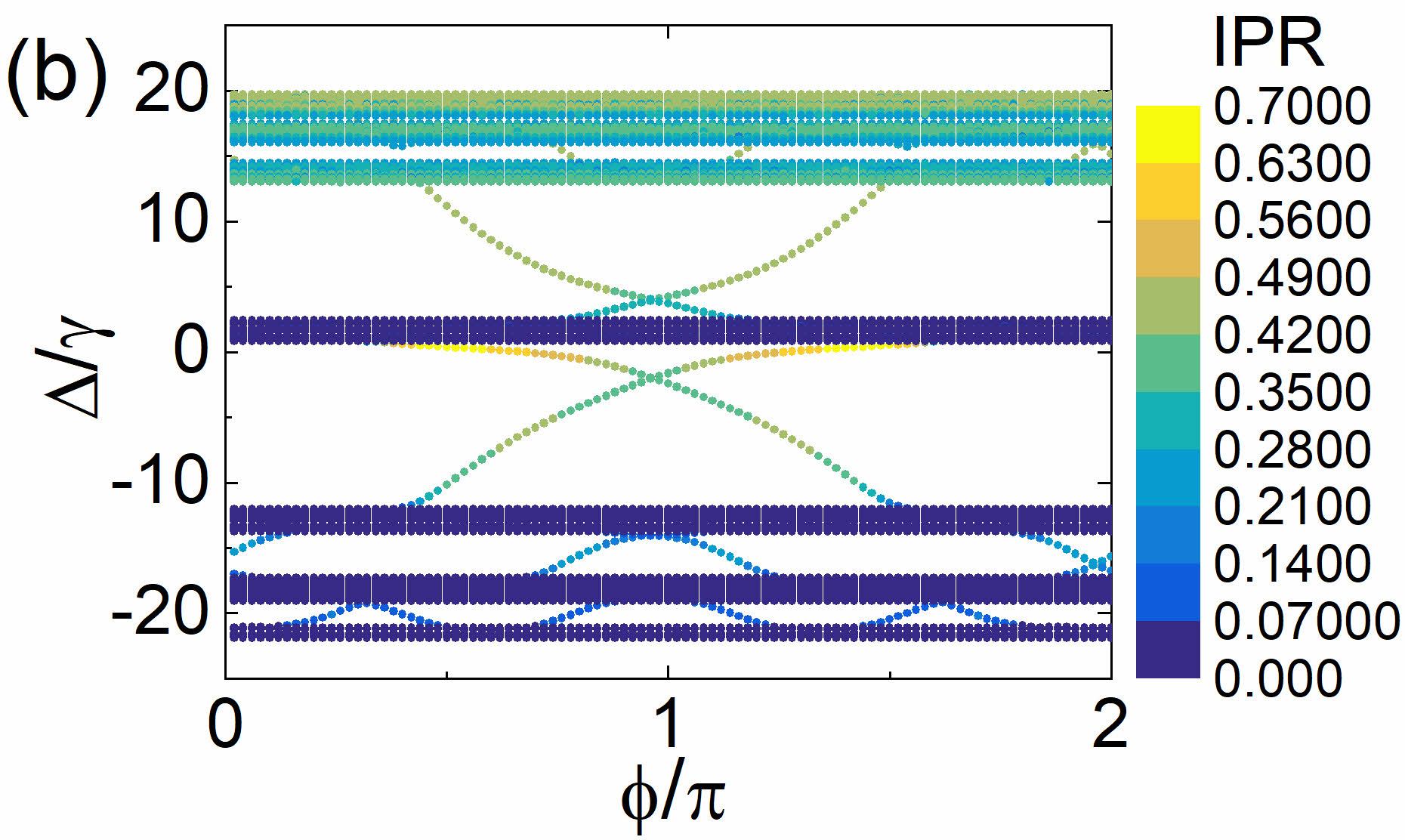}\label{incommd01eta03}
	}\\
	\subfloat{
		\includegraphics[width=0.4\linewidth]{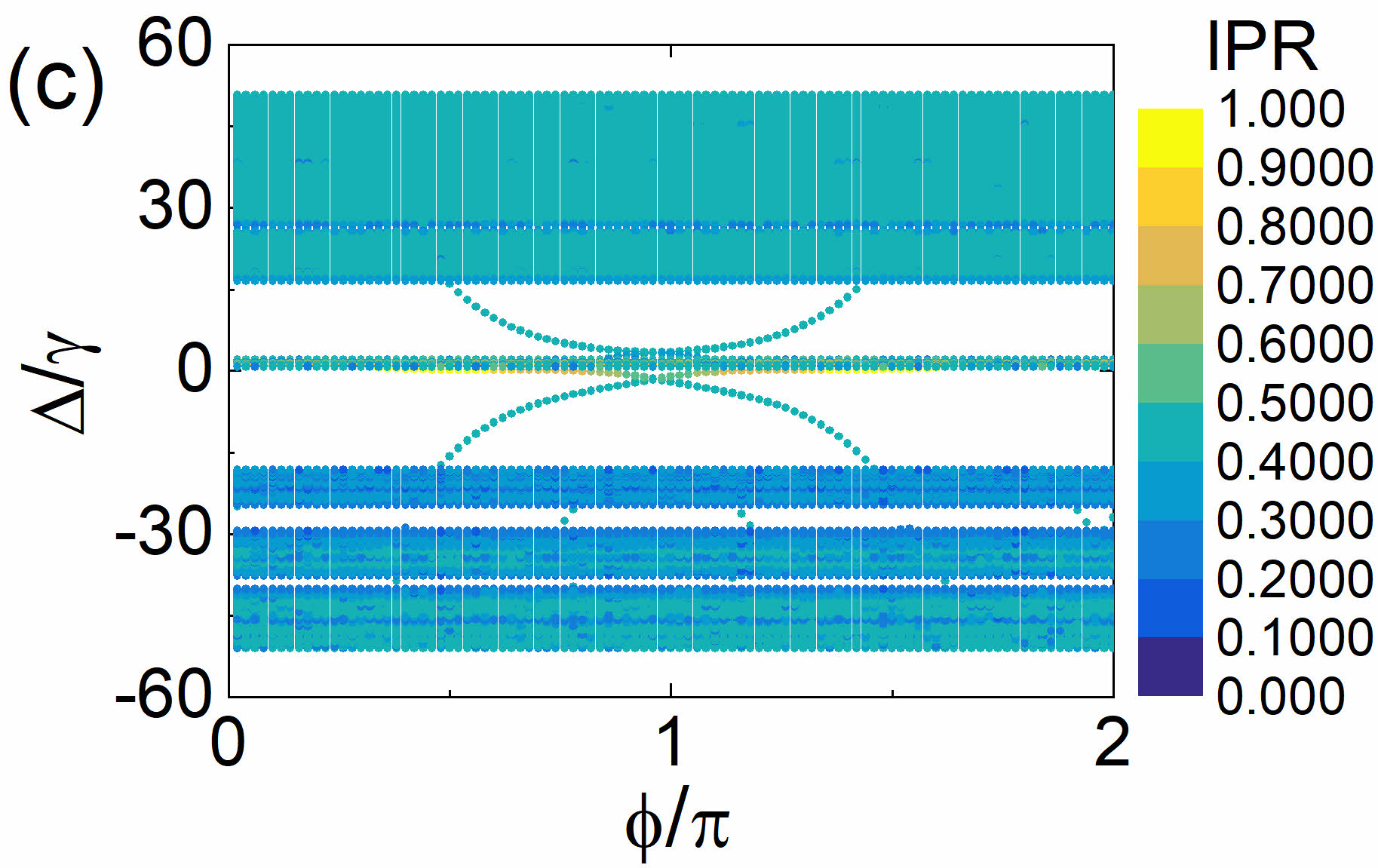}\label{incommd01eta05}
	}
	\subfloat{
		\includegraphics[width=0.4\linewidth]{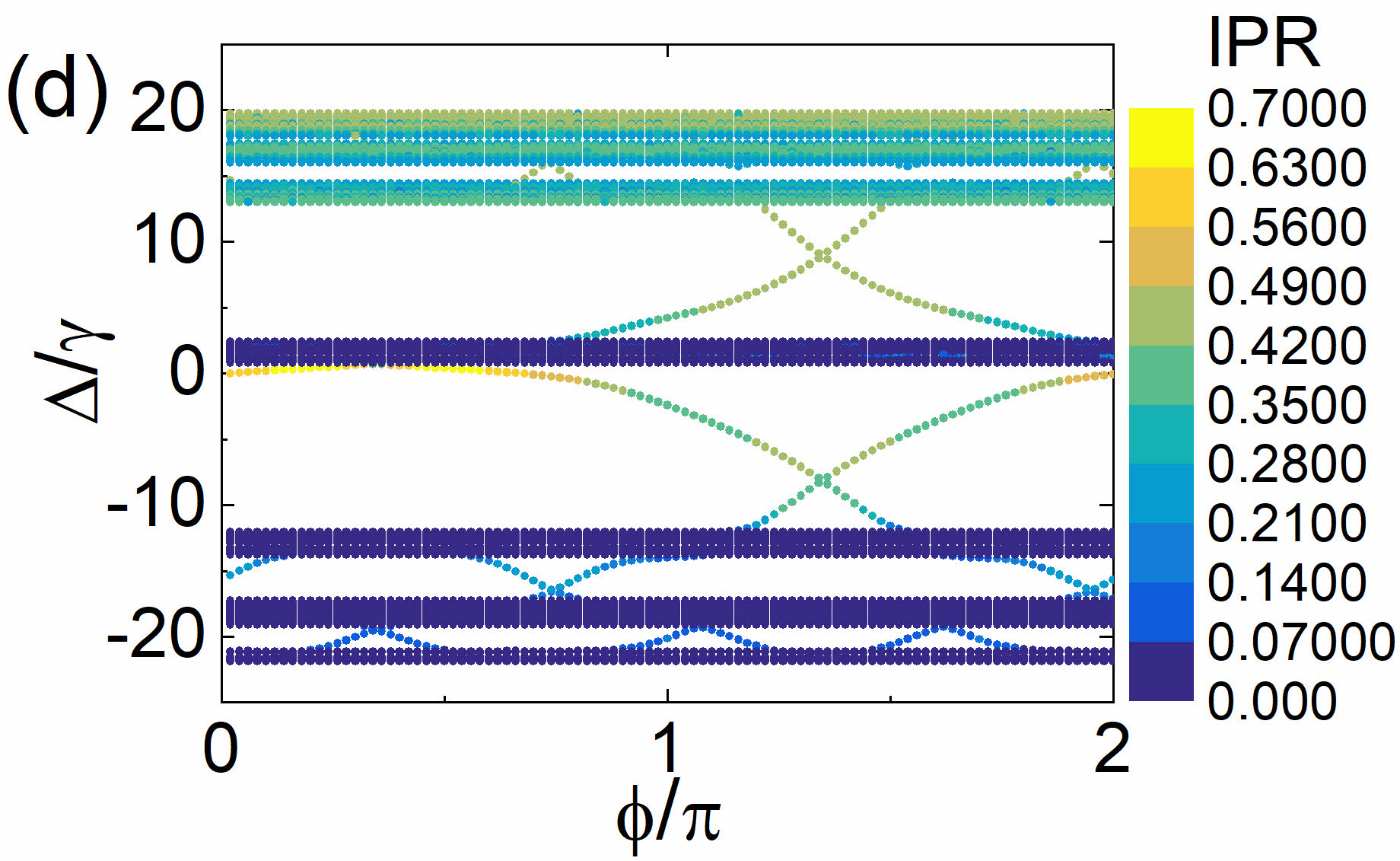}\label{incommd01eta03odd}
	}
	\caption{Longitudinal band structures of quasiperiodic lattices with $\beta=(\sqrt{5}-1)/2$ and $d=0.1\lambda_0$. (a) $\eta=0.1$ with 1000 atoms. (b) $\eta=0.3$ with 1000 atoms. (c) $\eta=0.5$ with 1000 atoms. (d) $\eta=0.3$ with 1001 atoms.}
	\label{longband}
\end{figure*}
 
The single excitation eigenstates of the 1D atomic chains can be classified into two categories according to the polarization directions of the excited states of atoms \cite{weberPRB2004}: transverse eigenstates if atoms are excited to the $|e_y\rangle$ or $|e_z\rangle$ states, and longitudinal ones if atoms are excited to $|e_x\rangle$ states. For the present system without periodicity, Bloch theorem is not applicable and the photonic band structure can only be calculated directly for a finite-sized chain. This can be done by calculating the eigenstates of the Hamiltonian in Eq. (\ref{Hamiltonian}) with respect to the wave function constructed as a linear combination of single-atom excited states, here taking longitudinal eigenstates as an example,
\begin{equation}
|\psi\rangle=\sum_{j=1}^{N}p_{j}|e_{j,x}\rangle,
\end{equation}
where $p_{i}$ is the expansion coefficient denoting the probability amplitude of each excited state $|e_{j,x}\rangle$ \cite{weberPRB2004,guerinPRL2016,kaiserFP2012,kaiserJMO2011}, whose physical significance, if treated in classical electrodynamics, is the normalized dipole moment of the $j$-th atom. This equation specifies a set of solutions in the form $E=\omega-i\Gamma/2$ ($\Gamma>0$) in the lower complex plane denoting to the eigenstates of the chain (Here we use the $e^{-i\omega t}$ convention for harmonic oscillations). $\omega$ amounts to the angular frequency of an eigenstate while $\Gamma$ refers to its radiative linewidth (decay rate), where the corresponding right eigenvector $|\mathbf{p}^R\rangle=[p_1p_2...p_j...p_N]$ then indicates the dipole moment distribution of an eigenstate in a classical interpretation. Moreover, to facilitate the analysis, we adopt the inverse participation ratio (IPR) of an eigenstate from its eigenvector as \cite{wangOL2018}
\begin{equation}
\mathrm{IPR}=\frac{\sum_{j=1}^{N}|p_j|^4}{[\sum_{j=1}^{N}|p_j|^2]^2}.
\end{equation}
The IPR can be used to indicate the degree of spatial confinement (localization) of an eigenstate \cite{Skipetrov2014,wangOL2018}. For instance, for an IPR approaches $1/M$, where $M$ is an integer, the corresponding eigenstate involves the excitation of $M$ atoms \cite{Skipetrov2014,wangOL2018}. 

\section{Band structures and topological edge states}\label{band}

\begin{figure*}[htbp]
	\centering
	\subfloat{
		\includegraphics[width=0.31\linewidth]{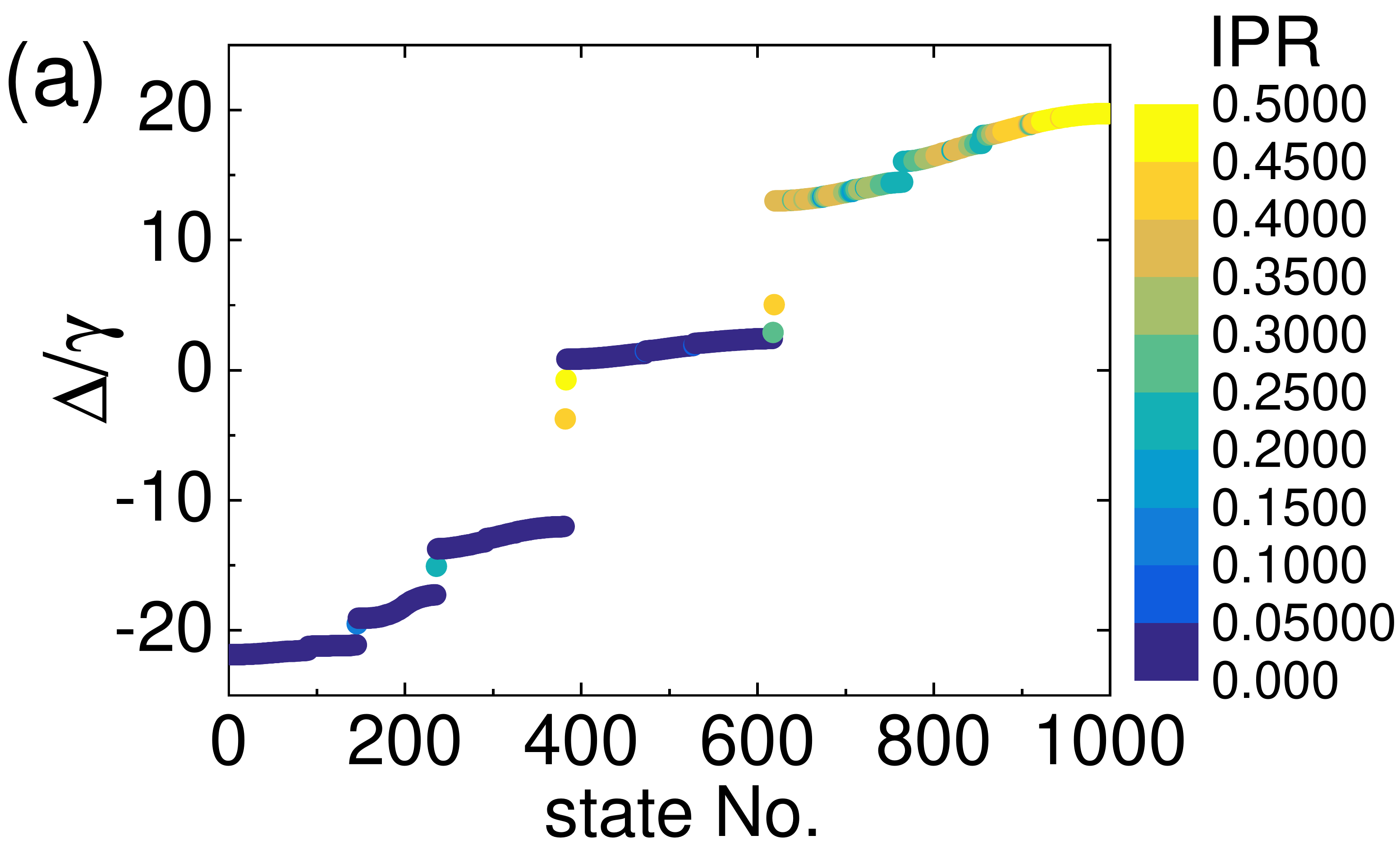}\label{incommd01eta0311pi}
	}
	\subfloat{
		\includegraphics[width=0.28\linewidth]{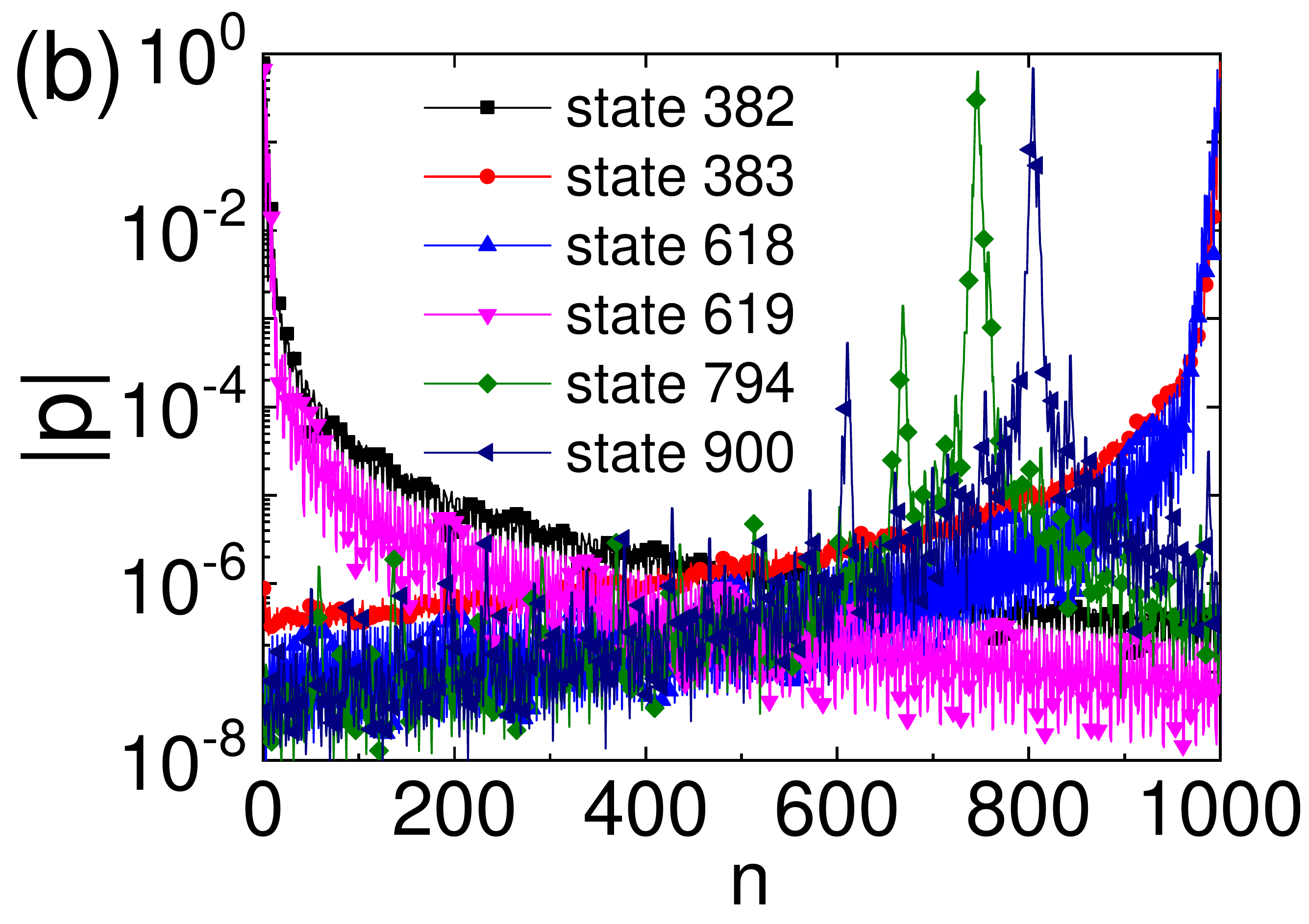}\label{incommd01eta0311pimodedipole}
	}
	\subfloat{
		\includegraphics[width=0.31\linewidth]{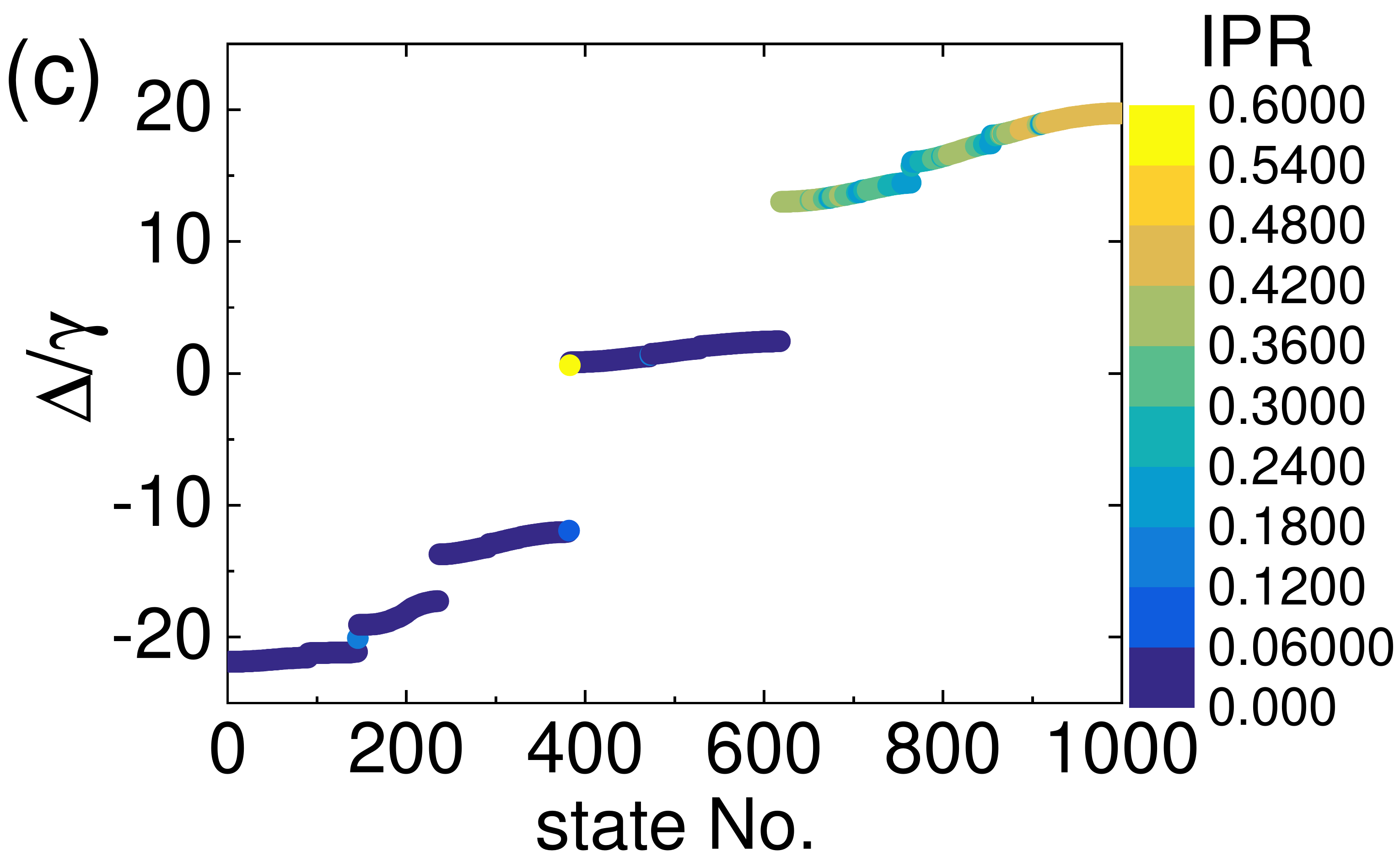}\label{incommd01eta0304pi}
	}\\
	\subfloat{
		\includegraphics[width=0.28\linewidth]{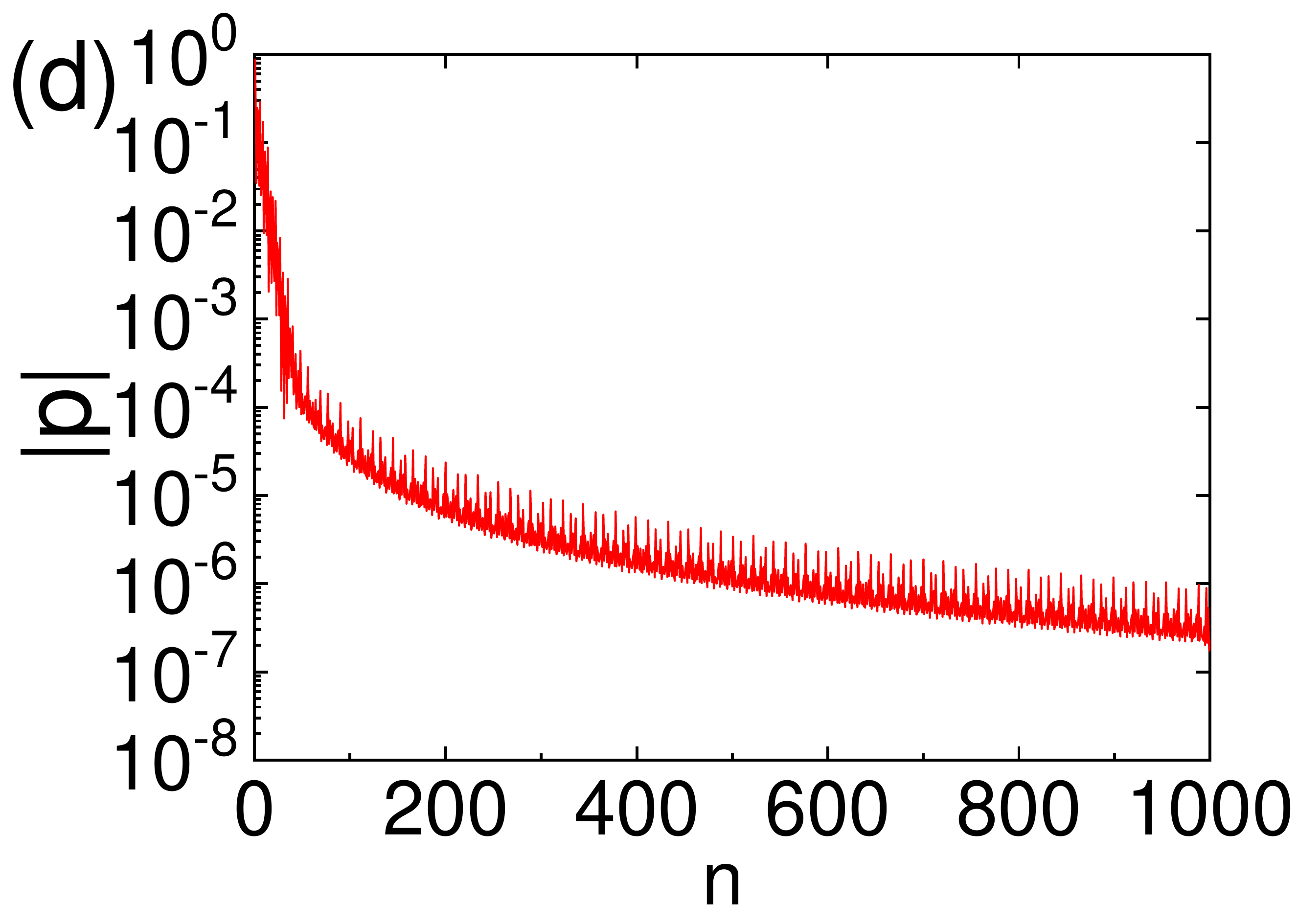}\label{incommd01eta0304pimodedipole}
	}
	\subfloat{
		\includegraphics[width=0.31\linewidth]{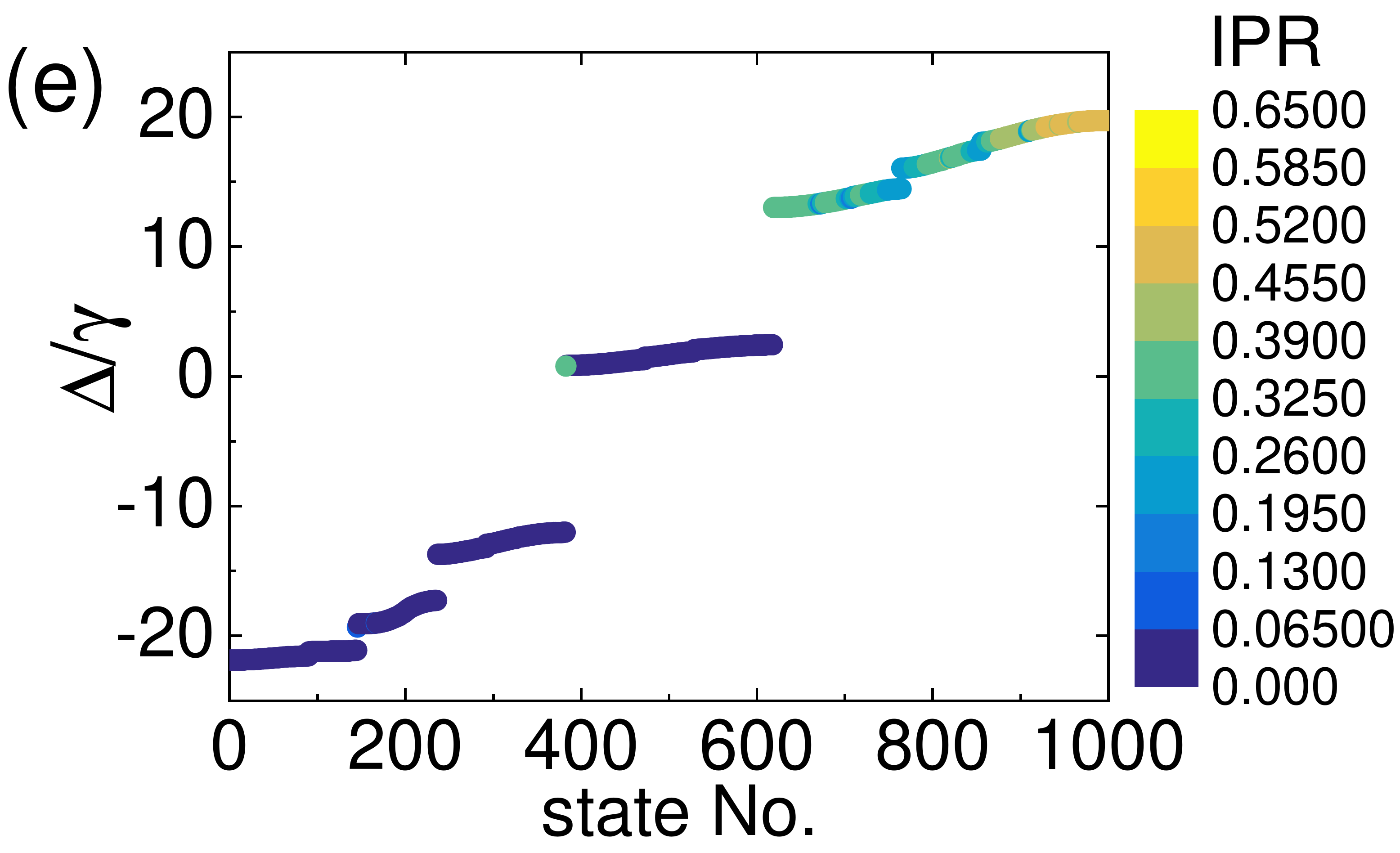}\label{incommd01eta0316pi}
	}
	\subfloat{
		\includegraphics[width=0.28\linewidth]{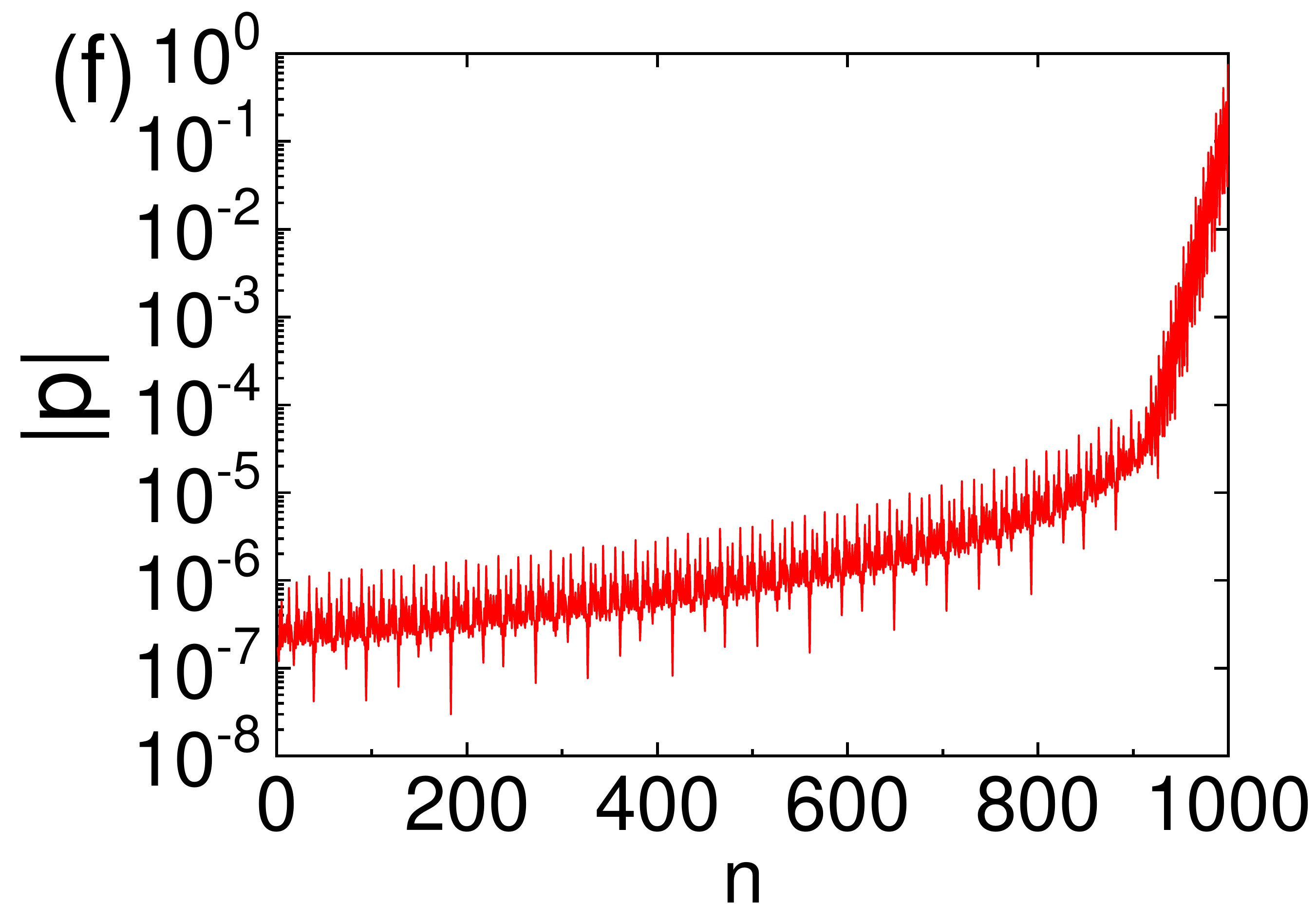}\label{incommd01eta0316pimodedipole}
	}
	\caption{Longitudinal eigenstate distribution at specific modulation phases. (a) $\phi=1.1\pi$. (b) Dipole moment distributions of six eigenstates. (c) $\phi=0.4\pi$. (d) Dipole moment distribution of the midgap state (state No. 383 in (c)). (e) $\phi=1.6\pi$. (f) Dipole moment distribution of the midgap state (state No. 383 in (e)). Here $d=0.1\lambda_0$ and $\eta=0.3$ and 1000 atoms are in the chain. }
	\label{longeigenstates}
\end{figure*}

In this paper, we focus on the quasiperiodic lattice with $\beta=(\sqrt{5}-1)/2$ that is most commonly investigated in the conventional AAH model. Similar to the conventional AAH model, the modulation phase $\phi$ here can be regarded as playing the role of momentum in a perpendicular synthetic dimension, and it is therefore straightforward to demonstrate the calculated eigenstate spectra (band structures) for the present system as a function of the modulation phase $\phi$ varying in the range from $0$ to $2\pi$.

In Figs. \ref{incommd01eta01} to \ref{incommd01eta05}, the longitudinal band structures of a lattice with $N=1000$ atoms and $d=0.1\lambda_0$ under different modulation amplitudes of $\eta=0.1, 0.3, 0.5$ are presented, respectively, where the color of the eigenstates indicates the value of eigenstate IPR. Here $\Delta$ represents the detuning of the eigenstate from the single atom resonance frequency, i.e., $\Delta=\mathrm{Re}E-\omega_0$, and is normalized against the linewidth $\gamma$ in these figures. Owing to the irrational nature of the interatomic distance modulation, the band structures thus break into a set of fractal bands and gaps (more precisely, for an infinitely long chain, the spectrum constitutes a Cantor set with Lebesgue measure zero \cite{kohmotoPRL1983b,kohmotoPRL1987,avilaAnnalsMath2009}), with several main gaps clearly visible, while many minigaps in the spectra can only be seen in an enlarged figure. For more details of the fractal spectra, one can refer to Fig. \ref{fractal_spectrum} in Appendix \ref{fractal_appendix}. 

With the increase of the modulation amplitude $\eta$, the main gaps become wider because stronger near-field dipole-dipole interactions between nearby atoms with smaller distances can give rise to larger frequency shift of eigenstates and thus open wider band gaps \cite{wangOE2017}. It is clearly observed that for all modulation amplitudes, there are midgap states residing in the band gaps. The midgap states in the two main gaps (e.g., the band gaps covering $2.8\lesssim\Delta/\gamma\lesssim13$ and $-12\lesssim\Delta/\gamma\lesssim 0.8$ in Fig. \ref{incommd01eta03}) are highly localized as indicated by their large IPR values. In fact, these states are topologically protected edge states  described by a nonzero Chern number, similar to the behavior of the conventional AAH model, as will be explained in Section \ref{topology}. It should be noted that, although the IPRs of the midgap states in the minigaps (e.g., the band gaps covering $-17\lesssim\Delta/\gamma\lesssim -13$) are not significantly large, these midgap states are still highly localized over the boundaries and topologically protected, as will also be discussed below. Moreover, it is found that with the increase of modulation amplitude, the IPRs of the eigenstates within the bulk bands are also increased which can reach 0.5 or even higher (especially for the case of $\eta=0.5$ where most of bulk eigenstates are localized), resulting in highly localized bulk states. This phenomenon is a consequence of localization transition, a similar behavior to the conventional off-diagonal AAH model at large modulations \cite{lahiniPRL2009,ganeshanPRL2013}. In addition, another feature to note is the even-odd effect presented in Fig. \ref{longband} for a lattice containing 1001 atoms, which shows a distinct distribution of midgap edge states as a result of the sublattice symmetry of the off-diagonal AAH model \cite{zengPRB2020,caoLPL2017}.

In Fig. \ref{longeigenstates}, more details on the midgap states are given. The eigenstate spectrum at $\phi=1.1\pi$ that contains a pair of midgap states in both main gaps is shown in Fig. \ref{incommd01eta0311pi}, where the two main band gaps along with several minigaps are more clearly observed. The state number is assigned according to the detuning of an eigenstate. In each of the two main gaps, there are two highly localized midgap states, and the state numbers are denoted by 382, 393, 618 and 619 respectively. The excitation probability amplitude distributions of these midgap states $p_j$ (or classically, normalized dipole moment distributions) are presented in Fig. \ref{incommd01eta0311pimodedipole}, which show that states No. 382 and No. 619 are highly localized over the left edge while states No. 383 and No. 618 are localized over the right edge. As a comparison, the dipole moment distributions of states No. 794 and 900 located in upper bulk bands are also given, implying that these eigenstates are localized in the bulk as a result of localization transition. Furthermore, the eigenstate spectra at the $\phi=0.4\pi$ and $\phi=1.6\pi$ are presented in Figs. \ref{incommd01eta0304pi} and \ref{incommd01eta0316pi}, which both consist of only one midgap state in the main gaps. For the $\phi=0.4\pi$ case, this midgap state localizes over the left edge [Fig. \ref{incommd01eta0304pimodedipole}] while for the $\phi=1.6\pi$ case, the midgap state is bound to the right edge [Fig. \ref{incommd01eta0316pimodedipole}]. In fact, by further investigating the eigenstate spectra at different modulation phases, it can be found that in each of the two main gaps, by varying the modulation phase $\phi$, the midgap edge states keep localized over the same edge as long as they remain in the band gap. Therefore, the midgap states localized over the same edge in the same band gap belong to the same mode when considering $\phi$ plays the role of an additional momentum. For the two main gaps, there are two edge modes traversing the spectral gap, one localized over the left edge and the other localized over the right edge. This property is a manifestation of the topological nature of the band gaps  \cite{krausPRL2012b}, as will be discussed below. In addition, it is noted that despite the even-odd effect for finite-size lattices, there is always one edge mode in the band gap that does not vary with the number of atoms (see Figs. \ref{incommd01eta03} and \ref{incommd01eta03odd}) \cite{guoOL2018}.

One issue worth mentioning is that the dipole moment distribution of edge state shows a fast decay near the edge and then a slower one in the bulk [Figs. \ref{incommd01eta0304pimodedipole} and \ref{incommd01eta0316pimodedipole}]. More accurately, it is identified that the topologically protected edge states decay exponentially from the edge with a short localization length while decaying algebraically in the long range. This phenomenon is a general result of algebraically (or power-law) decaying ($1/r^{\alpha}$, $\alpha>0$) dipole-dipole interactions that give rise to long-range hopping of excited states (photons), which is also observed in many similar systems with power-law interactions, for example, the 1D Kitaev model with power-law pairing \cite{vodolaPRL2014}, 1D Kitaev model with both power-law hopping and pairing \cite{viyuelaPRB2016,leporiNJP2017}, as well as 2D \textit{p}-wave superconductor models with power-law hopping or (and) pairing amplitudes \cite{viyuelaPRL2018,leporiPRB2018}. As a consequence, it should be borne in mind the definition of localization length of the edge state amounts to the length scale in which it decays exponentially near the edge. 

\begin{figure*}[htbp]
	\centering
	\subfloat{
		\includegraphics[width=0.4\linewidth]{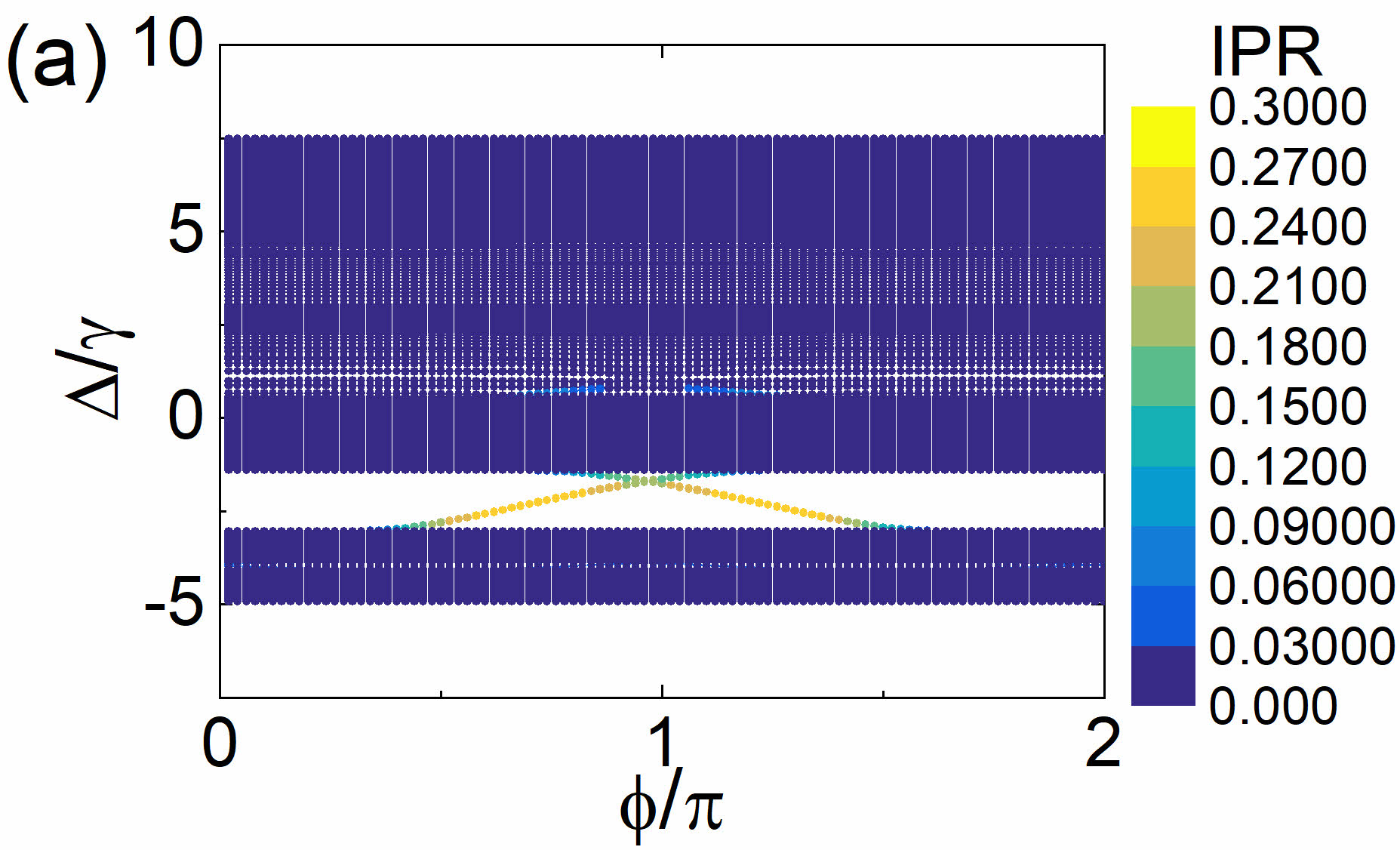}\label{transincommd01eta01}
	}
	\subfloat{
		\includegraphics[width=0.4\linewidth]{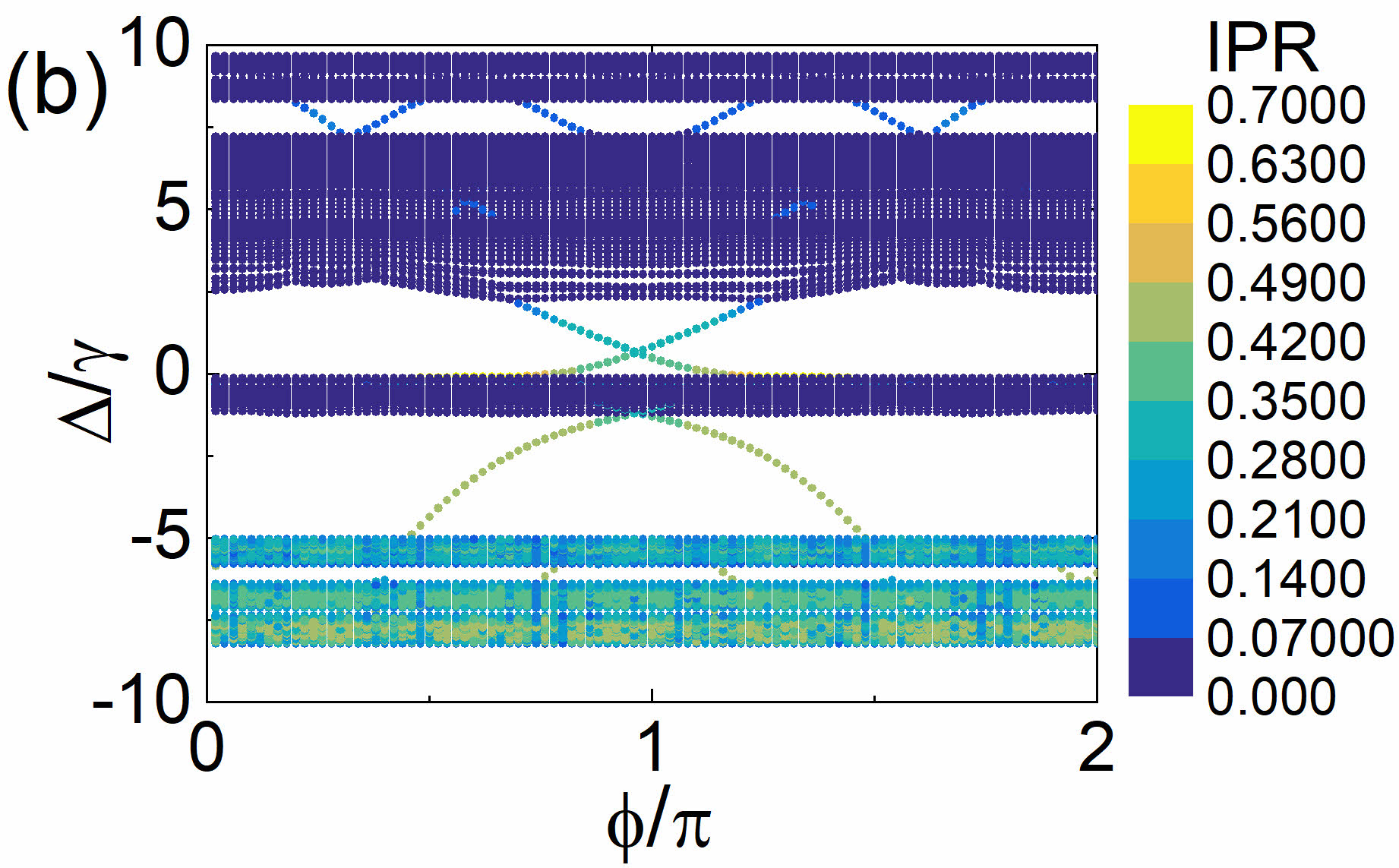}\label{transincommd01eta03}
	}
\\
	\subfloat{
		\includegraphics[width=0.4\linewidth]{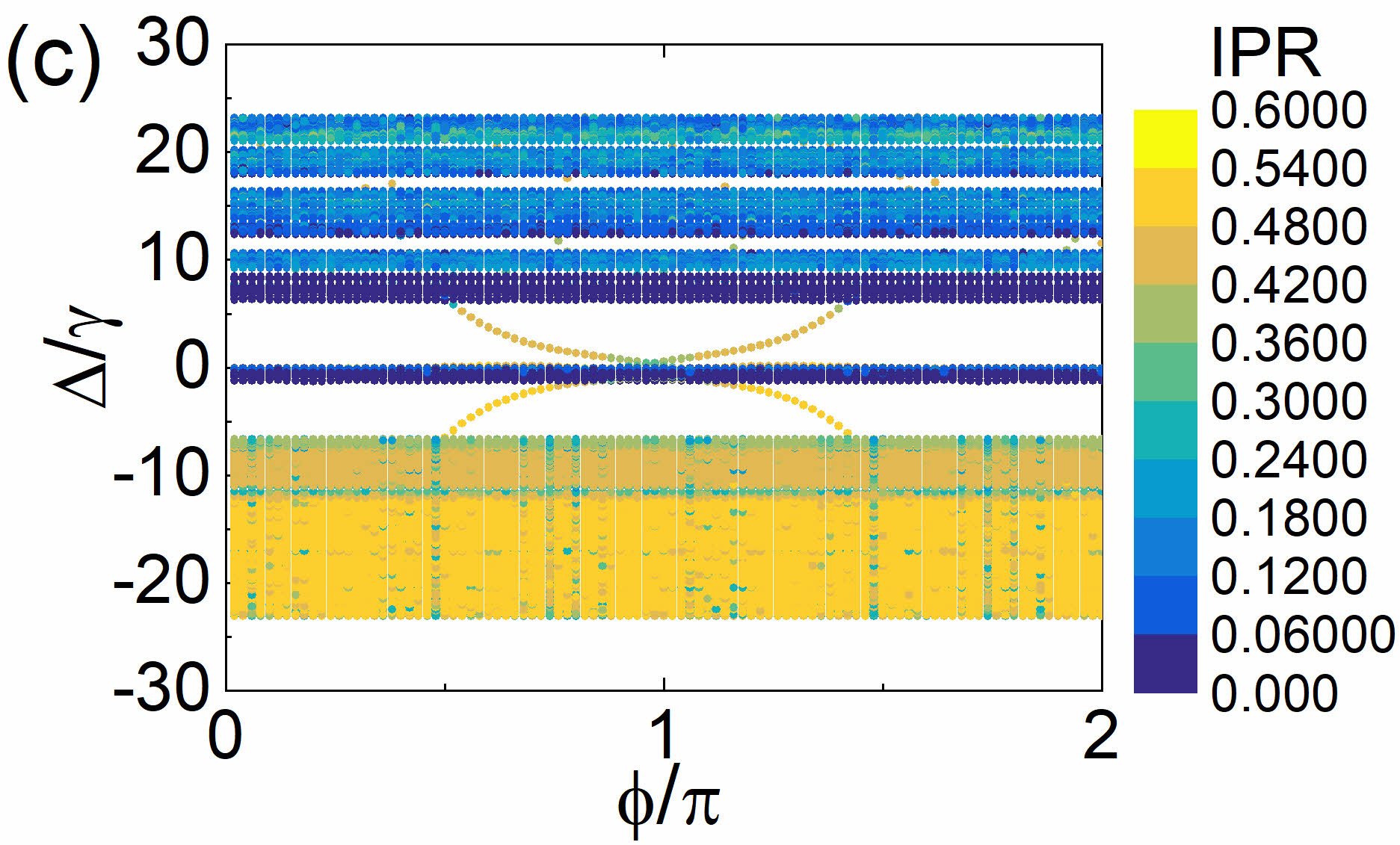}\label{transincommd01eta05}
	}
	\subfloat{
		\includegraphics[width=0.4\linewidth]{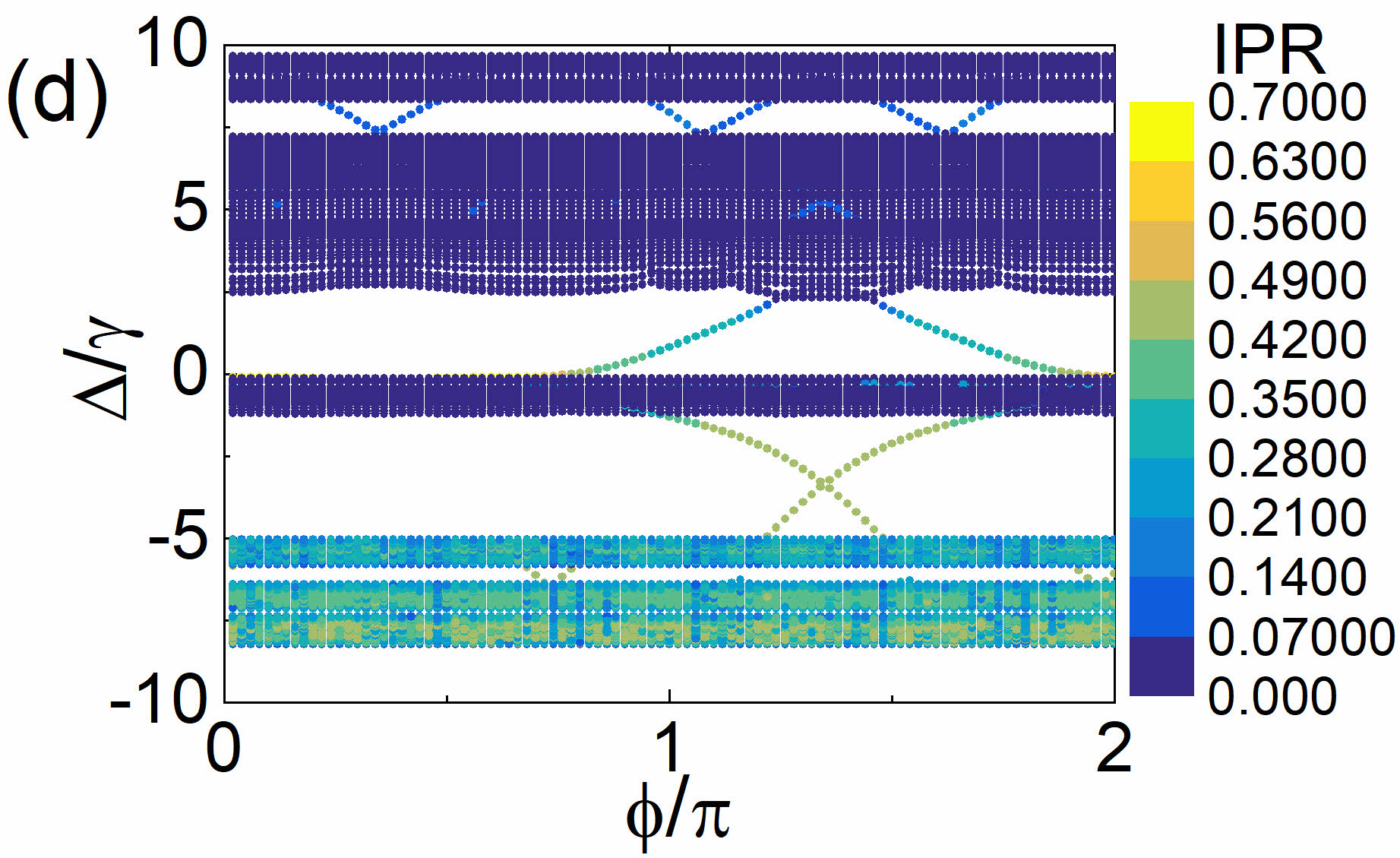}\label{transincommd01eta03odd}
	}
	\caption{Transverse band structures of quasiperiodic lattices with $\beta=(\sqrt{5}-1)/2$ and $d=0.1\lambda_0$. (a) $\eta=0.1$ with 1000 atoms. (b) $\eta=0.3$ with 1000 atoms. (c) $\eta=0.5$ with 1000 atoms. (d) $\eta=0.3$ with 1001 atoms.}
	\label{transband}
\end{figure*}

In a similar manner, we further calculate the band structures of transverse eigenstates, shown in Fig. \ref{transband} with the same parameters with the longitudinal case. The band gaps in the transverse band structures are substantially narrower as a consequence of weaker dipole-dipole interactions. This is because transverse eigenstates involve a far-field interaction term that decays very slowly with the distance $r$ as $1/r$. Such far-field interactions can result in very long-range hoppings of excited states and effectively reduce the strength of near-field interactions, leading to a reduction in the band gap width. In spite of these differences, the qualitative behavior is still quite similar to the conventional AAH model, including the fractal bands and gaps, midgap states and localization transition at large modulations, as in longitudinal band structures. 

\begin{figure*}[htbp]
	\centering
	\subfloat{
		\includegraphics[width=0.31\linewidth]{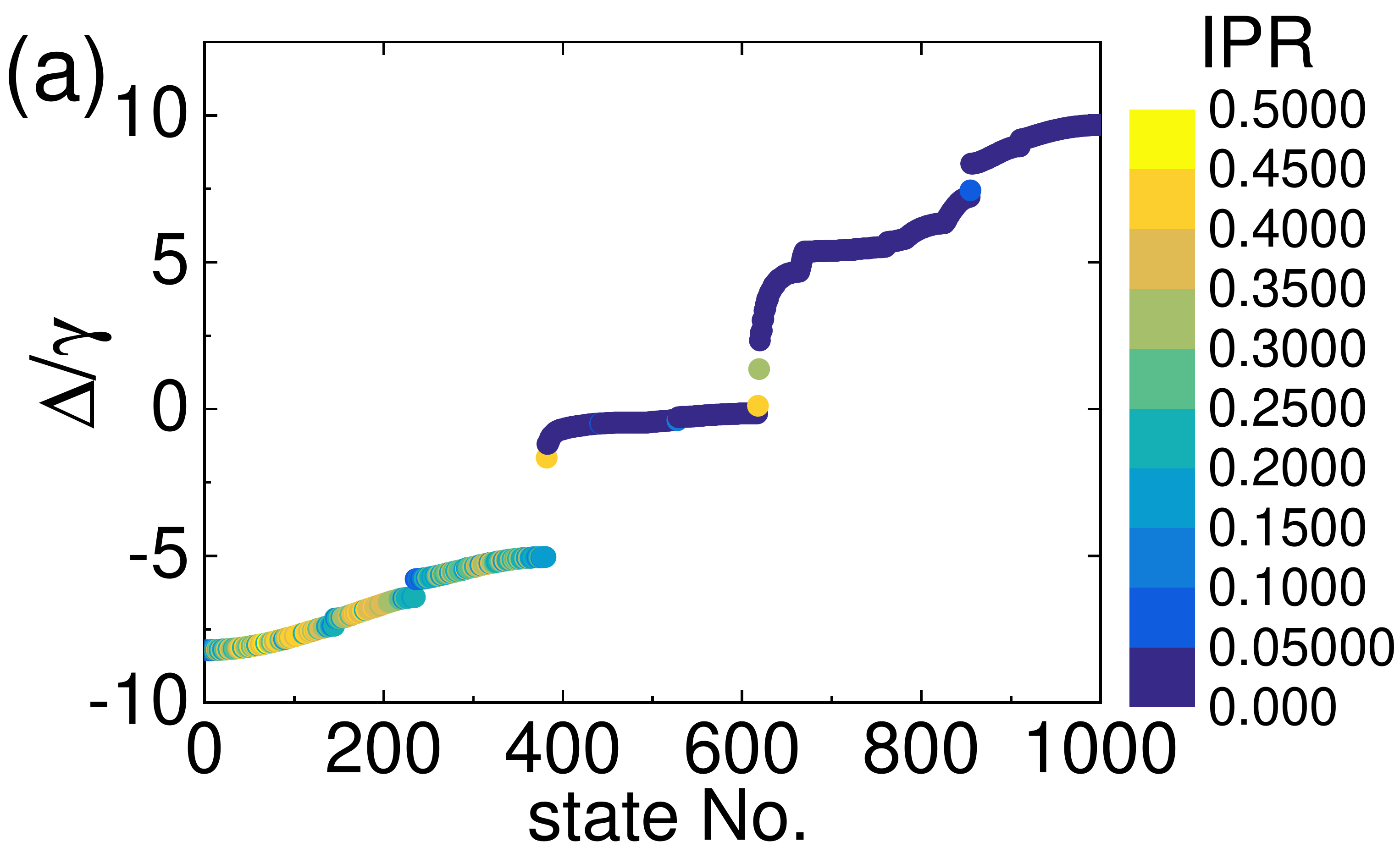}\label{transincommd01eta0311pi}
	}
	\subfloat{
		\includegraphics[width=0.28\linewidth]{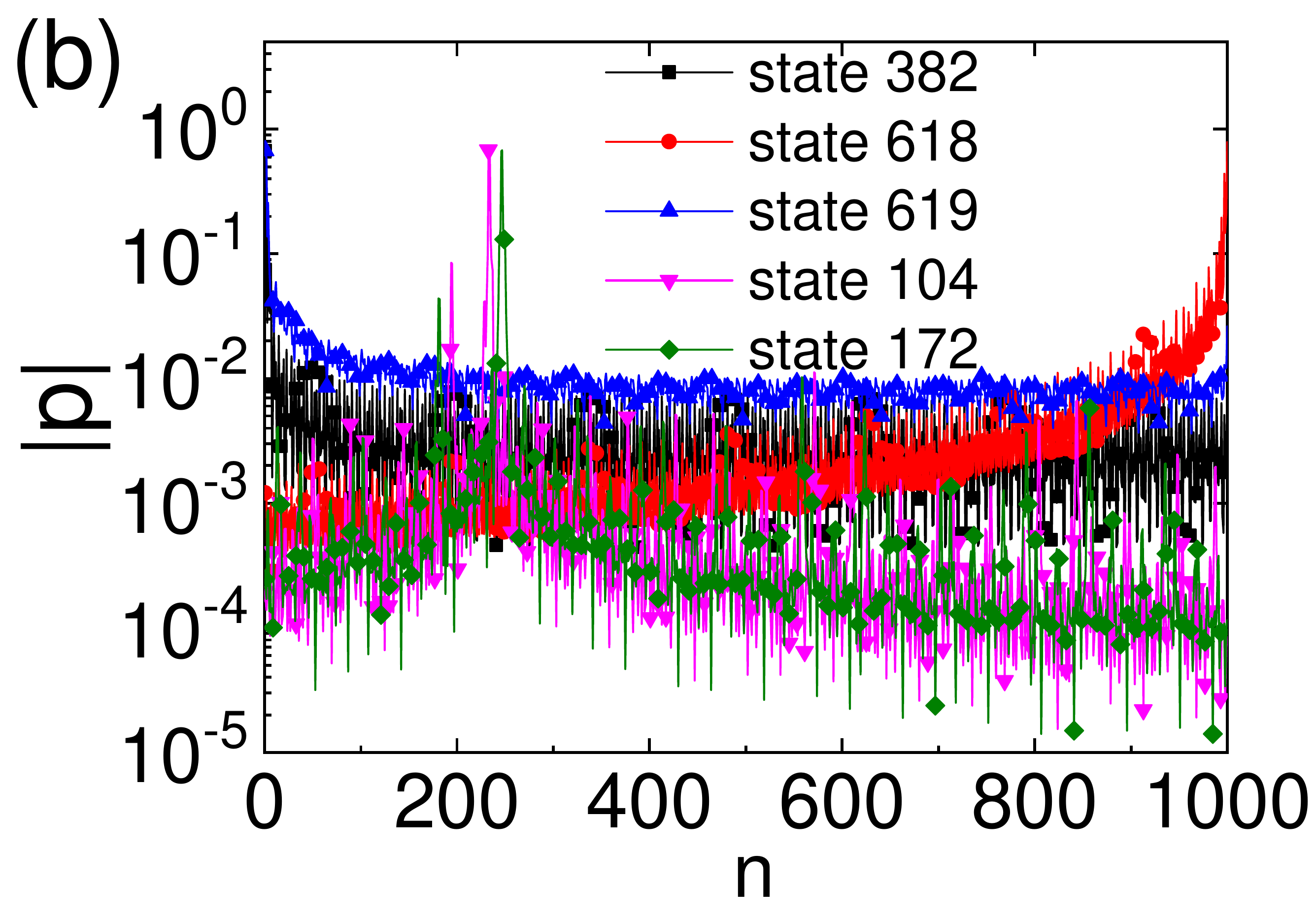}\label{transincommd01eta0311pimodedipole}
	}
	\subfloat{
		\includegraphics[width=0.31\linewidth]{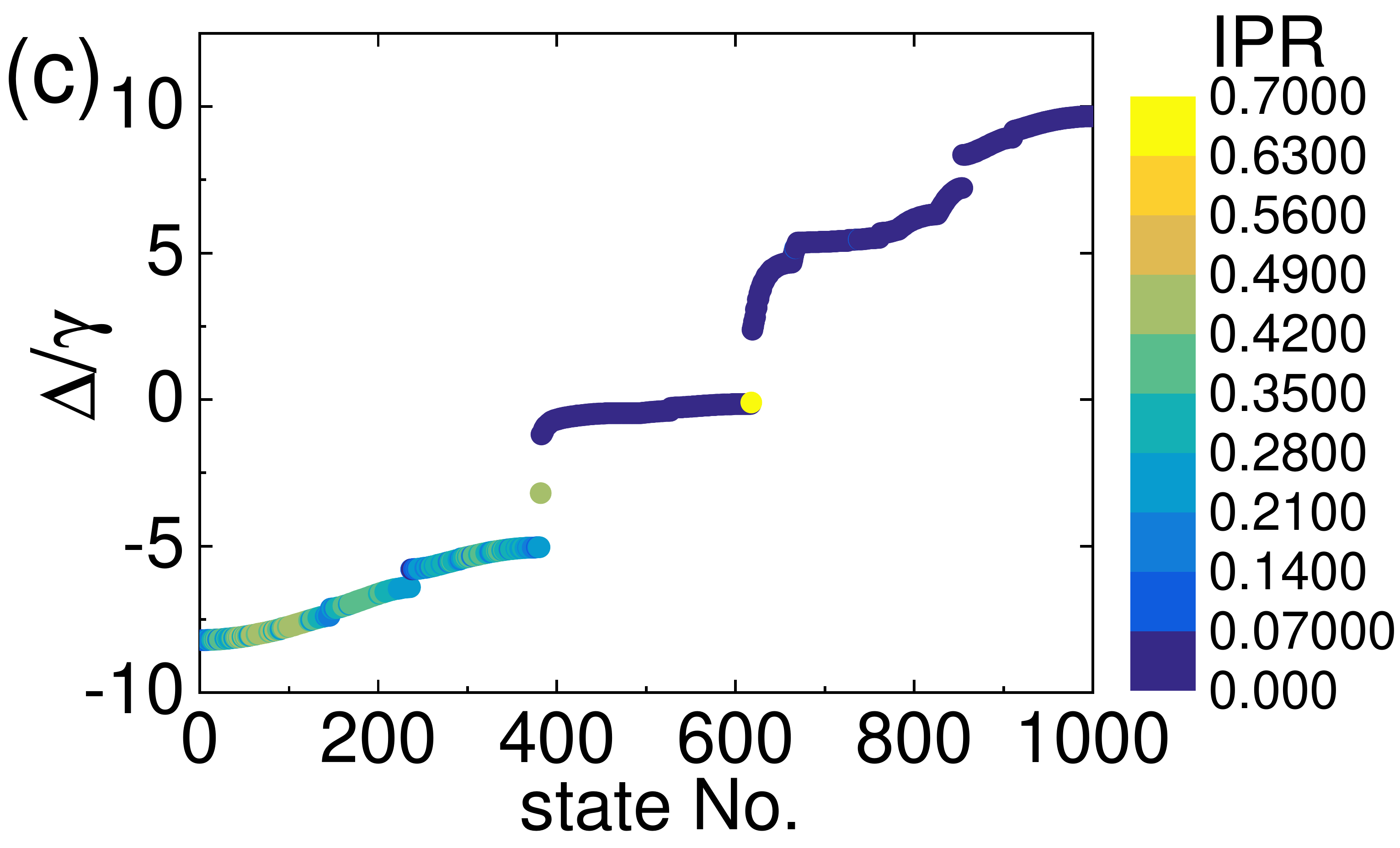}\label{transincommd01eta0306pi}
	}\\
	\subfloat{
		\includegraphics[width=0.28\linewidth]{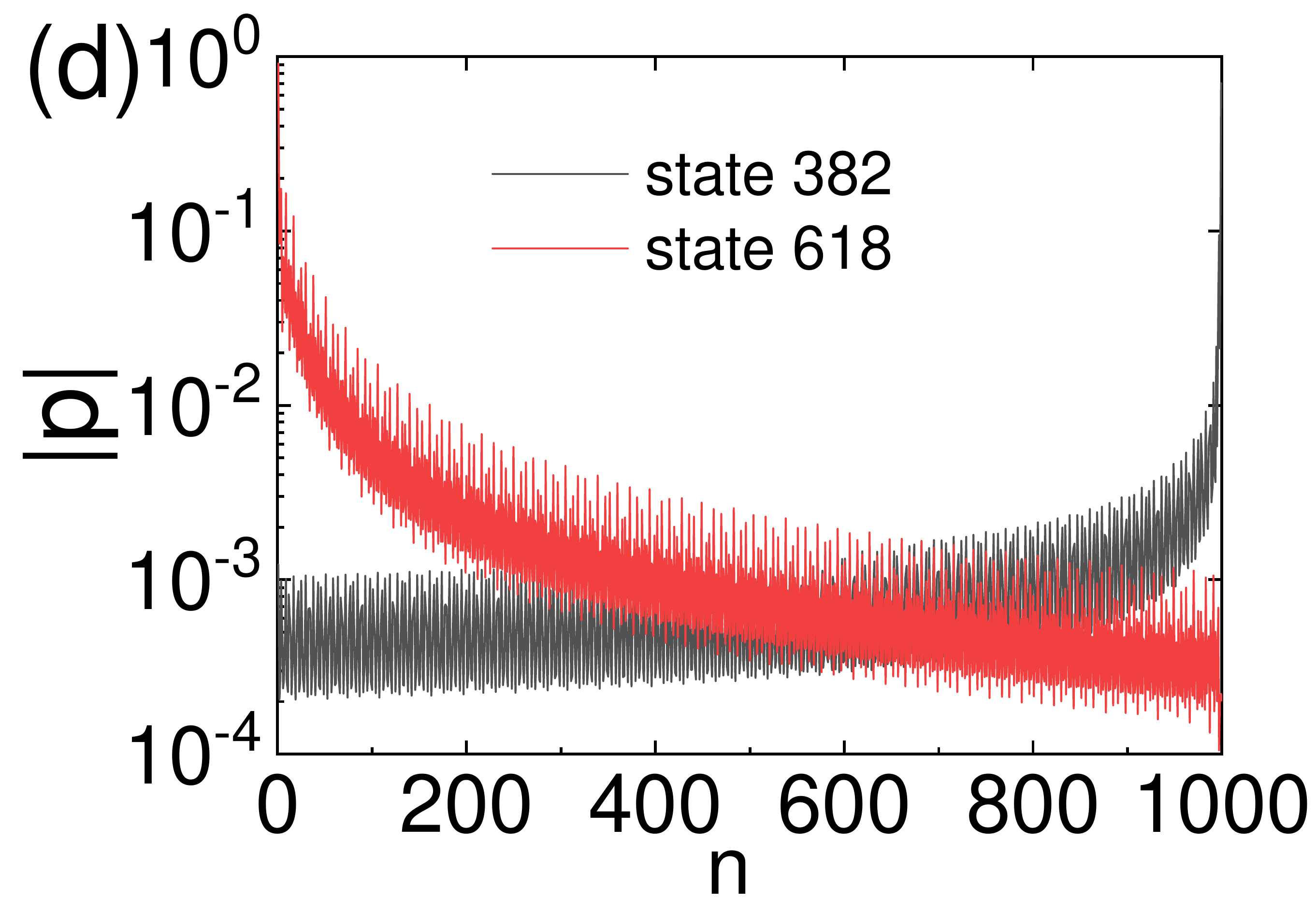}\label{transincommd01eta0306pimodedipole}
	}
	\subfloat{
		\includegraphics[width=0.31\linewidth]{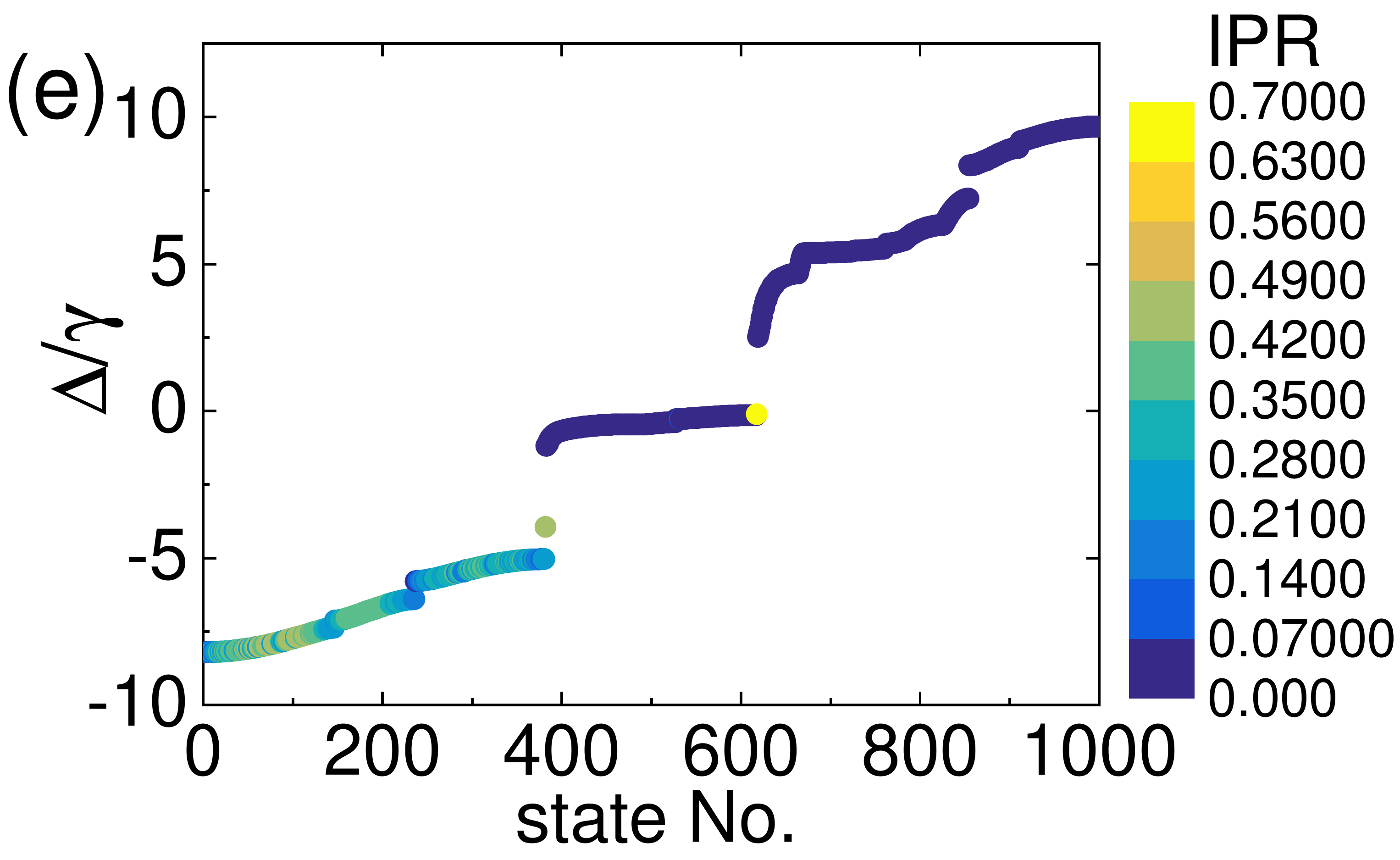}\label{transincommd01eta0314pi}
	}
	\subfloat{
		\includegraphics[width=0.28\linewidth]{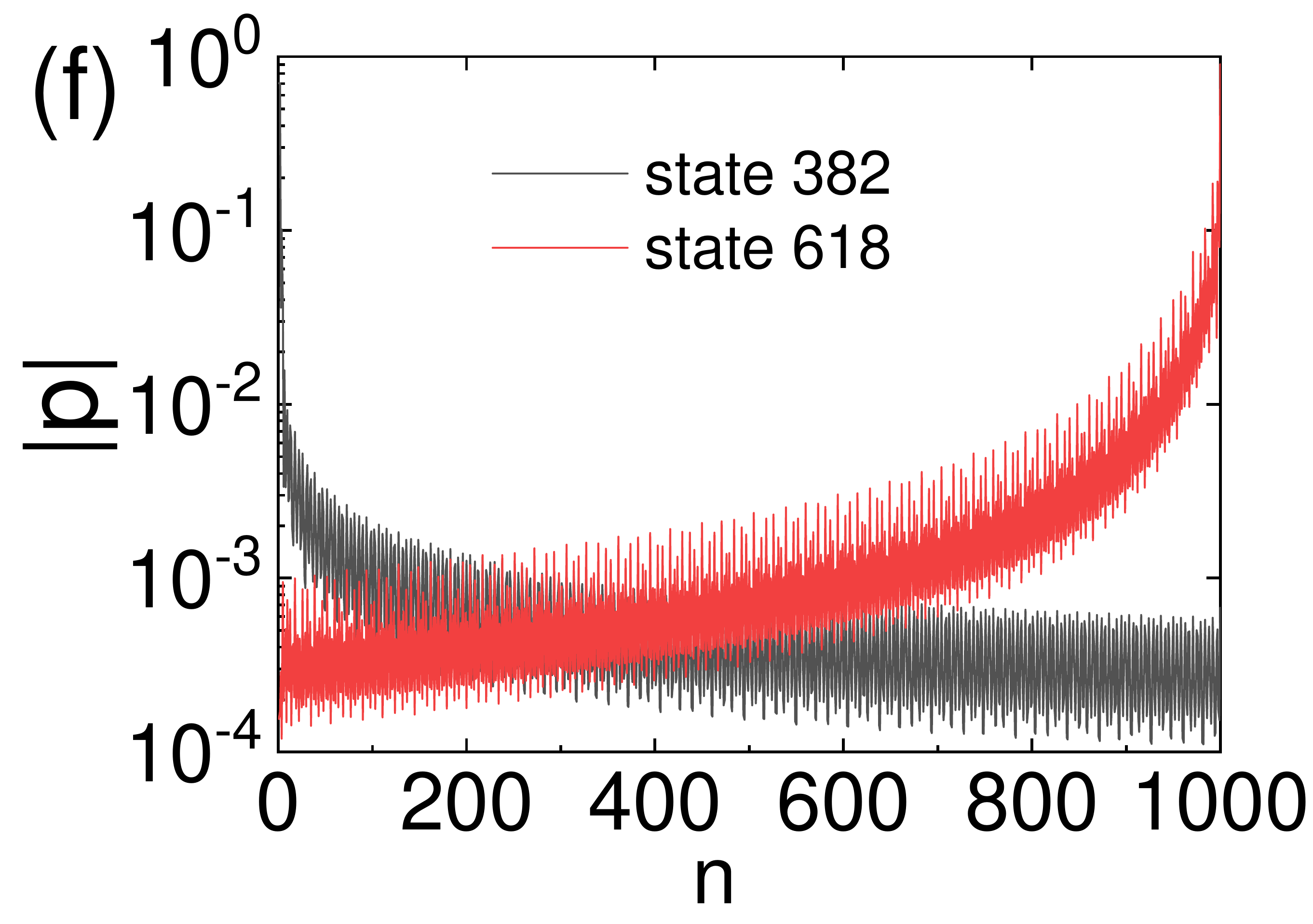}\label{transincommd01eta0314pimodedipole}
	}
	\caption{Transverse eigenstate spectra for different modulation phases. There are 1000 atoms in the chain with $d=0.1\lambda_0$ and $\eta=0.3$. (a) $\phi=1.1\pi$. (b) Dipole moment distributions of three midgap states, along with two arbitrarily selected trivially localized states. (c) $\phi=0.6\pi$. (d) Dipole moment distributions of the midgap states in (c). (e) $\phi=1.4\pi$. (f) Dipole moment distributions of two midgap states in (e).}
	\label{transeigenstates}
\end{figure*}

In Fig. \ref{transeigenstates}, transverse eigenstate spectra for several typical modulation phases are also presented. Firstly, Fig. \ref{transincommd01eta0311pi} shows the eigenstate spectrum at $\phi=1.1\pi$, along with the dipole moment distributions of midgap edge states with state numbers 382, 618 and 619 plotted in Fig. \ref{transincommd01eta0311pimodedipole}, in which states No. 104 and 172 are localized bulk states. The eigenstate spectra at $\phi=0.6\pi$ and $\phi=1.4\pi$ are presented in Figs. \ref{transincommd01eta0306pi} and \ref{transincommd01eta0314pi}, both of which contain one midgap state in each of the two main gaps. The two midgap gap states in the two main gaps are localized over the left and right edges, respectively, as indicated by Figs. \ref{transincommd01eta0306pimodedipole} and \ref{transincommd01eta0314pimodedipole}. Note the state numbers of the two midgap states under the two different modulation phases are the same, i.e., 382 and 618. This is indeed a signal of topological protection, as will be discussed below.

\section{Topological invariants}\label{topology}
Above results indicate that in spite of the long-range nature of dipole-dipole interactions beyond the nearest-neighbor approximation, the features of band structures of the present quasiperiodic lattice are largely similar to the conventional AAH model. We have also noticed several phenomena that hint the topological nature of the lattice, including the highly localized midgap edge states, quantized number of edge modes in the band gaps, fixed state number of midgap states. In this section, we proceed to a theoretical description of the band topology and demonstrate the validity of the bulk-boundary correspondence, in order to formally verify these midgap edge states are indeed topologically protected.

As mentioned before, in the conventional AAH model, the modulation phase $\phi$ plays the role of momentum (wavenumber) in a perpendicular synthetic dimension. On this basis, it is shown that this model can be mapped to the 2D Harper-Hofstadter model in the presence of a perpendicular magnetic field (the discrete lattice version of the Landau level problem) \cite{ozawa2018topological,krausPRL2012,krausPRL2012b}. To be more precisely, in such mapping, the modulation (quasi-)periodicity $\beta$ actually corresponds to the magnetic flux quanta per unit cell, and an irrational $\beta$ implies that an incommensurate magnetic field is imposed. The AAH Hamiltonian $\mathcal{H}(\phi)$ at a specific $\phi$ can be regarded as the $k$-th Fourier component of the 2D Harper-Hofstadter Hamiltonian, where $k=\phi/a$ and $a$ is the imaginary lattice constant in the synthetic dimension \cite{krausPRL2012b,zengPRB2020}. Therefore the conventional AAH model inherits nontrivial topological properties from a 2D quantum Hall (QH) system, and the midgap edges states are thus of topological origin from the robust chiral states in 2D integer quantum Hall effect (IQHE), without the need of a real magnetic flux and the breaking of the time-reversal symmetry \cite{krausPRL2012b,krausPRL2012}.

As a result, the band topology of the conventional AAH model can be well characterized by an integer known as the Chern number like that in 2D IQHE systems even when the modulation is incommensurate with the lattice, which satisfies the following ``Diophantine" equation  \cite{thoulessPRL1982,macdonaldPRB1984,danaJPC1985,amitPRB2018}:
\begin{equation} \label{gaplabel_eq}
\mathcal{N}=\mu+\nu\beta,
\end{equation}
in which $\mu$ is an integer, $\nu$ is the Chern number of the band gap and $\mathcal{N}$ is the normalized integrated density of states (IDOS) in a band gap. This equation is a general result following from the magnetic translational symmetry in a IQH system with 2D Bloch electrons subjected to rational (thus described by a rational $\beta$) magnetic fields by taking the irrational limit for $\beta$ \cite{danaJPC1985}. Note although a typical Diophantine equation is defined only for integers, by taking the Diophantine approximation of the irrational $\beta$ as $p/q$ with $p,q$ denoting sufficiently large coprime integers, we can still transform above equation into a typical Diophantine one and solve the topological integers. This is the reason why the above equation can be still dubbed a Diophantine equation for convenience \cite{krausPRL2012,danaPRB2014,cookmeyerPRB2020}. More details on the origin of this equation are presented in Appendix \ref{gaplabel_appendix}. In particular, it has been shown that for an irrational $\beta$ and a fixed $\mathcal{N}$, the above equation has only one set of solution $(\mu,\nu)$ \cite{danaPRB1985,krausPRL2012,danaPRB2014}. Then those band gaps with the same $\mathcal{N}$ and irrational $\beta$ can be labeled by the same set of integers $(\mu,\nu)$, independent of system details. In this sense, this equation is universally valid and thus dubbed the gap-labeling theorem \cite{simonAAM1982,luckPRB1989,bellissardCMP1989,liuPRB1992,fuPRB1997,babouxPRB2017,tanesePRL2014,rivoltaPRA2017,apigoPRMat2018,raiPRB2019,apigoPRL2019,levy2015topological,dareauPRL2017,wangOL2018}. This universality of these topological integers under an irrational $\beta$, which is robustly protected by the magnetic translational symmetry \cite{cookmeyerPRB2020}, is also the reason why in the presence of long-range interactions beyond the nearest-neighbor approximation, the gap-labeling theorem is still applicable, as will be implied by calculations in the following. On the other hand, the situation is vastly different for a rational $\beta$, in which there are infinite solutions for the equation at a fixed $\beta$ and therefore the topological properties is system-dependent \cite{thoulessPRL1982,macdonaldPRB1984,hatsugaiPRB1990,amitPRB2018}.

In the following we attempt to examine in detail the applicability of the gap-labeling theorem in describing the topology of band gaps and verify the bulk-boundary correspondence by identifying the topological edge states for the present system, which differently exhibits long-range dipole-dipole interactions and non-Hermitian nature (due to the coupling with the free-space modes). The normalized IDOS $\mathcal{N}$ of a band gap is simply given by the number of eigenstates below the gap divided by the total number of eigenstates. It is noted that the bands are invariant as a function of $\phi$ (namely, flatband) and therefore by only considering the eigenstate spectrum at a fixed $\phi$, one can readily obtain the normalized IDOS in a band gap of the whole band structure \cite{krausPRL2012,krausPRL2012b}. It is this flatband feature arising from the quasiperiodicity that allows us to associate the Chern number with this 1D system without resorting to the ``ancestor" 2D IQHE model with the synthetic dimension included \cite{krausPRL2012,krausPRL2012b}, which is formally required for periodic systems \cite{poshakinskiyPRL2014}.

Taking the longitudinal band structures as an example [Fig. \ref{incommd01eta03}], we can find the lower main gap (around $-12\lesssim\Delta/\gamma\lesssim 0.8$) has a normalized IDOS $\mathcal{N}\approx382/1000=0.382$. Therefore, we have $\mu=1,\nu=-1$ for this main gap. Similarly, the second main gap spanning $2.8\lesssim\Delta/\gamma\lesssim13$ has a normalized IDOS of $\mathcal{N}\approx618/1000=0.618$, which leads to $\mu=0$, $\nu=1$. Therefore, for bulk eigenstates, we have obtain the topological invariant, i.e., the Chern number of the lower (upper) main gap as $\nu=-1$ ($\nu=+1$). According to the bulk-boundary correspondence in 2D IQHE, for a gap Chern number of $\nu$, there should be exactly $|\nu|$ edge mode(s) on each edge, whose energy (frequency) traverses the gap when the modulation phase $\phi$ varies from $0$ to $2\pi$ \cite{krausPRL2012,krausPRL2012b,poshakinskiyPRL2014}. And the sign of gap Chern number determines the chirality (group velocity) of the edge modes \cite{poshakinskiyPRL2014}. Note the group velocity is defined over the modulation phase $\phi$, which acts as the wavenumber of the synthetic dimension. Considering the observed edge modes in the main gaps in Figs. \ref{longband} and \ref{longeigenstates}, we can conclude the bulk-boundary correspondence is valid for these two main band gaps. 

To further verify the gap-labeling theorem and bulk-boundary correspondence, the topological edge states in the minigaps with larger topological numbers are studied. A larger topological number indicates a narrower band gap \cite{piechonPRL1995}. Taking the low-energy minigap spanning the range of $-21\lesssim\Delta/\gamma\lesssim-19$ as an example, the normalized IDOS is $\mathcal{N}\approx146/1000=0.146$, leading to a solution of $\mu=2$, $\nu=-3$. For the second low-energy gap covering the range of $-17.3\lesssim\Delta/\gamma\lesssim-13.7$, we have $\mathcal{N}\approx236/1000=0.236$, giving rise to a topological characterization of $\mu=-1$, $\nu=2$. Regarding the edge eigenstates, we can easily find the number of edge modes on each edge that traverse the band gaps exactly corresponds to the gap Chern number. In Appendix \ref{moreedgestate} [Figs. \ref{state236pi01pi11} and \ref{state236pi08pi18}], the dipole moment distributions of representative midgap edge states belonging to the two left and right edge modes at the $\mathcal{N}\approx0.236$ gap are presented, selected from the longitudinal eigenstate spectra of lattices with specific modulation phases denoted in the figure legends. Similarly, the the dipole moment distributions of representative midgap edge states belonging to the three left and right edge modes at the $\mathcal{N}\approx0.146$ gap are also plotted in Appendix \ref{moreedgestate} [Figs. \ref{state146pi02pi04}-\ref{state146pi16pi18}]. In addition, we can confirm that the gap-labeling theorem can also describe the band topology of transverse band structures successfully and the obtained gap Chern number correctly predicts the number of edge modes traversing the gap. In Appendix \ref{moreedgestate} [Fig. \ref{transversestatesmallgaps}], similar to the longitudinal case, the the dipole moment distributions of representative midgap edge states selected from the $\mathcal{N}\approx0.236$ and $\mathcal{N}\approx0.146$ gaps are given. It is noted that these transverse edge states have larger localization length as a consequence of more significant long-range interactions due to the far-field term decaying as $1/r$ \cite{wang2018topological,wangPRB2018b}. Therefore, on the basis of the consistency between edge mode number and gap Chern number, we can confirm that in the present system, the bulk-boundary correspondence is valid.

\begin{figure}[htbp]
	\centering
	\subfloat{
		\includegraphics[width=\linewidth]{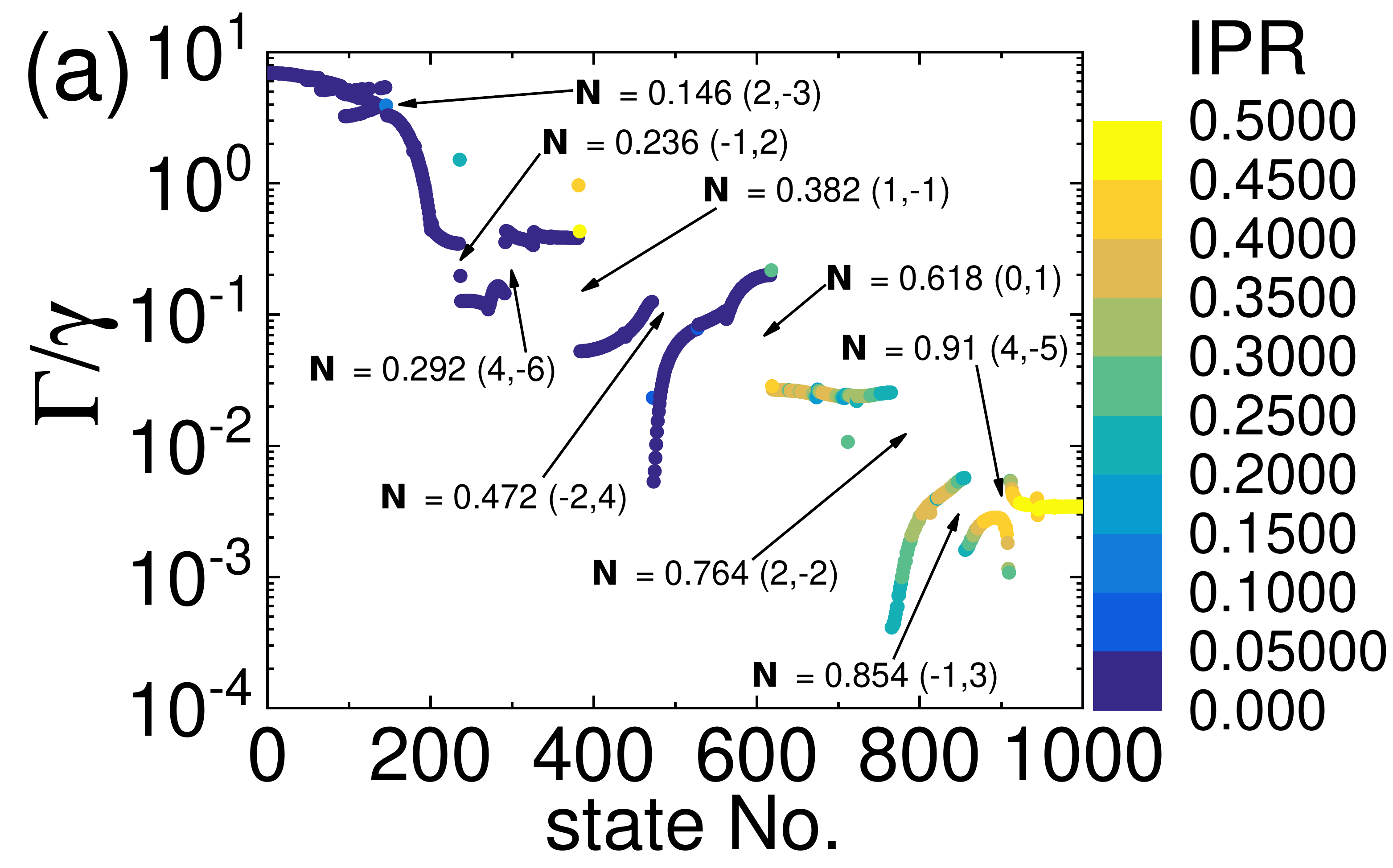}\label{incommd01eta0311pi_imag}
	}\\
	\subfloat{
		\includegraphics[width=\linewidth]{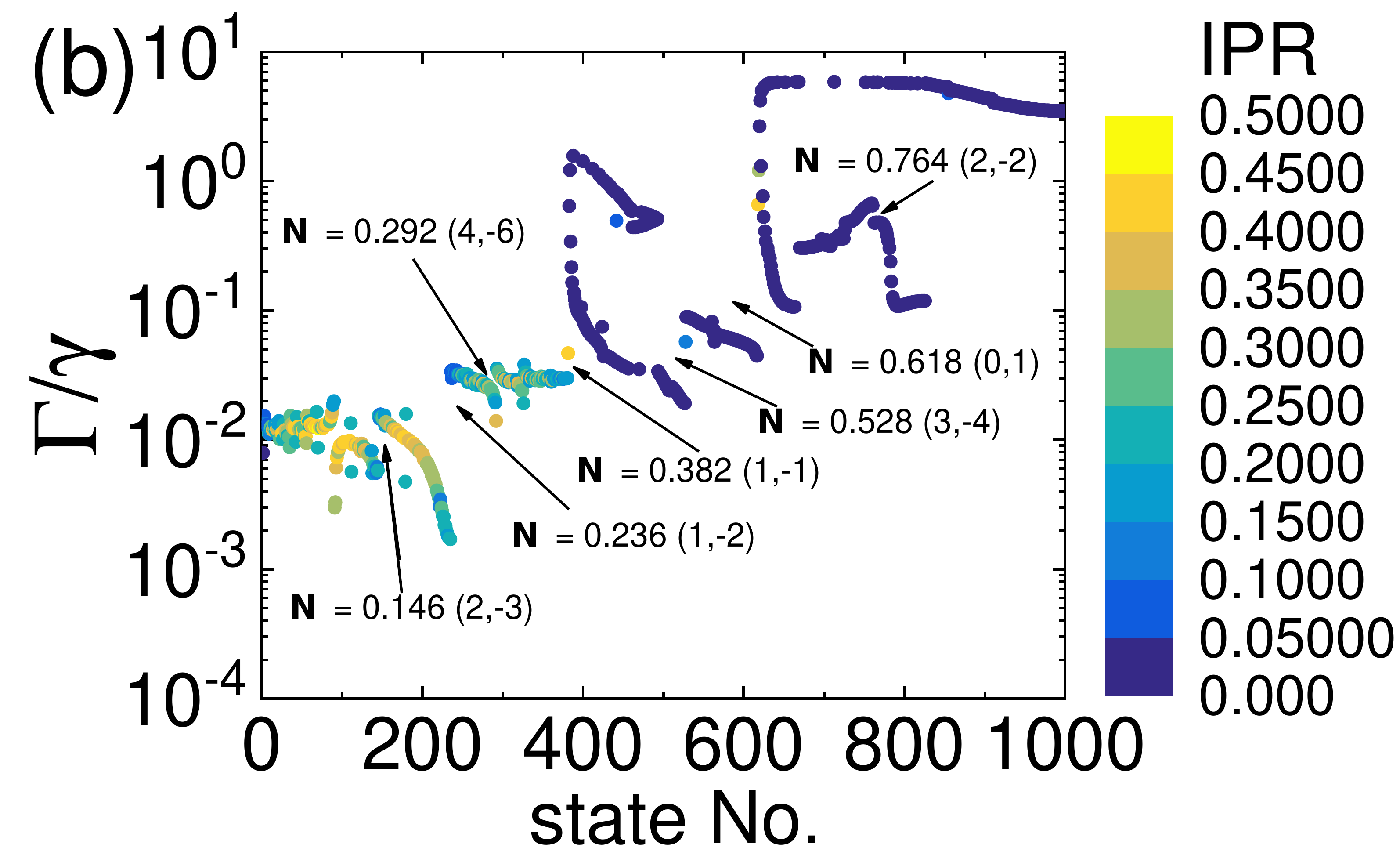}\label{transincommd01eta0311pi_imag}
	}

	\caption{The imaginary parts $\Gamma$ in the logarithmic scale of (a) longitudinal and (b) transverse band structures with the normalized IDOS and topological numbers of typical band gaps being labeled. Here the modulation phase is arbitrarily selected to be $\phi=1.1\pi$. There are 1000 atoms in the chain with $d=0.1\lambda_0$ and $\eta=0.3$.}
	\label{imagpart}
\end{figure}

The present system is in essence non-Hermitian \cite{perczelPRA2017,bettlesPRA2017,yelinPRL20172,heOL2020,heIJHMT2020}. Recent theoretical progresses in non-Hermitian topological physics, including the topological classification of non-Hermitian systems \cite{gongPRX2018,liuPRB2019,kawabataPRX2019}, theoretical investigation \cite{borgnia2019nonhermitian,songPRL2019} and experimental observation \cite{xiao2019observation,ghatak2019observation} of non-Hermitian bulk-boundary correspondence including non-Hermitian skin effect (NHSE) \cite{yaoPRL2018a,yaoPRL2018b,kunstPRL2018}, etc., has received a lot of attention. Therefore, it would be instructive to further investigate the imaginary parts of the band structures, which are shown in Figs. \ref{incommd01eta0311pi_imag} and \ref{transincommd01eta0311pi_imag} under a specific modulation phase $\phi=1.1\pi$ for longitudinal and transverse eigenstates, respectively. It is observed that the imaginary spectra also exhibit clear band gaps, which, as we have verified, can also be characterized by the gap-labeling theorem. The normalized IDOS and topological numbers $(\mu,\nu)$ of typical band gaps are also marked in the figure. An impressive feature of the longitudinal spectra is that due to the logarithmic scale minigaps with large gap Chern numbers can be distinguished, e.g., a gap with a Chern number of $\nu=-6$ (at $\mathcal{N}\approx0.292$) is identified in Fig. \ref{incommd01eta0311pi_imag}. We can also find there are 12 midgap states in this gap (not shown here), confirming the BBC again. Another interesting feature is that, in the longitudinal band structure, the topological edge states at high IDOS gaps are highly subradiant (e.g., the topological edge state at the $\mathcal{N}\approx0.91$ gap can reach $\Gamma/\gamma\sim7.8\times10^{-4}$), and in the transverse band structure, the topological edge states at low IDOS gaps are also highly subradiant (e.g., the topological edge state at the $\mathcal{N}\approx0.146$ gap can reach $\Gamma/\gamma\sim0.005$). Due to the significant suppression of the collective decay rate, these topological edge states therefore provide an appealing route for achieving highly protected and long-lived optical states, which are promising for controlling the emission of external quantum emitters \cite{yelinPRL20172} and high-efficiency robust quantum storage \cite{guimondPRL2019,zhangCommPhys2019}.

It is known that such long lifetimes of emergent subradiant eigenstates are usually due to the collective interference effect among a group of nearby atoms, especially stemming from near-field dipole-dipole interactions between pairs of atoms \cite{markelJOSAB1995,Schilder2016,grossPhysRep1982,wangJOSAB2020,fengPRA2014,kornovanPRA2019,zhangPRL2019,zhangPRResearch2020,wangARHT2020}. An observation of the orientations of atomic dipole moments in these subradiant eigenstates can further validate this argument. More precisely, when the induced dipoles of a pair of atom oscillate out of phase, their electric fields interference destructively in the far-field, leading to very small (radiative) decay rates ($\Gamma\ll\gamma$) \cite{markelJOSAB1995,Schilder2016,grossPhysRep1982,wangJOSAB2020,fengPRA2014,kornovanPRA2019,zhangPRL2019,zhangPRResearch2020}. 
This phenomenon of highly subradiant states is due to the resonant dipole–dipole interactions among closely distributed atoms, which especially exhibit a $r^{-3}$ behavior at small interatomic distances, as mentioned above and by many previous works \cite{Schilder2016,grossPhysRep1982,wangJOSAB2020,fengPRA2014,kornovanPRA2019,zhangPRL2019,zhangPRResearch2020}. Hence the decay rate sensitively depends on the interatomic distances. In the transverse case, the distribution of imaginary eigenfrequencies narrows down compared to the longitudinal case as a result of the presence of far-field dipole-dipole interactions. Such eigenstates with small decay rates disappear when interatomic distances increase to be comparable with or even larger than the wavelength (which are not shown here and one can refer to Refs. \cite{bettlesPRA2015,wangJOSAB2020}).

\section{Effects of Long-range dipole-dipole interactions}
\begin{figure*}[htbp]
	\centering
	\subfloat{
		\includegraphics[width=0.31\linewidth]{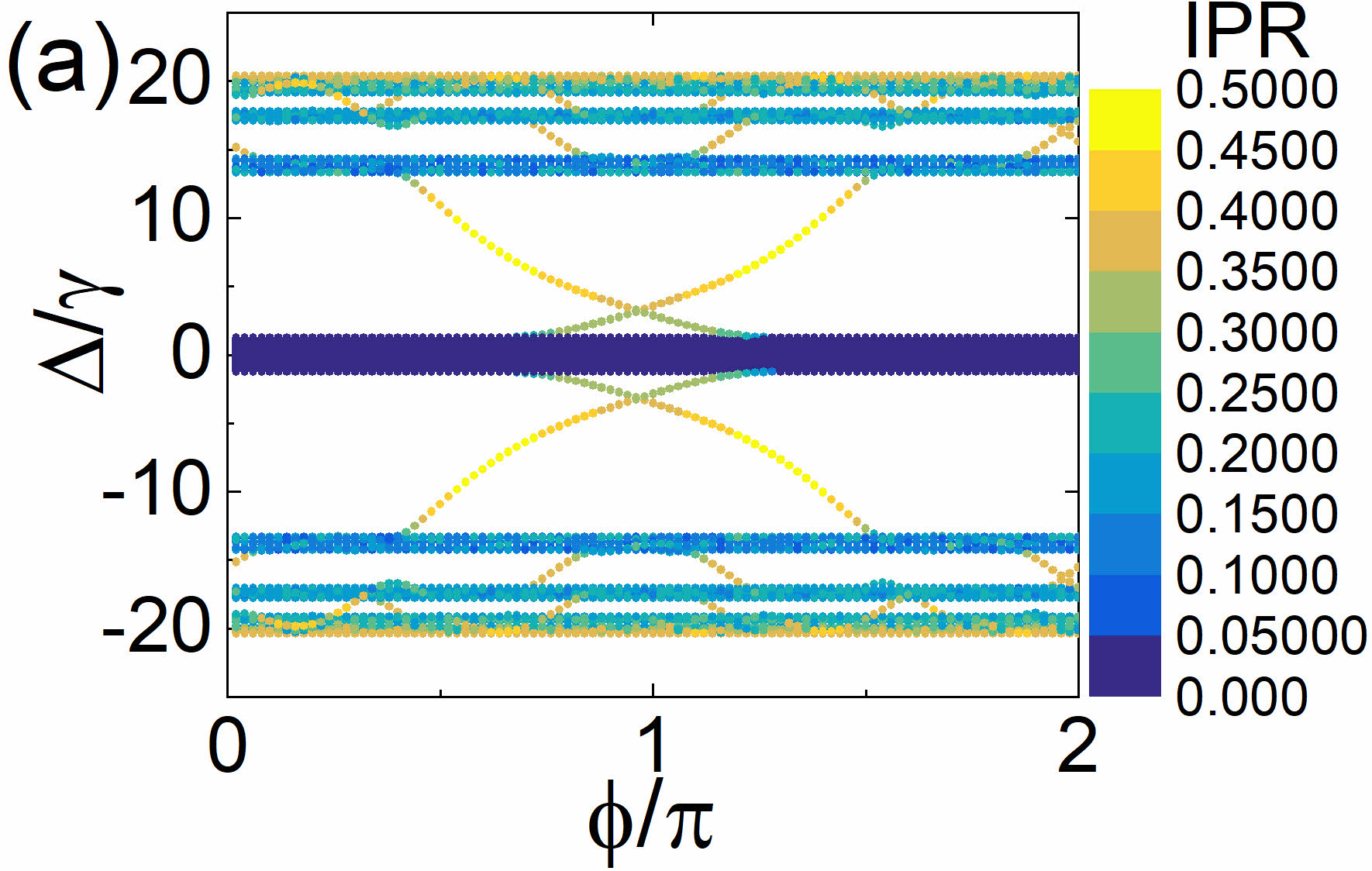}\label{incommd01eta03NN}
	}
	\subfloat{
		\includegraphics[width=0.31\linewidth]{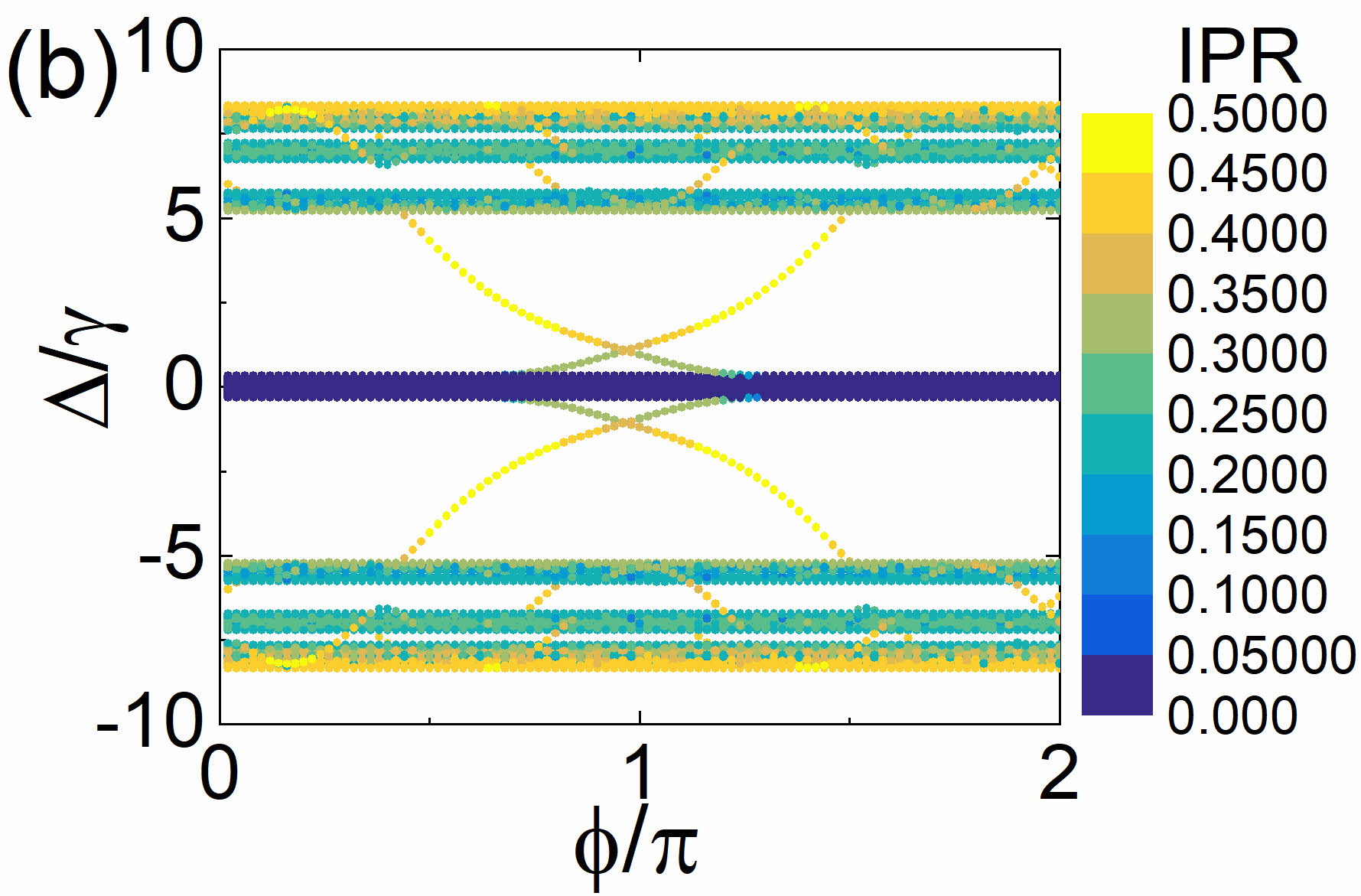}\label{transincommd01eta03NN}
	}
	\subfloat{
		\includegraphics[width=0.31\linewidth]{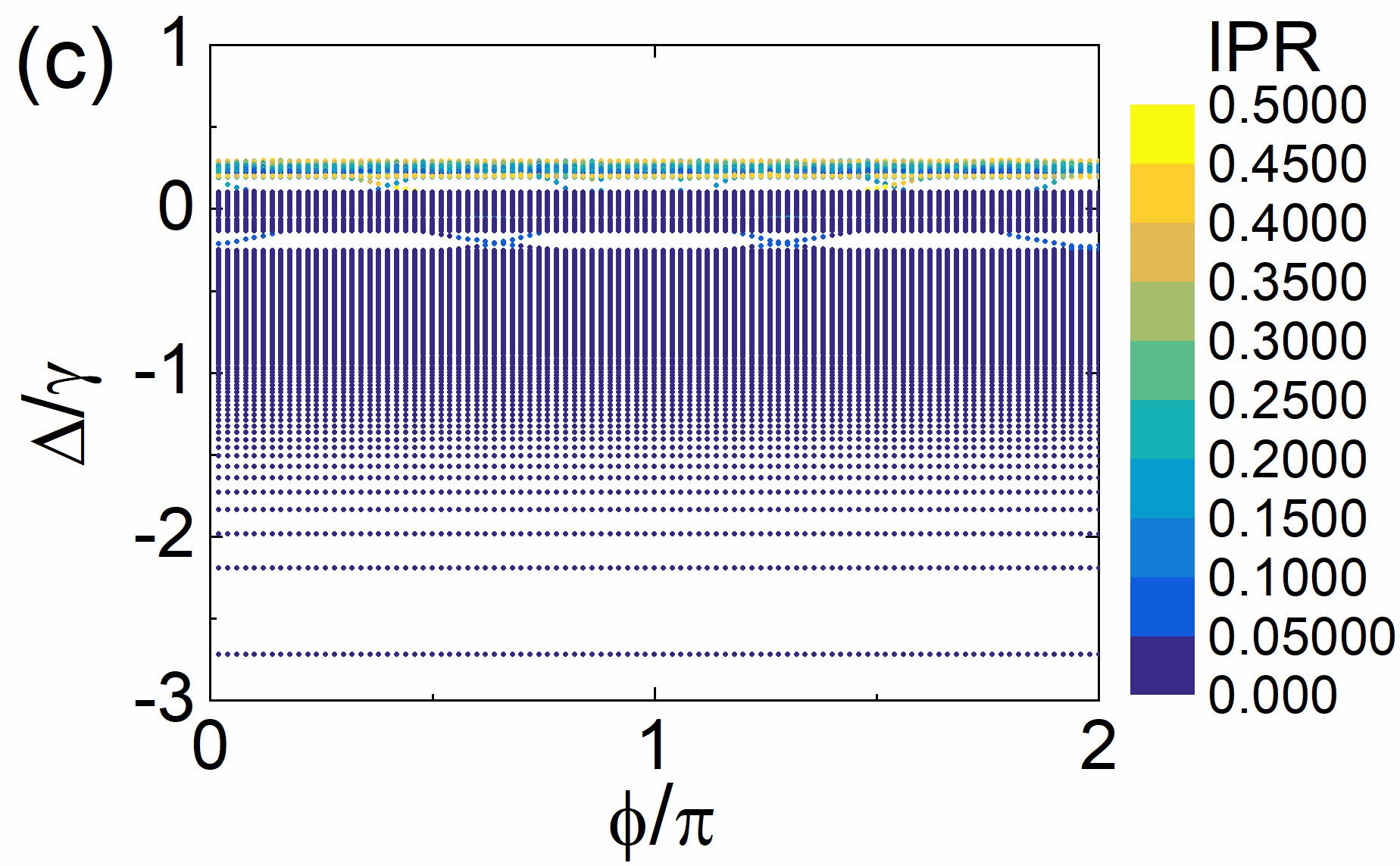}\label{transincommd05eta03}
	}\\	
	\subfloat{
		\includegraphics[width=0.31\linewidth]{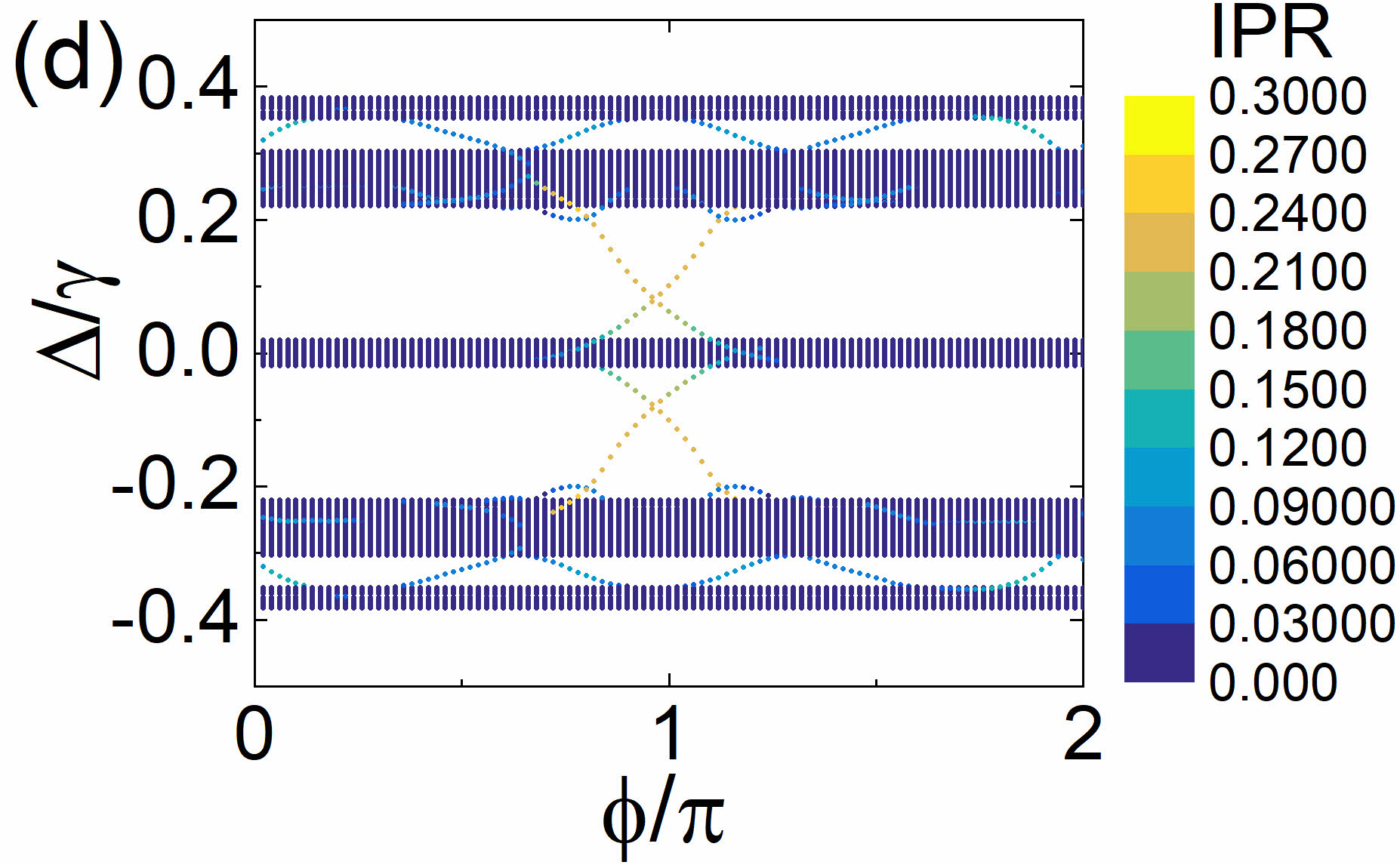}\label{transincommd05eta03NN}
	}
	\subfloat{
		\includegraphics[width=0.31\linewidth]{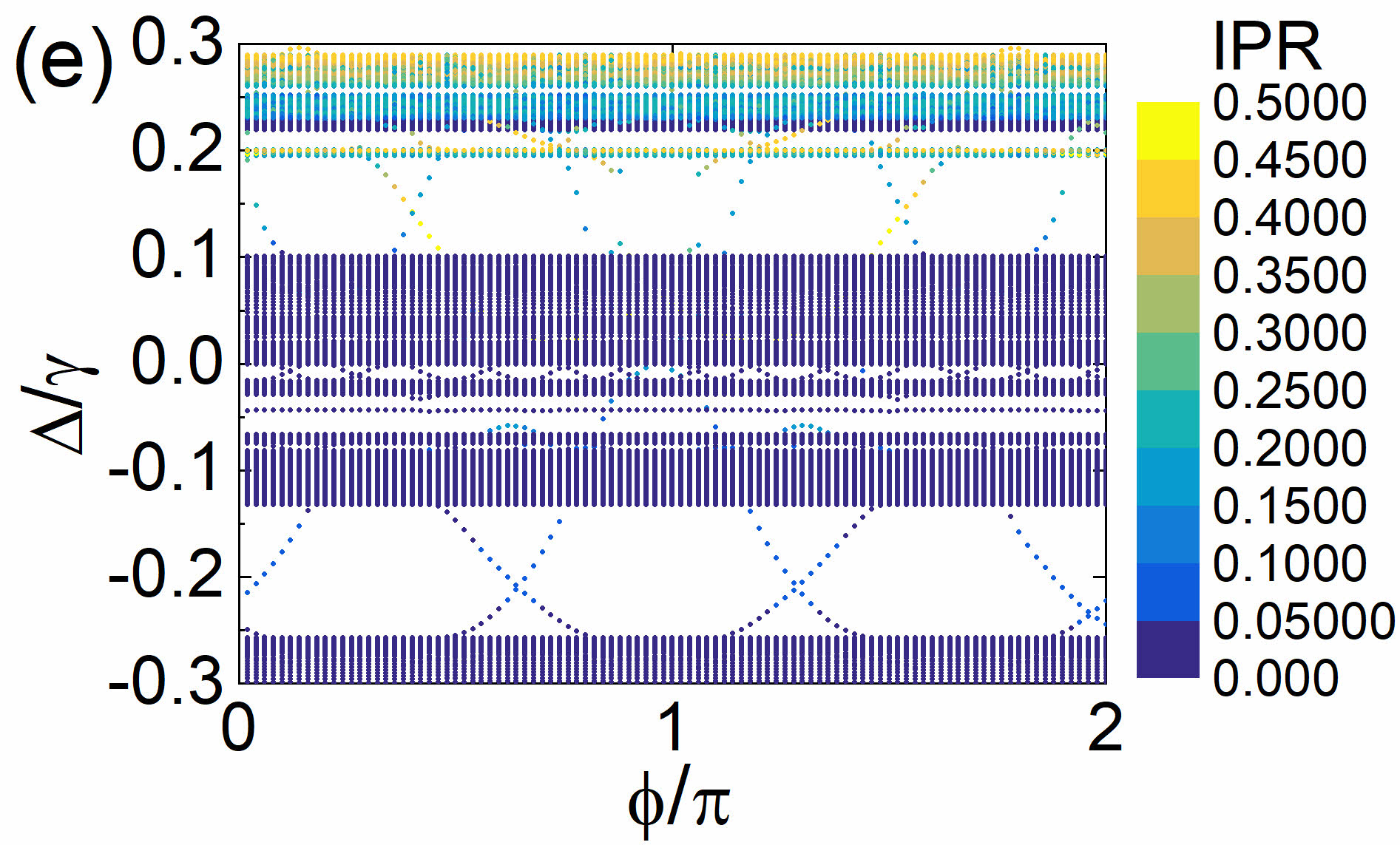}\label{transincommd05eta03upperband}
	}	
	\subfloat{
		\includegraphics[width=0.31\linewidth]{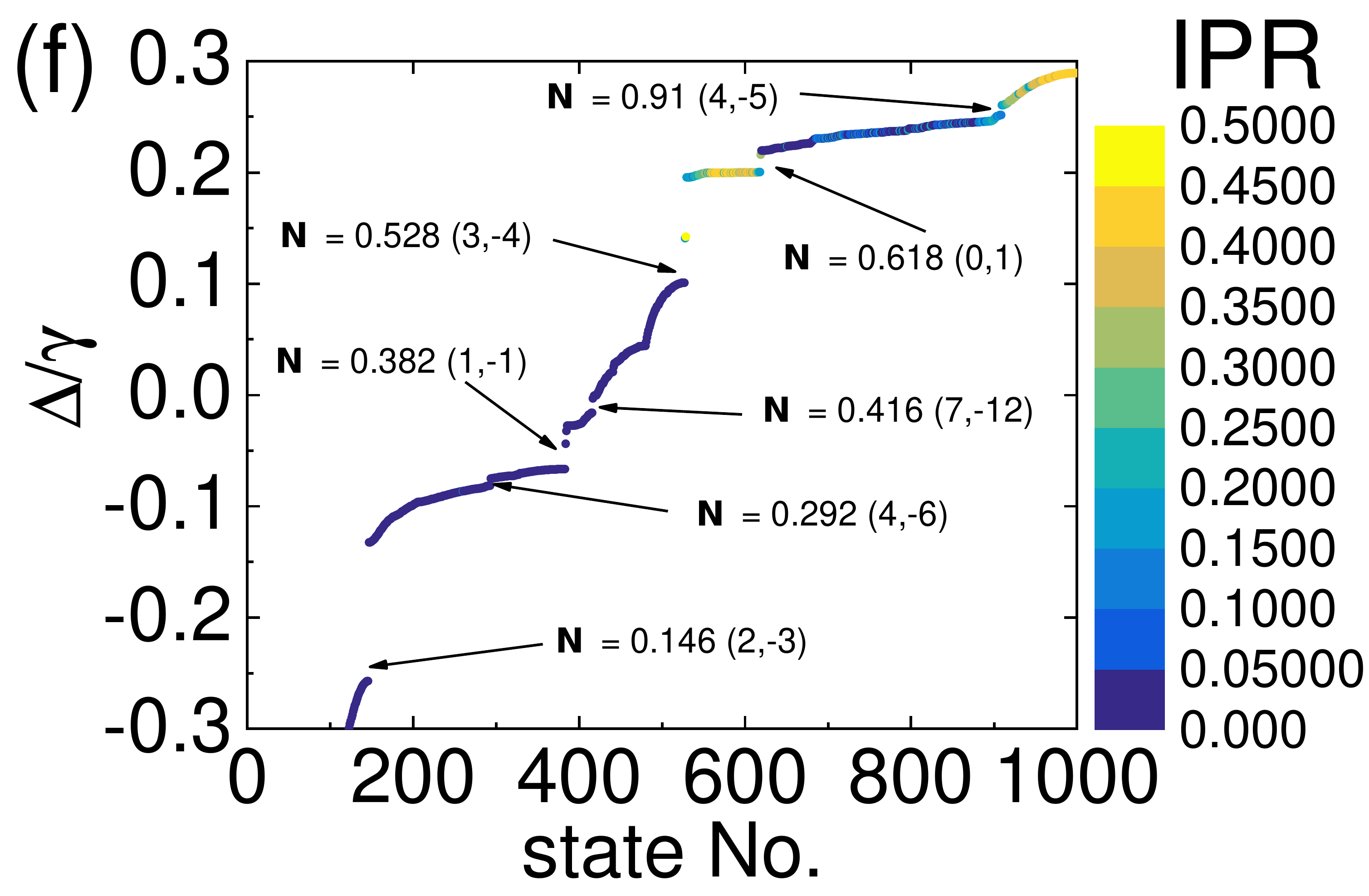}\label{transincommd05eta03pi04}
	}
	
	\caption{Role of long-range dipole-dipole interactions. (a-b) NN band structures as a function of the synthetic dimension wave vector $\phi$ for (a) longitudinal and (b) transverse eigenstates under $d=0.1\lambda_0$. (c-d) Band structures for transverse eigenstates under $d=0.5\lambda_0$ considering (c) full dipole-dipole interactions and (d) only NN interactions. (e) Enlarged display of the upper-in-frequency part ($\Delta>-0.3$) of the transverse band structures shown in (c). (f) The upper-in-frequency part ($\Delta>-0.3$) of transverse eigenstate distribution at $\phi=0.4\pi$ extracted from (e). The IDOS $\mathcal{N}$ and topological invariants $(\mu,\nu)$ for those clearly observable band gaps are indicated in this figure. Other system parameters are fixed as $\eta=0.3$ and $N=1000$.}
\end{figure*}

For completeness, the effects of long-range dipole-dipole interactions are further discussed with more details in this section.  We would like to emphasize that long-range dipole-dipole interactions indicate those interactions beyond the nearest-neighbor (NN) approximation that is usually assumed in conventional AAH model \cite{tanesePRL2014,levy2015topological,bandresPRX2016,dareauPRL2017,babouxPRB2017,krausPRL2012,krausPRL2012b,ganeshanPRL2013,verbinPRL2013}, namely, next-nearest-neighbor (NNN) and beyond. Long-range interactions usually play a nontrivial role in topological phases of matter \cite{bettlesPRA2017,wangPRB2018b,zhangCommPhys2019}. For instance, recently it was found that for certain lattices, long-range electromagnetic interactions can induce new class of topological corner states \cite{liNaturephton2019}. Our previous work also predicted a topological phase transition due to long-range interactions and then a breakdown of bulk-boundary correspondence due to the persistence of edge states in a topologically ``trivial" phase \cite{wangPRB2018b}. Nevertheless, these works were focused on periodic systems, while the role of long-range interactions in the topological properties of quasiperiodic systems remains elusive.

For doing this, the band structures for the 1000-atom lattice with $d=0.1\lambda_0$ and $\eta=0.3$ under the NN approximation are first calculated, as presented in Figs. \ref{incommd01eta03NN} and \ref{transincommd01eta03NN} for longitudinal and transverse eigenstates, respectively. After a comparison with the band structures considering full dipole-dipole interactions (Figs. \ref{incommd01eta03} and \ref{transincommd01eta03}), it is found that at small atomic distances ($kd\ll1$), the role of long-range interactions is not prominent because NN hoppings mainly due to strong near-field dipole-dipole interactions ($1/r^2$ and $1/r^3$ terms) dominate.
However, it is revealed that even for such small lattice constants the long-range dipole-dipole interactions lead to an asymmetric band structure. More precisely, the band structure under NN is exactly symmetric with respect to $\Delta=0$ and the IPR distribution of eigenstates is also symmetric \cite{wang2018topological}. The asymmetric spectrum is also found in other modified AAH model with high-order hopping terms beyond the NN approximation \cite{biddlePRB2011}.

Moreover, this asymmetry is most prominent at large atom distances for transverse eigenstates, as presented in Figs. \ref{transincommd05eta03} and \ref{transincommd05eta03NN}, which compare the band structures considering full dipole-dipole interactions and only NN interactions for a lattice with $d=0.5\lambda_0$ that is typically encountered in cold atom experiments.  In this circumstance with $kd>1$, the contribution of long-range hoppings (NNN and beyond)  mainly due to the slowly decaying far-field dipole-dipole interactions ($1/r$ term) cannot be neglected when compared with that of NN hoppings, as demonstrated by our previous works \cite{wangPRB2018b,wang2018topological}. Nevertheless, we note that these relatively significant long-range hoppings do not alter the topological properties of the present system qualitatively \cite{zhangCommPhys2019}. To demonstrate this, in Fig. \ref{transincommd05eta03upperband}, the upper-in-frequency part of the band structure of Fig. \ref{transincommd05eta03} ($\Delta>-0.3$) is presented in an enlarged fashion, from which midgap edge modes are clearly observed. Furthermore, the number of edge modes in a gap is consistent with the gap Chern number obtained from the gap-labeling theorem. In particular, the eigenstate distribution at $\phi=0.4\pi$ is given in Fig. \ref{transincommd05eta03pi04}, where the normalized IDOS $\mathcal{N}$ and topological integers $(\mu,\nu)$ are labeled for several typical gaps. Therefore, it can be confirmed that the gap-labeling theorem can give a suitable topological characterization of the present system and the bulk-boundary correspondence is valid, even when long-range interactions are prominent and NN approximation severely breaks down \cite{wangPRB2018b}. In Appendix \ref{largedistance}, we further exhibit the case of $d=\lambda_0$, in which similar conclusions can be drawn [Fig. \ref{d1transband}]. In some sense, the topological properties can remain unaffected as long as these long-range interactions do not close the band gaps, irrespective of the precise form of the interactions \cite{tanesePRL2014,luckPRB1989,levy2015topological,danaPRB2014}.

\section{Conclusions}
In this work, topological optical states in 1D quasiperiodic cold atomic chains are studied, which can be regarded as an extension of the off-diagonal AAH model, despite the existence of long-range dipole-dipole interactions in the Hamiltonian. The discrete band structures are investigated for finite chains beyond the NN approximation. It is found that the present system indeed supports nontrivial topological states localized over the boundaries. It is demonstrated that, for both longitudinal and transverse eigenstates, the present system inherits the topological properties of two-dimensional integer quantum Hall systems, and the spectral position (for both real and imaginary frequencies) and number of these edge states are governed by the gap-labeling theorem for quasicrystals and protected by the nonzero Chern number. These results indicate the validity of bulk-boundary correspondence in spite of long-range dipole-dipole interactions that can lead to asymmetric band structures. Due to the fractal nature of the spectrum, the present system readily provides a large number of topological gaps and optical states. Moreover, it is noted that a substantial proportion of these topologically nontrivial states are highly subradiant and thus are promising for controlling the emission of external quantum emitters and robust quantum state storage. Note that the optical states in a two-level cold atomic system are intrinsically quantum, since an individual atom cannot be excited twice \cite{massonPRResearch2020}, and our theoretical treatment also allows us to investigate the topological states in the single photon level \cite{yelinPRL20172,perczelPRL2020,zhangCommPhys2019}. This work thus provides useful implications for the design of efficient interfaces between quantum states of light and matter.

The proposed quasiperiodic cold atom chain is within reach of current quantum simulation techniques, e.g., by superimposing two optical lattices with incommensurate wavelengths \cite{roatiNature2008,schreiberScience2015,rajagopalPRL2019} or applying the cut-and-project procedure to a 2D optical lattice \cite{singhPRA2015}. Subwavelength lattice can be realized by using some ``magic" wavelength optical lattices with long-wavelength dipole transitions. For example, the transition between triplet states $^3P_0$ and $^3D_1$ of $^{84}$Sr gives emission at the wavelength of $\lambda=2.6~\mathrm{\mu m}$. One can use the optical lattice formed by lasers at 412.8 nm to trap the atoms, which achieves a subwavelength lattice spacing 206.4 nm, i.e, $d/\lambda\approx0.08$ \cite{olmosPRL2013}. Moreover, cutting-edge developments in the one-by-one assembling of atoms based on arrays of optical tweezers make the fabrication of such aperiodic atomic chains feasible \cite{endresScience2016,lahayeScience2016,bernienNature2017}. Nanophotonic atom lattices using dielectric photonic crystals \cite{gonzalezNaturephoton2015} or plasmonic nanoparticle arrays \cite{gullansPRL2012} also provide possible routes.

\begin{acknowledgments}
This work is supported by the National Natural Science Foundation of China (Grants No. 51636004 and No. 51906144), Shanghai Key Fundamental Research Grant (Grants No. 18JC1413300 and No. 20JC1414800), China Postdoctoral Science Foundation (Grants No. BX20180187 and No. 2019M651493), Open Fund of Key Laboratory of Thermal Management and Energy Utilization of Aircraft of Ministry of Industry and Information Technology (Grant No. CEPE2020015), and the Foundation for Innovative Research Groups of the National Natural Science Foundation of China (Grant No. 51521004).
\end{acknowledgments}

\appendix
\section{On the fractal nature of the eigenstate spectrum}\label{fractal_appendix}

\begin{figure*}[htbp]
	\centering
	\includegraphics[width=1\linewidth]{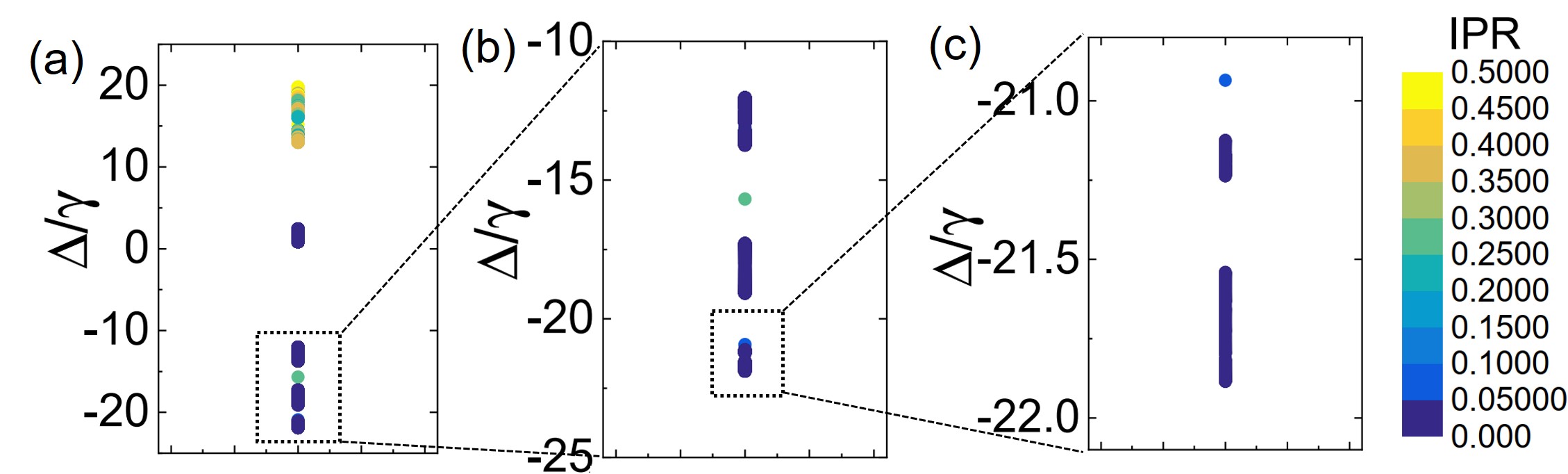}
	
	\caption{The fractal nature of the spectrum of longitudinal eigenstates. (a) The spectrum in the range of $-25\leq\Delta/\gamma\leq25$ in which the lowest band is identified and enlarged in (b). (b) The spectrum in the range of $-25\leq\Delta/\gamma\leq-10$ in which the lowest band is identified and enlarged in (c). (c) The spectrum in the range of $-22.5\leq\Delta/\gamma\leq-20.5$. System parameters are $\eta=0.3$, $\phi=0$, $N=3000$, $\beta=(\sqrt{5}-1)/2$ and $d=\lambda_0$. }
	\label{fractal_spectrum}
\end{figure*}

The fractal nature of the eigenstate spectrum is manifested as the fact that whenever we enlarge the spectrum, we can find finer features that resemble the ones in larger scales, especially the emergence of smaller and smaller band gaps \cite{silva2019phononic,kohmotoPRL1987}. In Fig. \ref{fractal_spectrum}, we can clearly see the self-similarity of the spectrum, which means a part of the spectrum well resembles the entire spectrum. More specifically, in all these figures, there are three main bands. Note the midgap state with a high-IPR in Fig. \ref{fractal_spectrum}(b) cannot be regarded as an additional band since it is only a single edge state that does not persist in the band. Moreover, if the atomic chain is infinitely long, it is possible to continue with this process to an infinite order, in which each band can be further split into three narrower bands. In this sense, the spectrum is evidently fractal. However, since here we only consider the case of $N=3000$ atoms, this process cannot be further employed. Other methods like the one that calculates the scaling of total bandwidth with the length of the chain can also be used to further demonstrate this fractality, which tends to be zero as the length of the chain goes to infinity \cite{vasconcelosPRB1998,kohmotoPRL1983b,albuquerquePhysrep2003}.

\section{The gap-labeling theorem}\label{gaplabel_appendix}
In this appendix, we present more introductory details on the gap-labeling theorem [Eq. (\ref{gaplabel_eq})], which are well-established and extensively discussed in many previous works.

Thouless, Kohmoto, Nightingale and den Nijs (TKNN) \cite{thoulessPRL1982} considered particular models of Bloch electrons in ``rational" magnetic fields with flux $\varphi=\varphi_0p/q$ per unit cell, where $\varphi_0$ is the quantum of magnetic flux and $p, q$ are coprime integers. They
showed that a magnetic band $b$, which arises from the magnetic translational symmetry, is characterized by an integer, here denoted by $\nu_b$, giving the contribution $\nu_be^2/h$ of the band to the quantized Hall conductance of the system in linear-response theory, in which $e$ and $h$ are the elementary charge and Planck's constant respectively. This integer is a Chern topological
invariant for the band \cite{avronPRL1983,simonPRL1983} and satisfies the Diophantine
equation \cite{thoulessPRL1982,macdonaldPRB1984,danaJPC1985,danaPRB2014}:
\begin{equation}
p\nu_b+q\mu_b=1.
\end{equation} 
This equation thus constitutes the topological description of the IQHE in a 2D periodic potential. 
Then, by summing this equation over $n$ magnetic bands, we have 
\begin{equation}
\bar{\beta}\nu+\mu=\bar{\mathcal{N}}
\end{equation}
with $\bar{\beta}=p/q$ and $\bar{\mathcal{N}}=n/q$ for the gap between $n$-th and $(n+1)$-th magnetic bands \cite{danaJPC1985}. In this sense, Eq. (\ref{gaplabel_eq}) in the main text can be regarded as the limiting case of above equation by taking $p\rightarrow\infty$ and $q\rightarrow\infty$, to make $\bar{\beta}\rightarrow\beta$ become an irrational number, and the $\bar{\mathcal{N}}$ approaches $\mathcal{N}$ as the IDOS, since in this irrational case a magnetic band reduces to an infinitely degenerate level as a result of Cantor set feature of the spectrum \cite{danaPRB2014}. In other words, this is due to the fact that the spectrum is a Cantor set with Lebesgue measure zero \cite{kohmotoPRL1983b,kohmotoPRL1987,avilaAnnalsMath2009}. In such a 2D IQHE system, $\nu e^2/h$ is the quantum Hall conductance of the system \cite{danaJPC1985,kunzPRL1986,macDonaldPRB1983} and $\mu e$ is the charge per unit cell that is transported when the periodic potential is dragged adiabatically by one lattice constant \cite{kunzPRL1986,macDonaldPRB1983}.




As introduced by Dareau \textit{et al} \cite{dareauPRL2017}, the gap-labeling theorem enables the topological classification of these gaps and plays for quasiperiodic systems a similar role to that of Bloch theorem for
periodic ones \cite{bellissardCMP1989,bellissardRMP1992}. More precisely, Bloch theorem labels the
eigenstates of a periodic system with a quasimomentum
and identifies topological invariants (Chern numbers) expressed in terms of a Berry curvature. This labeling is robust as long as the lattice translational symmetry is preserved. Similarly, the gap-labeling theorem permits to associate integer-valued topological invariants to each gap, which are \textit{K}-theory invariants. Note they are not strictly speaking Chern numbers which describe the topology of smooth Riemannian manifolds. This is because quasiperiodic systems cannot be ascribed to such a smooth manifold. Nevertheless there may exist an interpolation between both situations that could establish a link between Chern numbers and above topological numbers appearing in the gap-labeling theorem \cite{krausPRL2012}. As demonstrated by Bellissard and coworkers \cite{bellissardCMP1989,bellissardRMP1992,bellissard1992gap}, these integers can be given both a topological meaning and invariance properties akin in nature to Chern numbers but not expressible in terms of a curvature.

As we have mentioned earlier in the main text, due to the quasiperiodic modulation, the AAH model interestingly possesses nontrivial topological properties that can be mapped to the 2D quantum Hall system, without the need to apply a magnetic field. The modulation phase $\phi$ plays the role of momentum in a perpendicular synthetic dimension, leading to a dimensional extension. Therefore, the topological properties of our system can also be characterized by this gap-labeling theorem.

\section{Illustrations of selected edge states}\label{moreedgestate}
\begin{figure}[htbp]
	\centering
	\subfloat{
		\includegraphics[width=0.4\linewidth]{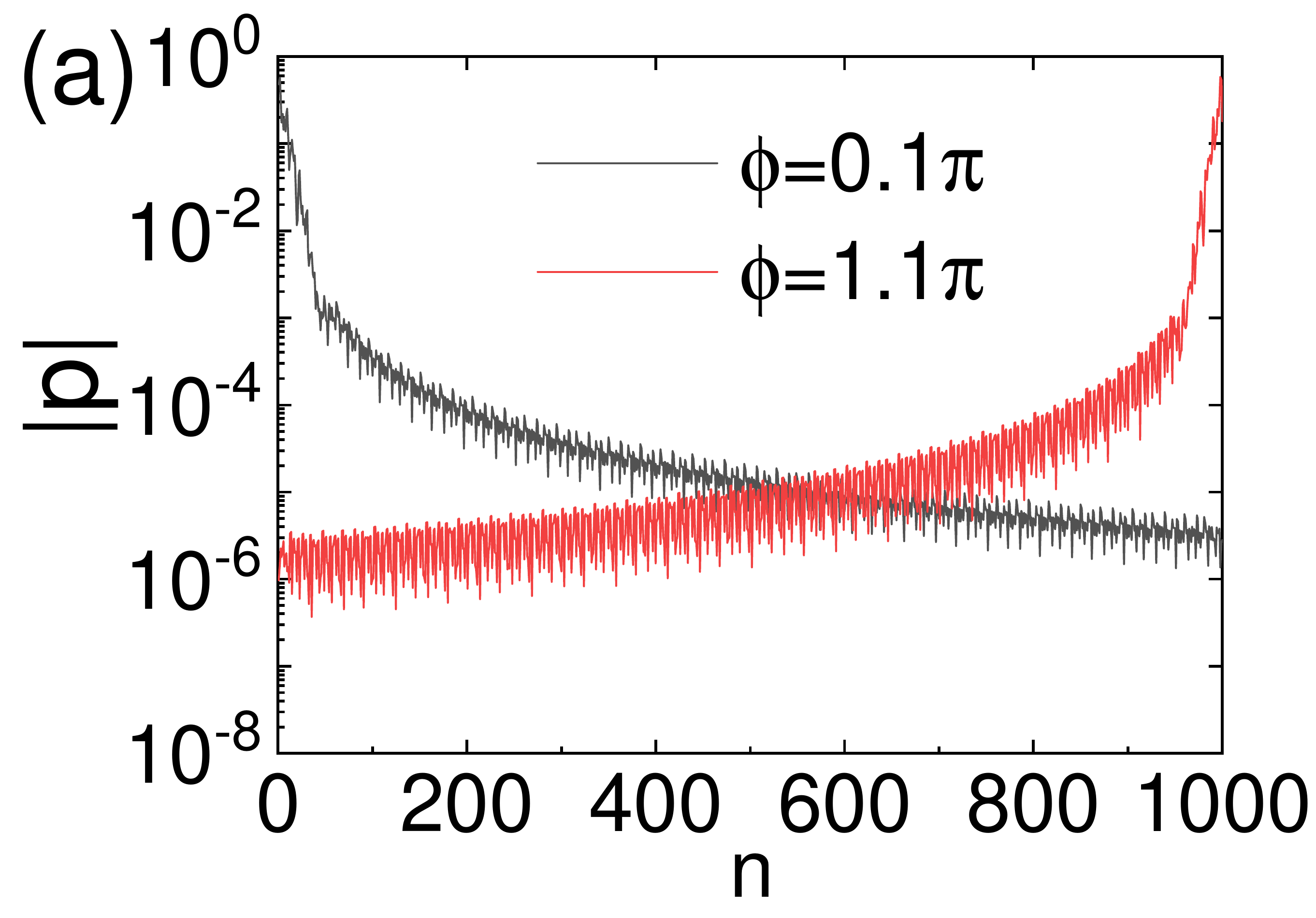}\label{state236pi01pi11}
	}
	\subfloat{
		\includegraphics[width=0.4\linewidth]{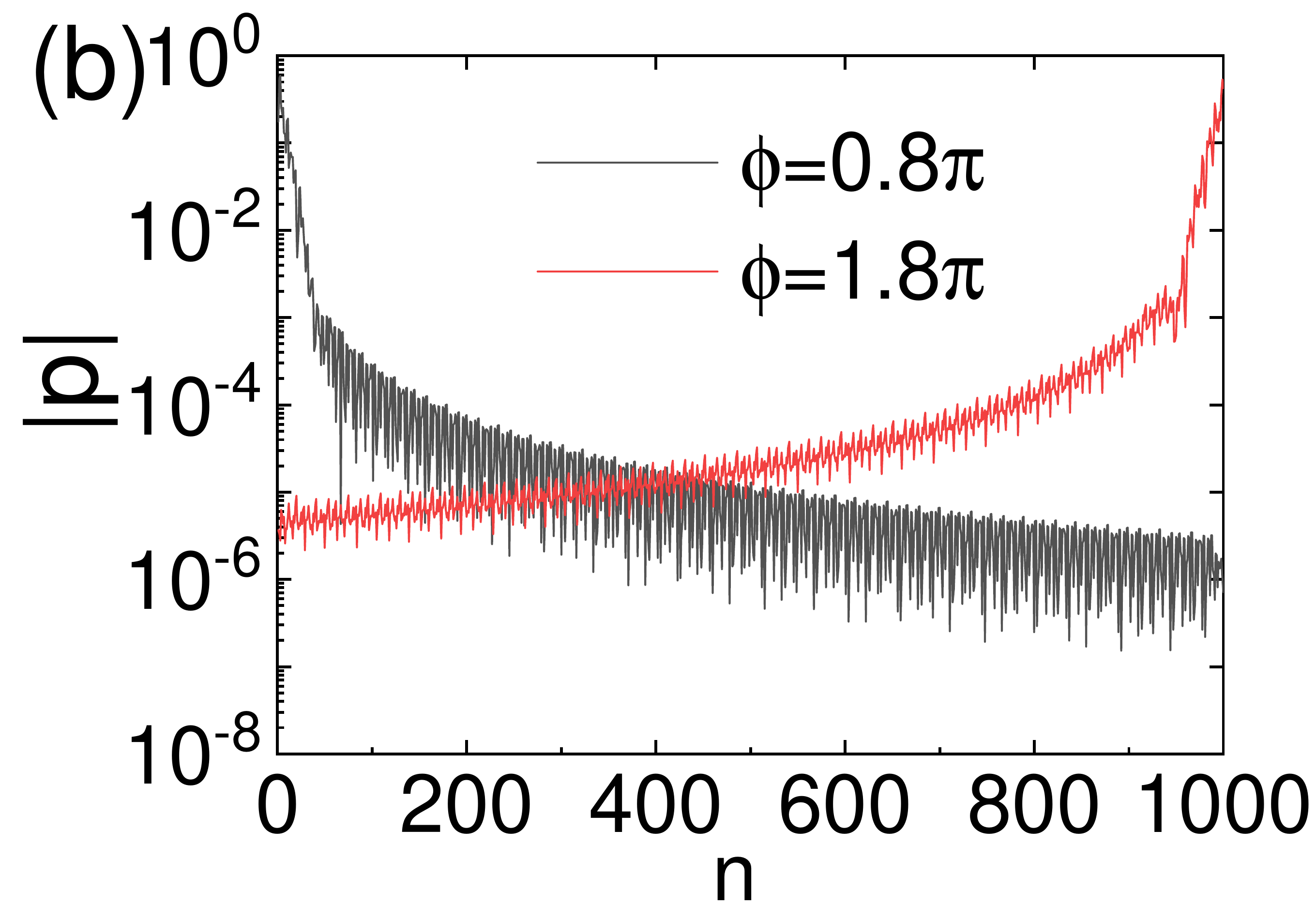}\label{state236pi08pi18}
	}
	\hspace{0.01in}
	\subfloat{
		\includegraphics[width=0.33\linewidth]{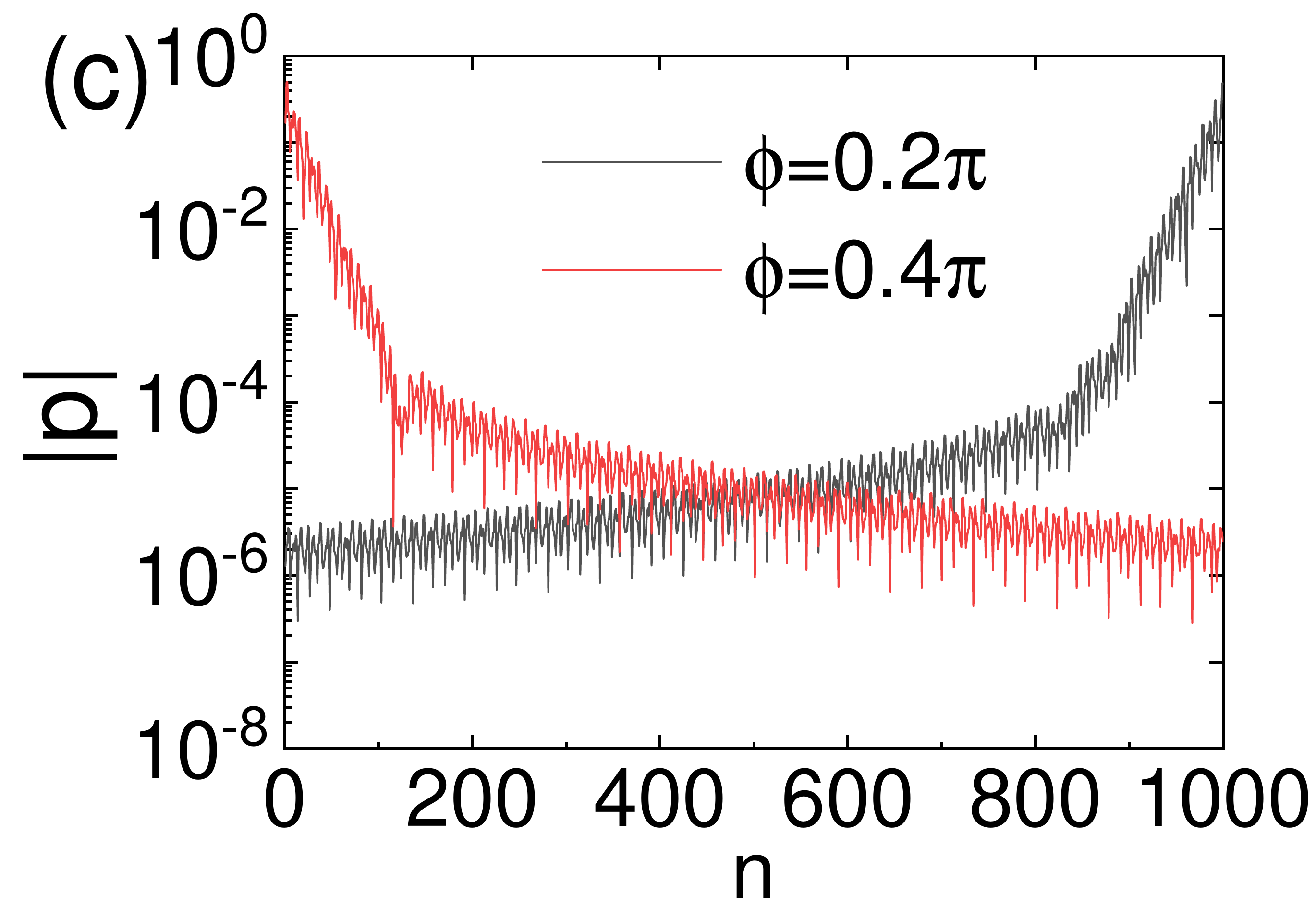}\label{state146pi02pi04}
	}
	\subfloat{
		\includegraphics[width=0.33\linewidth]{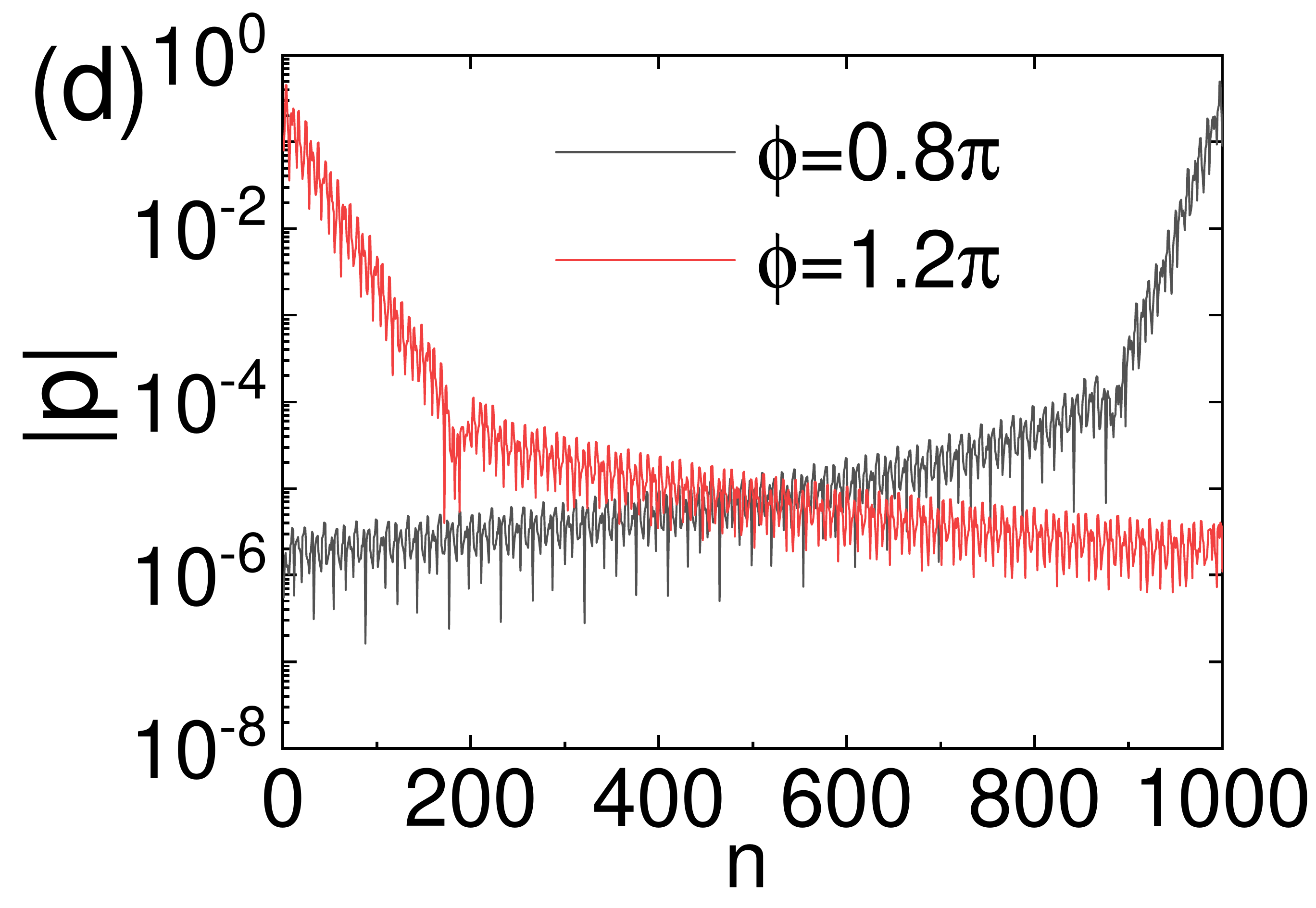}\label{state146pi08pi12}
	}
	\subfloat{
		\includegraphics[width=0.33\linewidth]{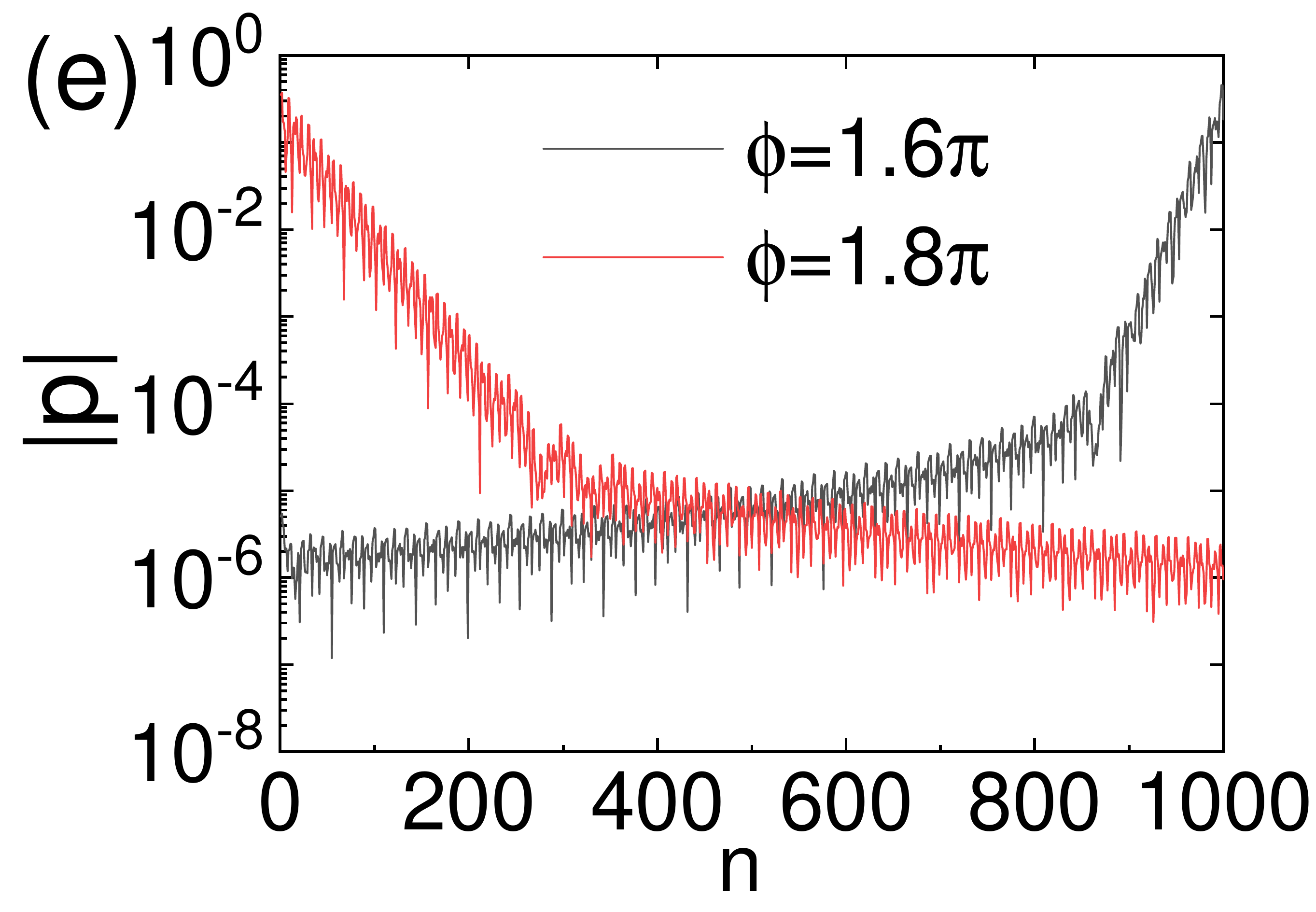}\label{state146pi16pi18}
	}
	
	\caption{Dipole moment distributions of topological edge states in two minigaps in the longitudinal band structure. (a-b) The $\mathcal{N}=0.236$ gap. (a) Left edge state under the modulation of $\phi=0.1\pi$ and right edge state under $\phi=1.1\pi$. (b)  Left edge state under the modulation of $\phi=0.8\pi$ and right edge state under $\phi=1.8\pi$. (c-e) The $\mathcal{N}=0.146$ gap. (c) Right edge state under the modulation of $\phi=0.2\pi$ and left edge state under $0.4\pi$. (d) Right edge state under the modulation of $\phi=0.8\pi$ and left edge state under $\phi=1.2\pi$ (e)  Right edge state under the modulation of $\phi=1.6\pi$ and left edge state under $\phi=1.8\pi$. There are 1000 atoms in the chain with $d=0.1\lambda_0$ and $\eta=0.3$. }
	\label{longstatesmallgaps}
\end{figure}

\begin{figure}[htbp]
	\centering
	\subfloat{
		\includegraphics[width=0.4\linewidth]{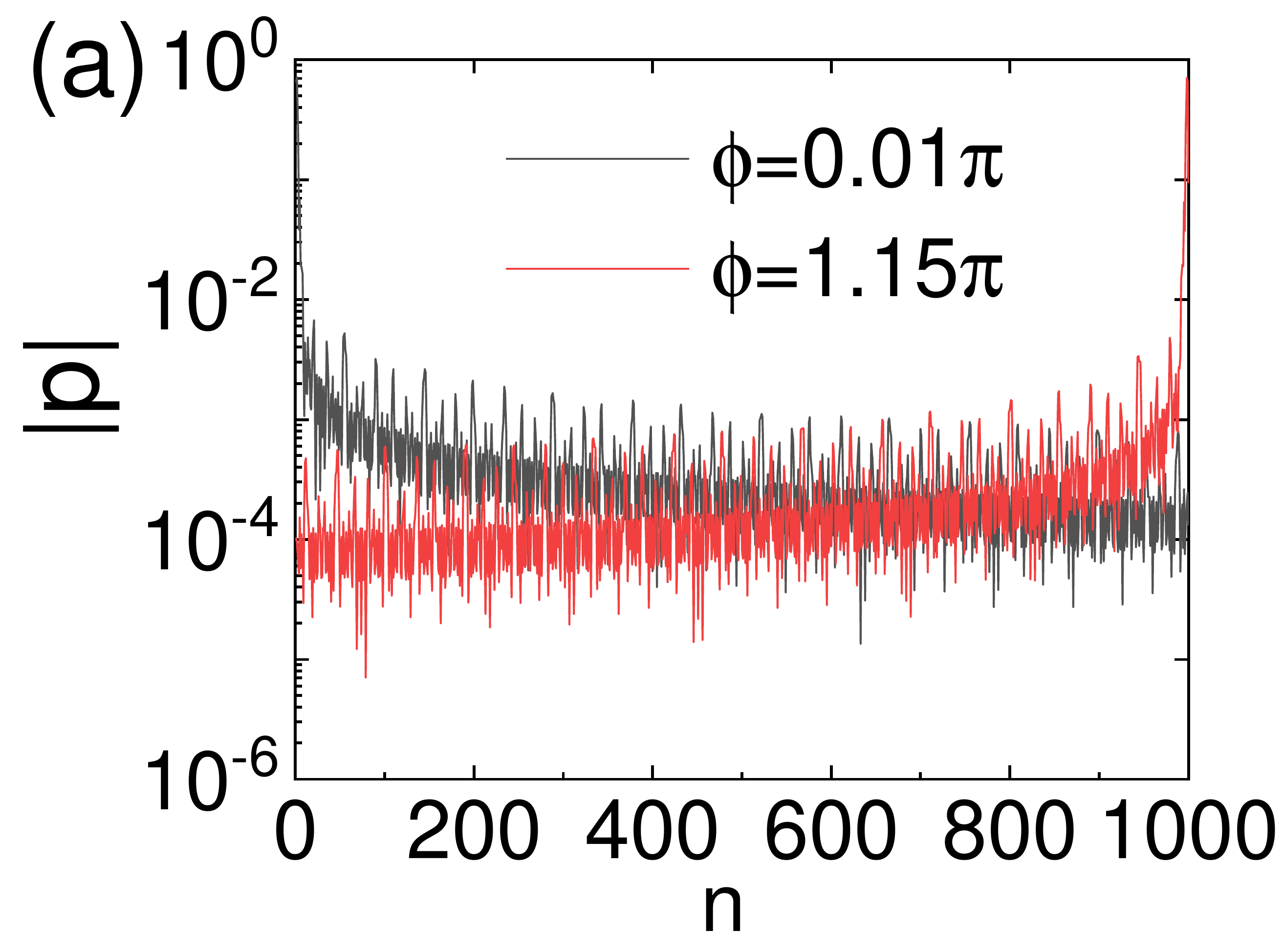}\label{transstate236pi001pi115}
	}
	\subfloat{
		\includegraphics[width=0.4\linewidth]{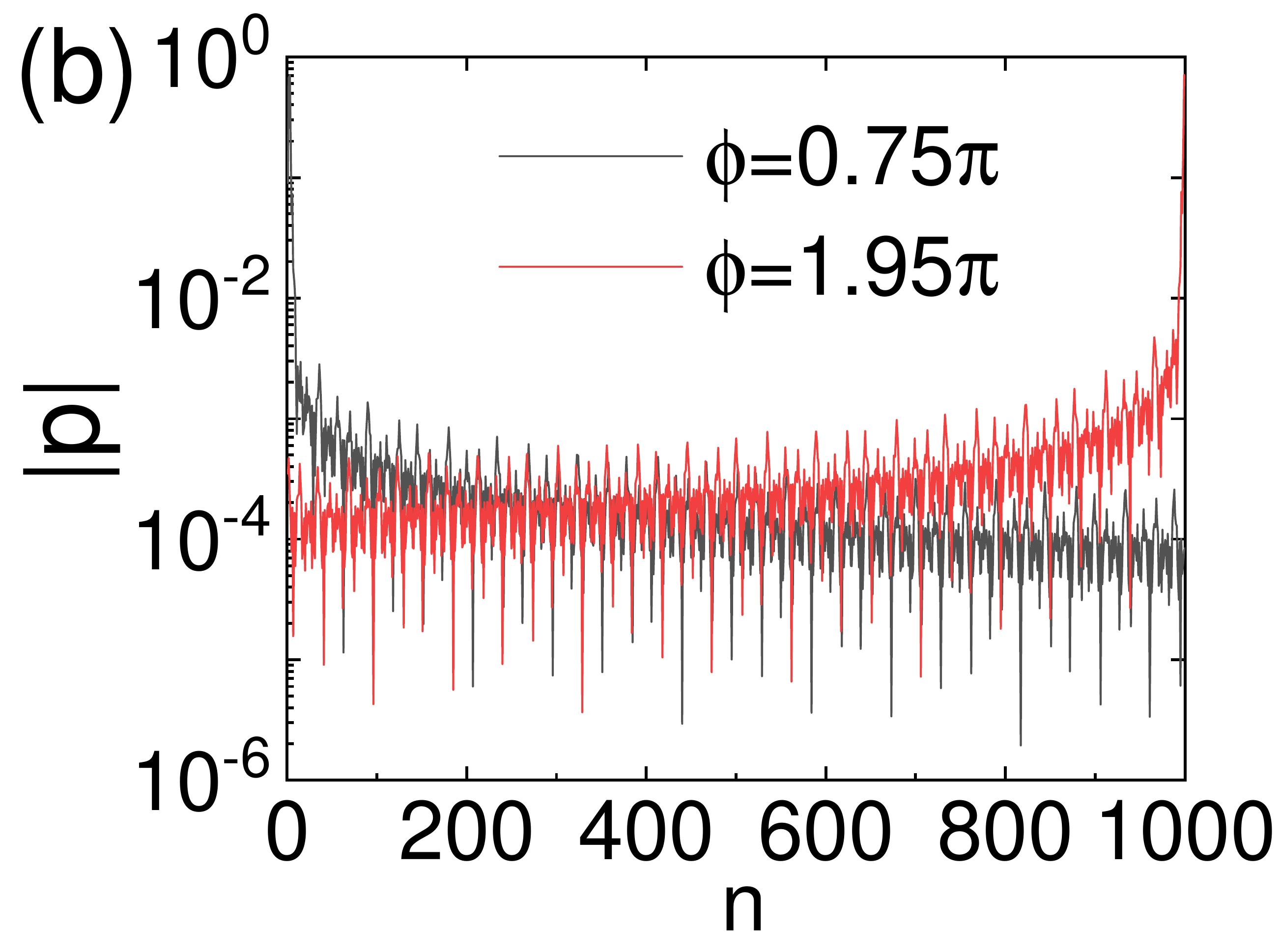}\label{transstate236pi075pi195}
	}
	\hspace{0.01in}
	\subfloat{
		\includegraphics[width=0.33\linewidth]{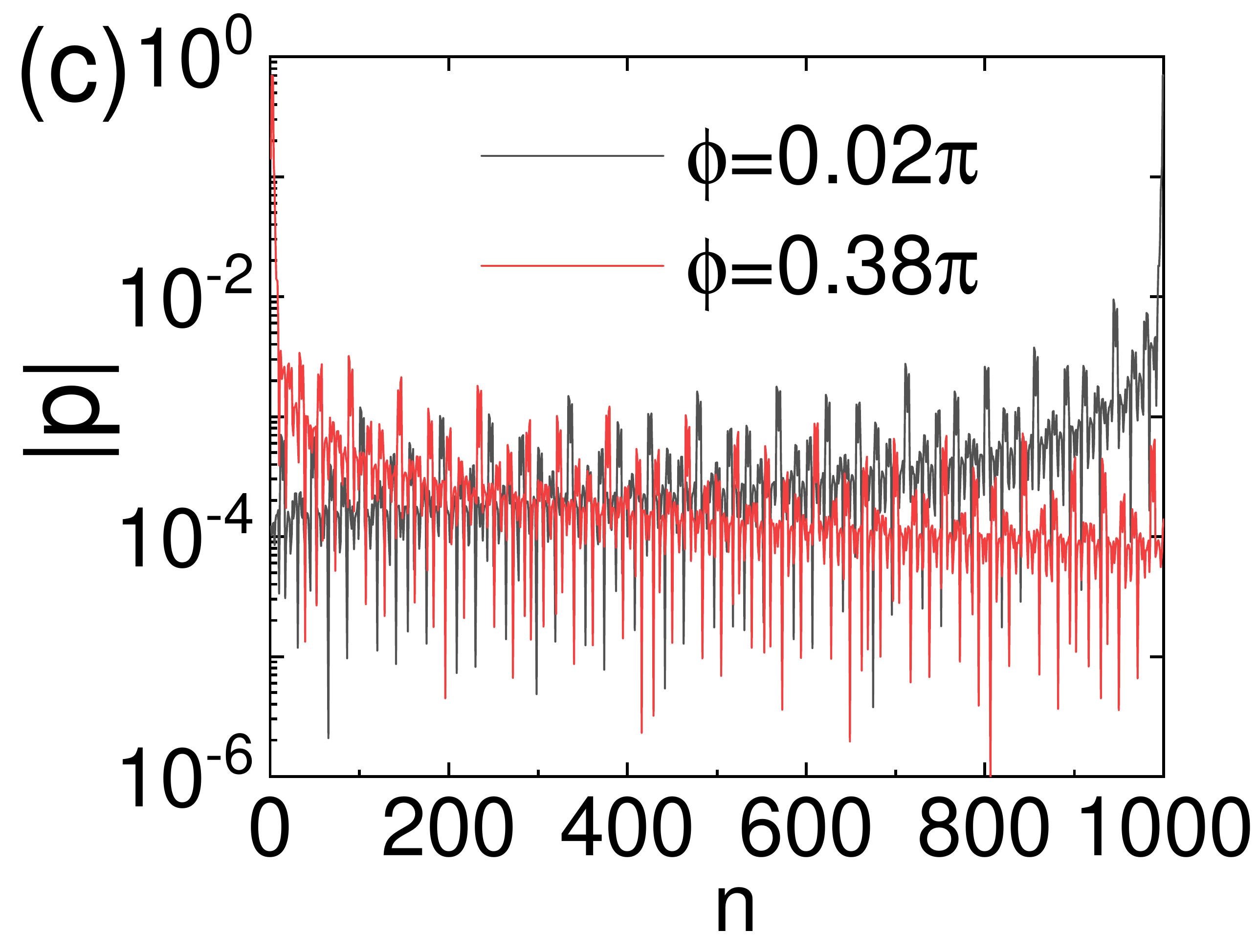}\label{transstate146pi002pi038}
	}
	\subfloat{
		\includegraphics[width=0.33\linewidth]{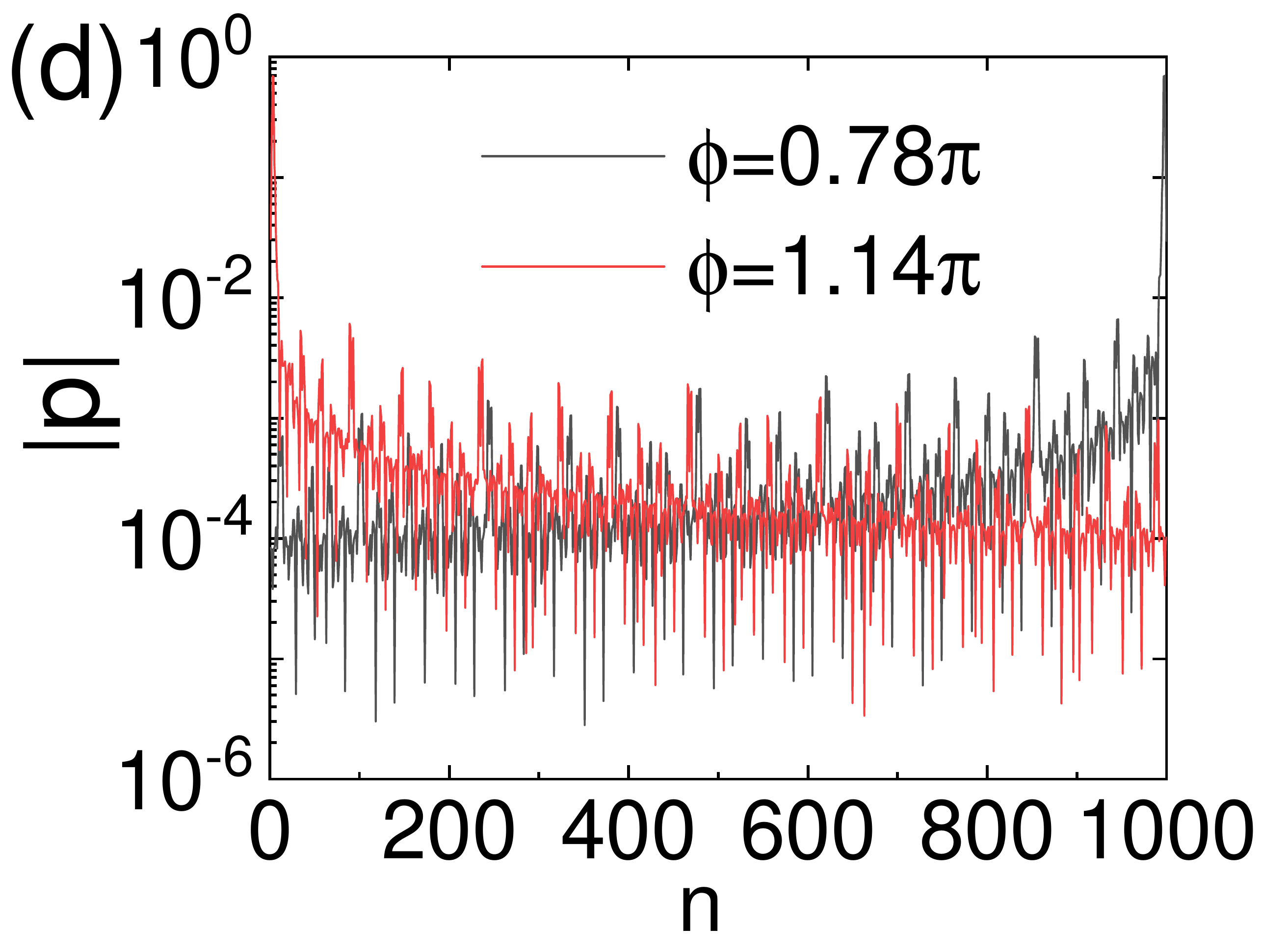}\label{transstate146pi078pi114}
	}
	\subfloat{
		\includegraphics[width=0.33\linewidth]{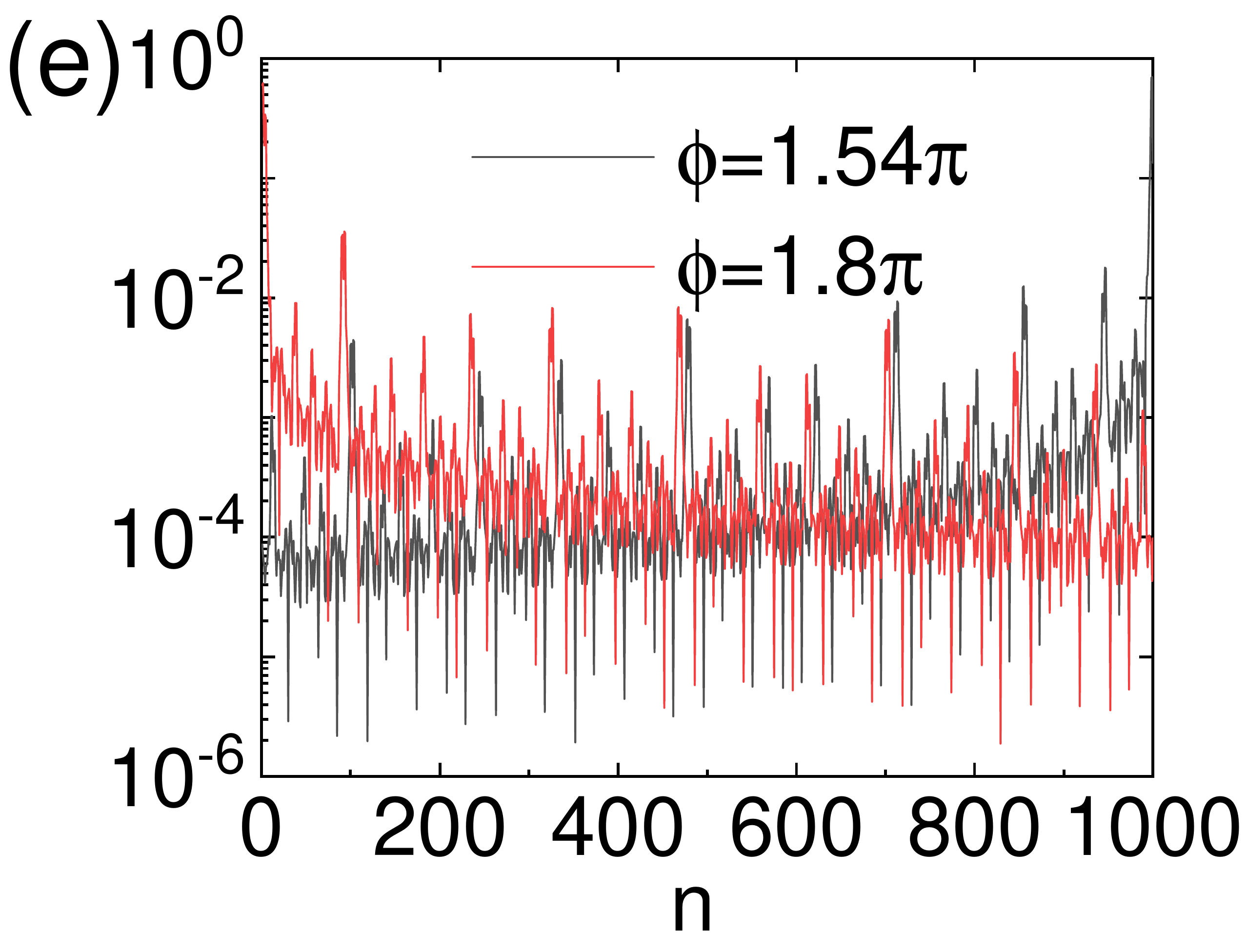}\label{transstate146pi154pi18}
	}
	
	\caption{Dipole moment distributions of topological edge states in two minigaps in the transverse band structure. (a-b) The $\mathcal{N}=0.236$ gap. (a) Left edge state under the modulation of $\phi=0.01\pi$ and right edge state under $\phi=1.15\pi$. (b) Left edge state under the modulation of $\phi=0.75\pi$ and right edge state under $\phi=1.95\pi$. (c-e) The $\mathcal{N}=0.146$ gap. (c) Right edge state under the modulation of $\phi=0.02\pi$ and left edge state under $\phi=0.38\pi$. (d) Right edge state under the modulation of $\phi=0.78\pi$ and left edge state under $\phi=1.14\pi$. (e)  Right edge state under the modulation of $\phi=1.54\pi$ and left edge state under $\phi=1.8\pi$. There are 1000 atoms in the chain with $d=0.1\lambda_0$ and $\eta=0.3$. }
	\label{transversestatesmallgaps}
\end{figure}

In Figs. \ref{state236pi01pi11} and \ref{state236pi08pi18}, the dipole moment distributions of representative midgap edge states belonging to the two left and right edge modes at the $\mathcal{N}\approx0.236$ gap are presented, selected from the longitudinal eigenstate spectra of lattices with specific modulation phases that is given in the figure legends. There are $N=1000$ atoms in the chain with $d=0.1\lambda_0$ and $\eta=0.3$. Similarly, the dipole moment distributions of representative midgap edge states belonging to the three left and right edge modes at the $\mathcal{N}\approx0.146$ gap are also plotted in Figs. \ref{state146pi02pi04}-\ref{state146pi16pi18}. In Fig. \ref{transversestatesmallgaps}, similar to the longitudinal case, the the dipole moment distributions of representative midgap edge states selected from the $\mathcal{N}\approx0.236$ and $\mathcal{N}\approx0.146$ gaps are given. Here we would like to point out that the topologically protected edge states actually decay exponentially from the edge with a short localization length while decaying algebraically in the long range (namely when far from the edge), which is due to the power-law decaying dipole-dipole interactions, as demonstrated in previous work \cite{wangPRB2018b} as well as the work presented by Dong \textit{et al} for a 2D Fibonacci lattice \cite{dongPRB2009}, among many others. 

\section{Large interatomic distances}\label{largedistance}

\begin{figure*}[htbp]
	\centering
	\subfloat{
		\includegraphics[width=0.4\linewidth]{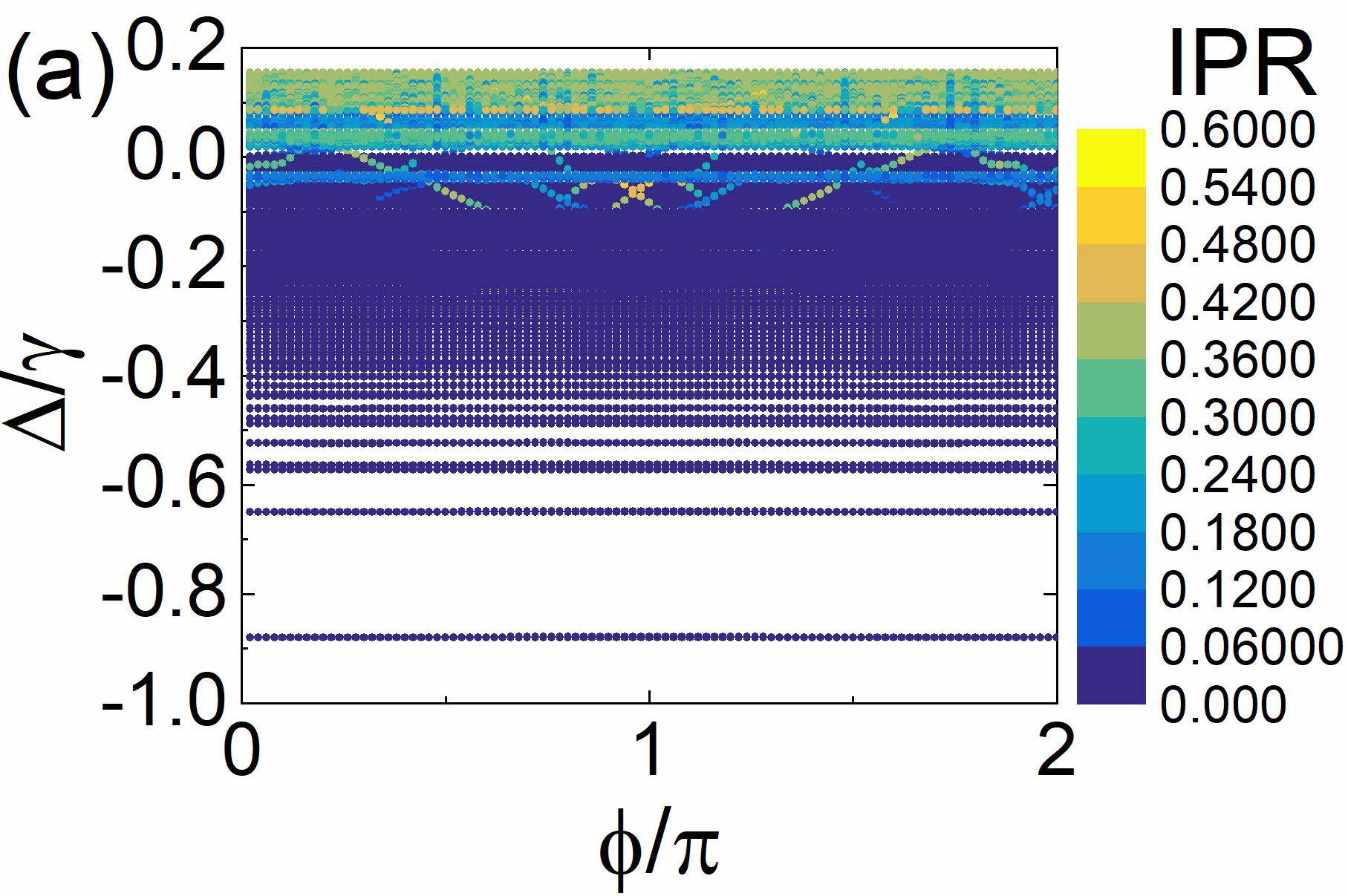}\label{incommd1eta03}
	}
	\subfloat{
		\includegraphics[width=0.4\linewidth]{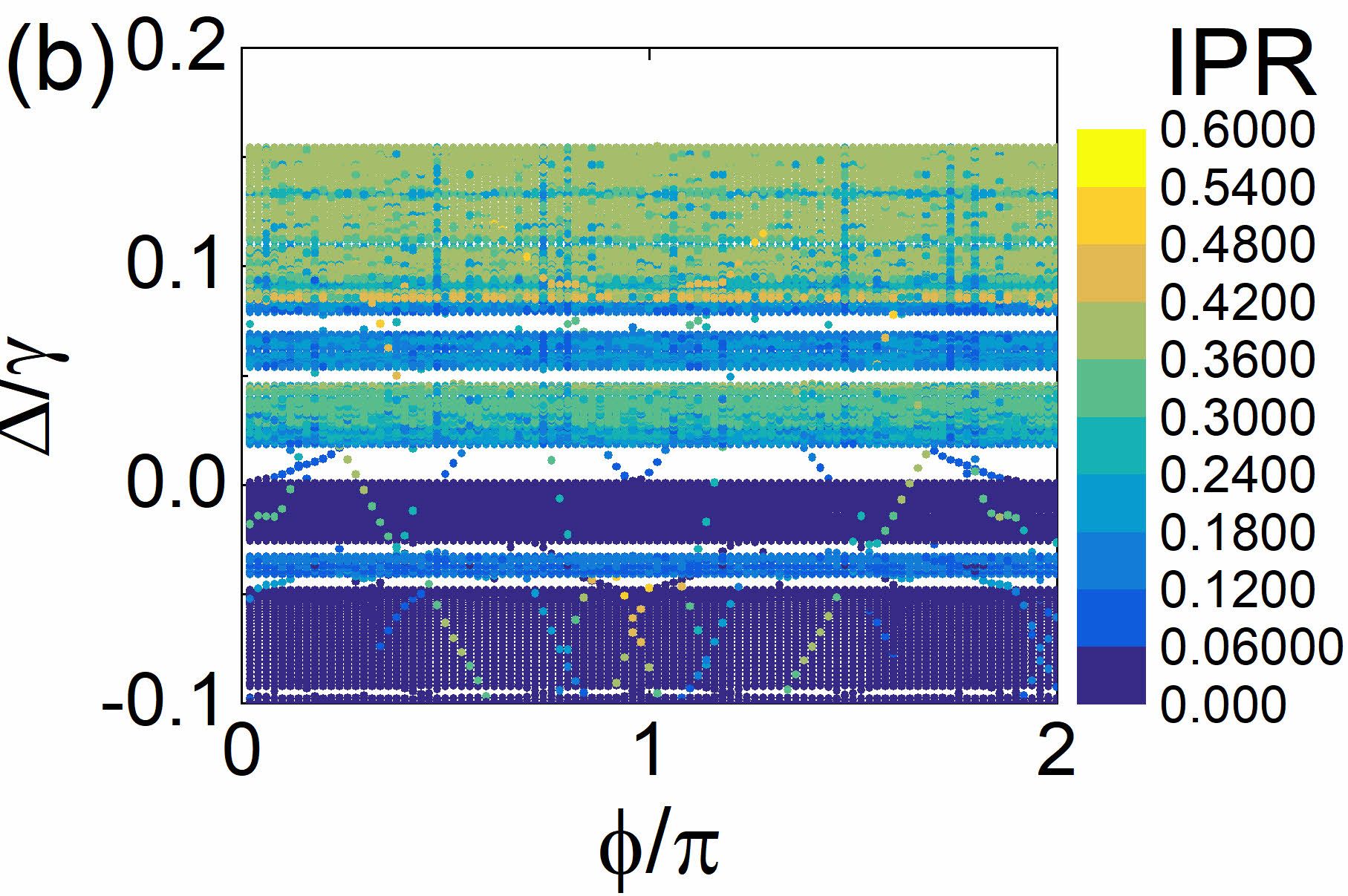}\label{incommd1eta03upperband}
	}\\
	\subfloat{
		\includegraphics[width=0.4\linewidth]{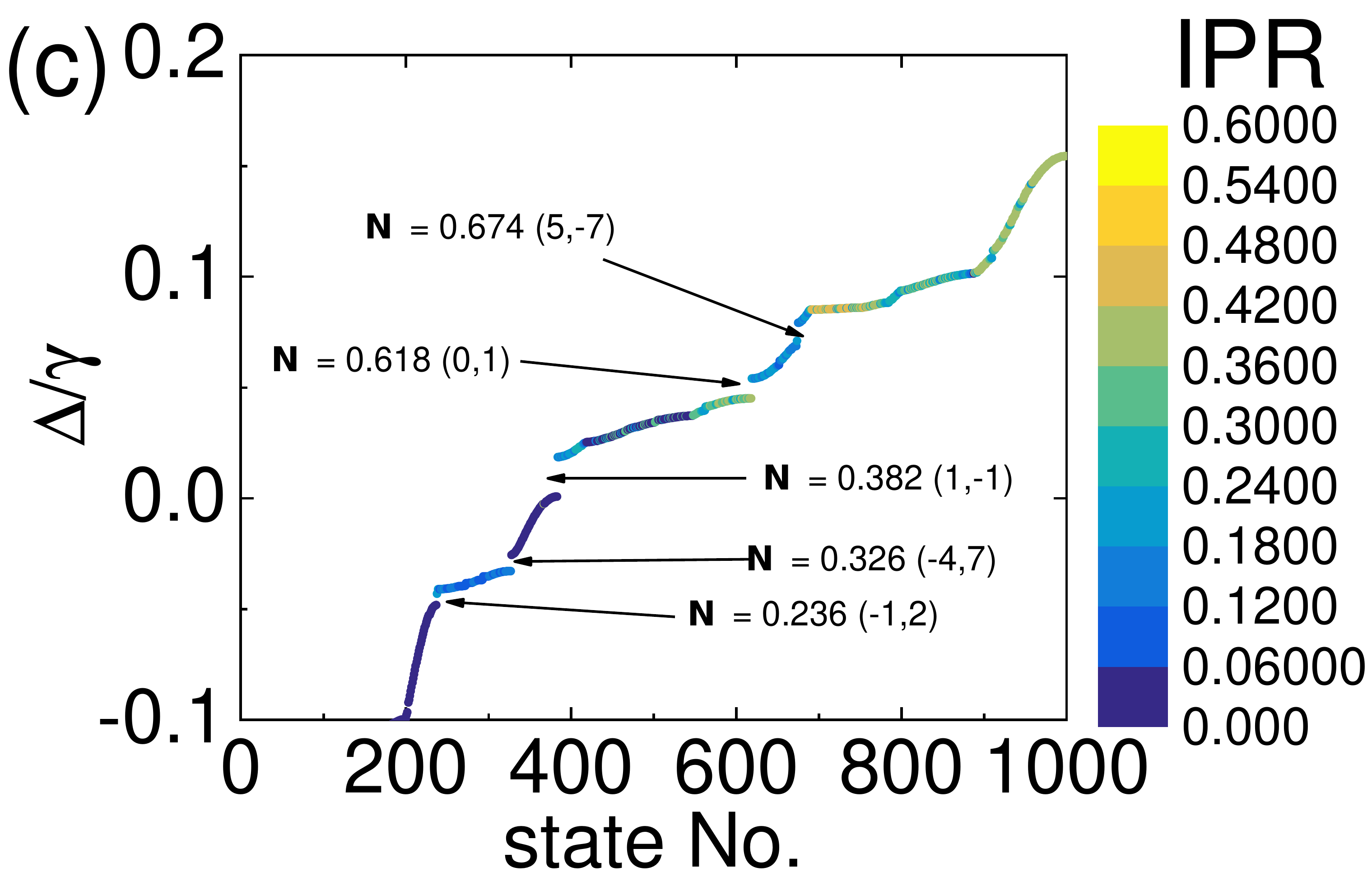}\label{transincommd1eta03pi02}
	}
	\subfloat{
		\includegraphics[width=0.4\linewidth]{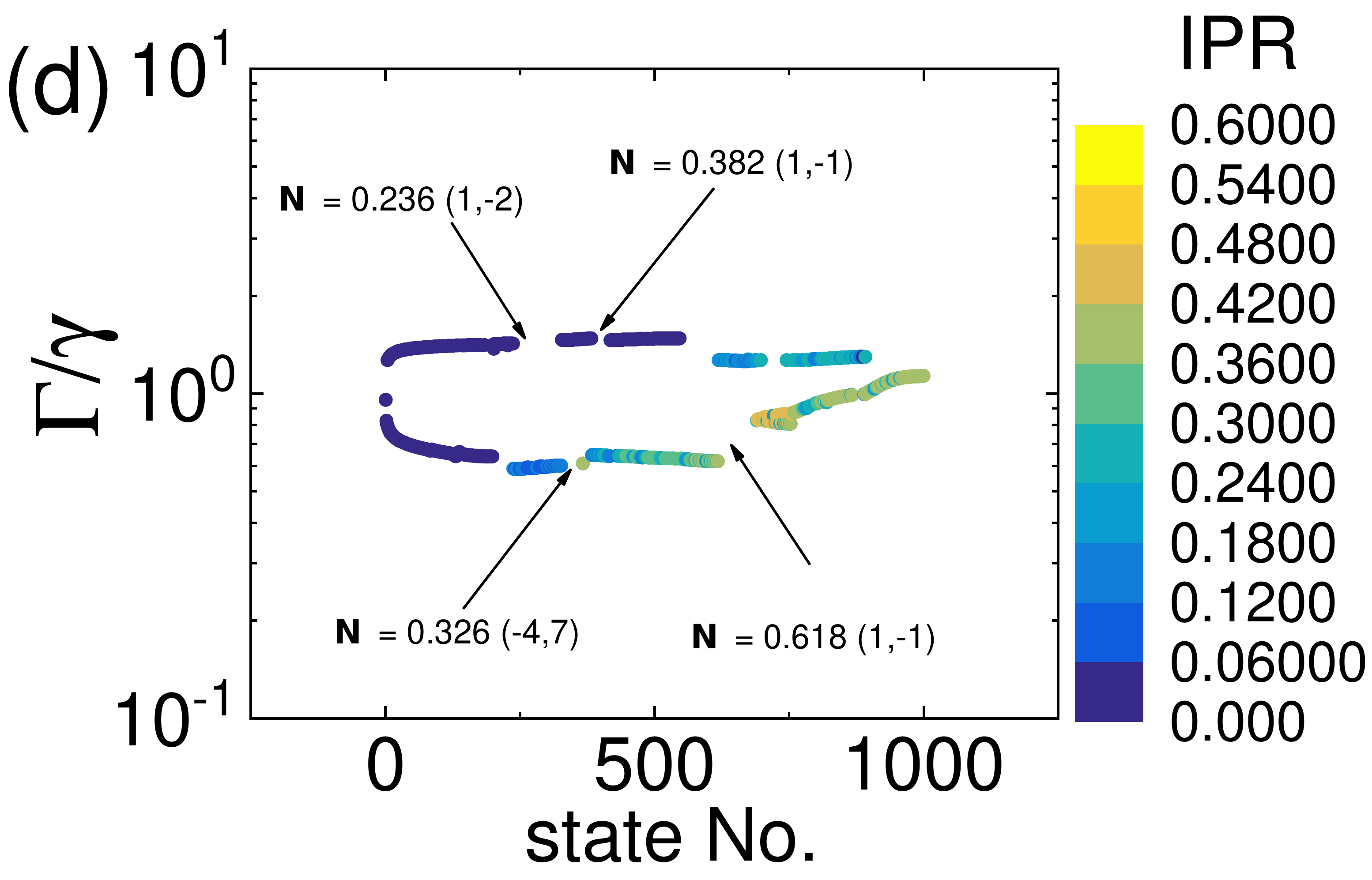}\label{transincommd1eta03pi02_imag}
	}
	\caption{The quasiperiodic lattice with $\beta=(\sqrt{5}-1)/2$ and $d=\lambda_0$. (a) Transverse band structure. (b) Enlarged display of the upper-in-frequency part ($\Delta/\gamma>-0.1$) of the transverse band structure shown in (a). (c-d) Real (c) and imaginary (d) parts of the transverse eigenstate spectrum for an arbitrarily chosen modulation phase ($\phi=0.2\pi$). Note in (c), the upper-in-frequency part of the band structure is given and in (d) the logarithmic scale is used for better discernibility.  Other system parameters are fixed as $\eta=0.3$ and $N=1000$.}
	\label{d1transband}
\end{figure*}
To further address the role of long-range dipole-dipole interactions, we increase the average lattice constant $d$ to a frequently-encountered situation where it is comparable to the wavelength of the atomic dipole transition, namely $d=\lambda_0$, as given in Fig. \ref{d1transband}.

As mentioned in the main text, in this situation, the contribution of long-range hoppings cannot be neglected compared with that of NN hoppings. In Fig. \ref{incommd1eta03}, the transverse band structure is plotted with its upper-in-frequency part presented in Fig. \ref{incommd1eta03upperband} ($\Delta>-0.1$) in an enlarged fashion, from which midgap edge modes are clearly observed. Furthermore, the number of edge modes in a gap is consistent with the gap Chern number obtained from the gap-labeling theorem. More specifically, the eigenstate distribution at $\phi=0.2\pi$ is given in Fig. \ref{transincommd1eta03pi02}, where the normalized IDOS $\mathcal{N}$ and topological integers $(\mu,\nu)$ are labeled for several typical gaps. In addition, the imaginary part of the band structure is also provided in the logarithmic scale with several topological gaps identified in Fig. \ref{transincommd1eta03pi02_imag}, similar to those in Fig. \ref{imagpart}. Therefore, it can be asserted that these relatively significant long-range hoppings do not alter the topological properties of the present system qualitatively.

However, we have to say that in such large interatomic distances, the band gap becomes very small, making edge states hardly observable in experiment. As a result, in order to experimentally detect the topological edge states, small interatomic distances are desirable. This can be achieved by using the state-of-art optical lattice technologies as discussed in the main text. 

\bibliography{aah_model}
\end{document}